\documentclass[universe,review,submit,pdftex,moreauthors]{Definitions/mdpi}

\firstpage{1} 
\pubvolume{1}
\issuenum{1}
\articlenumber{0}
\pubyear{2023}
\copyrightyear{2023}
\datereceived{ } 
\daterevised{ } % Comment out if no revised date
\dateaccepted{ } 
\datepublished{ } 
\hreflink{https://doi.org/} % If needed use \linebreak
\def\aap{A\&A}
\def\apj{ApJ}

\def\mnras{MNRAS}
\def\apjl{ApJ}

\def\aj{AJ}
\def\prd{PRD}

\def\jcap{JCAP}

\def\physrep{Phys. Rep.}
\def\pasa{PASA}

\def\prd{Phys. Rev. D}
\def\prl{Phys. Rev. Lett.}

\def\nat{Nature}
\def\araa{ARAA}

\Title{Fermionic dark matter: physics, astrophysics, and cosmology}

\TitleCitation{Fermionic dark matter}

\Author{C. R. Arg\"uelles $^{1,2,3,\dagger,*}$\orcidA{}, E. A. Becerra-Vergara $^{2,3,4,\dagger,*}$\orcidB{}, J. A. Rueda $^{2,3,5,6,7\dagger,*}$\orcidC{} and R. Ruffini$^{2,3,8\dagger,*}$\orcidD{}}

\AuthorNames{C. R. Argu\"uelles, E. A. Becerra-Vergara, J. A. Rueda and R. Ruffini}

\AuthorCitation{Argu\"uelles, C. R.; Becerra-Vergara, E. A.; Rueda, J. A.; Ruffini, R.}
\address{%
$^{1}$ \quad Instituto de Astrof\'isica de La Plata, UNLP-CONICET, Paseo del Bosque s/n B1900FWA La Plata, Argentina\\
$^{2}$ \quad ICRANet, Piazza della Repubblica 10, I-65122 Pescara, Italy\\
$^{3}$ \quad ICRA, Dip. di Fisica, Sapienza Universit\`a di Roma, P.le Aldo Moro 5, I-00185 Rome, Italy\\
$^{4}$ \quad  FIELDS, Departamento de Ciencias Básicas, Universidad de Investigación y Desarrollo, 680002 Bucaramanga, Santander, Colombia\\
$^{5}$ \quad ICRANet-Ferrara, Dip. di Fisica e Scienze della Terra, Universit\`a degli Studi di Ferrara, Via Saragat 1, I-44122 Ferrara, Italy\\
$^{6}$ \quad Dip. di Fisica e Scienze della Terra, Universit\`a degli Studi di Ferrara, Via Saragat 1, I-44122 Ferrara, Italy\\
$^{7}$ \quad INAF, Istituto de Astrofisica e Planetologia Spaziali, Via Fosso del Cavaliere 100, I-00133 Rome, Italy\\
$^{8}$ \quad INAF, Viale del Parco Mellini 84, I-00136 Rome, Italy}

\corres{Correspondence: carguelles@fcaglp.unlp.edu.ar (C.R.A.); jorge.rueda@icra.it (J.A.R.); eduar.becerra@icranet.org (E.A.B-V.); ruffini@icra.it (R.R.) }

\firstnote{These authors contributed equally to this work.}
\abstract{The nature of dark matter (DM) is one of the most relevant questions in modern astrophysics. We present a brief overview of recent results that inquire into a possible fermionic quantum nature of the DM particles, focusing mainly on the interconnection between the microphysics of the neutral fermions {and the macrophysical structure of galactic halos, including their formation both in the linear and non-linear cosmological regimes. We discuss the general relativistic Ruffini-Arg\"uelles-Rueda (RAR) model of fermionic DM in galaxies, its applications to the Milky Way, the possibility that the Galactic center harbors a DM core instead of a supermassive black hole (SMBH), the S-cluster stellar orbits with an in-depth analysis of the S2's orbit including precession, the application of the RAR model to other galaxy types (dwarf, elliptic, big elliptic and galaxy clusters), and universal galaxy relations. All the above focusing on the model parameters constraints, most relevant to the fermion mass. We also connect the RAR model fermions with particle physics DM candidates, self-interactions, and galactic observables constraints. The formation and stability of core-halo galactic structures predicted by the RAR model and their relation to warm DM cosmologies are also treated. Finally, we briefly discuss how gravitational lensing, dynamical friction, and the formation of SMBHs can also probe the DM nature.}
}


\begin{document}

%%%%%%%%%%%%%%%%%%%%%%%%%%%%%%%%%%%%%%%%%%%%%%%%%%%%%%%%%%%%%
%%%%%%%%%%%%%%%%%%%%%%%%%%%%%%%%%%%%%%%%%%%%%%%%%%%%%%%%%%%%%
\section{Introduction}
%%%%%%%%%%%%%%%%%%%%%%%%%%%%%%%%%%%%%%%%%%%%%%%%%%%%%%%%%%%%%%%%%%%%%%%%%%%%%%%%%%%%%%%%%%%%%%%%%%%%%%%%%%%%%%%%%%%%%%%%%%

The main evidence for the existence of dark matter (DM) is implied by its gravitational effects in a plethora of astrophysical and cosmological environments, including the cosmic microwave background (CMB), baryon acoustic oscillations (BAO), galactic structures (its formation, evolution, and morphology), gravitational lensing, stellar streams, and many others. However, understanding its nature and precise overall mass distribution in galactic structures within a particle DM paradigm are still open questions.

Several attempts have been made to explain this phenomenon through ordinary matter, including active neutrinos \cite{1980PhRvL..45.1980B, 1984Natur.310..637H} or macroscopic objects such as MACHOS \cite{2000ApJ...542..281A}. However, a microscopic origin of the DM particles regarding a new particle species not included in the Standard Model remains the most likely hypothesis \cite{2005PhR...405..279B, 2018Natur.562...51B}.
  
More recent cosmological observations obtained in the last three decades have favored the adoption of the $\Lambda$CDM paradigm \cite{1999Sci...284.1481B, 2020PhRvD.101h3504I}: in the standard scheme, the DM is assumed to be produced in thermal equilibrium via weak interactions with the primordial plasma and modeled as collisionless after its decoupling from the other particle species. In this scenario, the DM decoupling is assumed to occur at a temperature smaller than the DM rest mass, so the distribution corresponds to non-relativistic particles. The traditional DM candidate within this paradigm is the so-called weakly interacting massive particle (WIMP), typically a heavy neutral lepton with masses around $100$ GeV/c$^2$ \cite{2018Natur.562...51B}. Nonetheless, many other DM candidates exist either of bosonic or fermionic nature, such as axion-like particles or sterile neutrinos, respectively, with somewhat different early Universe decoupling regimes relative to WIMPs, though still in agreement with current cosmological observables (see, e.g., \cite{2016PhR...643....1M,2017JCAP...01..025A} for extensive reviews in the case of axions and sterile neutrinos respectively).

Besides their different early Universe physics, these different DM particles imply different outcomes in the non-linear regime of structure formation, which may allow favoring one candidate over the other. These non-linear processes typically correspond to the gravitational collapse of primordial self-gravitating DM structures such as DM halos, in which the quantum nature of the DM particles (either bosonic or fermionic) can cause distinguishable patterns which data can test. Their effects on the precise shape and stability of the DM density profiles, including the distinctive quantum effects (e.g., quantum pressure) through the central regions of the halos, may open new important avenues of research in the field. Indeed, in \cite{2014NatPh..10..496S}, it has been shown that these interesting \textit{quantum imprints} in the DM halos exist for bosonic DM, and in \cite{2021MNRAS.502.4227A} for fermionic DM.

A key motivation to include the quantum nature of the particles in the study of DM halos is the limited inner spatial resolution obtained within cosmological N-body simulations for such formed structures. It encompasses huge uncertainties in (1) the central DM mass distributions; (2) the relation/effects with the supermassive black hole (SMBH) at the center of large galaxies; and (3) the relationship with the baryonic matter on inner-intermediate halo scales, among others. Indeed, using classical (rather than quantum) massive pseudo-particles as the matter building blocks in N-body simulations within the $\Lambda$CDM cosmology does not allow testing quantum pressure effects in DM halos.

Increasing attention has been given in the last decade to the study of DM halos in terms of quantum particles, given they may alleviate/resolve many of the drawbacks still present in the traditional CDM paradigm on small-scales (i.e., typically below $\sim 10$ kpc scales). The bulk of these models is comprised of the following three categories: 

(i) Ultralight bosons with masses\footnote{Particle masses can be considerably larger up to $\sim 10^{-3}$ eV if self-interactions among the bosons are allowed \cite{2017PhRvD..95f3515S,2021PhRvD.103l3551C}.} $m_bc^2\sim 1$--$10\times 10^{-22}$ eV, known as ultralight DM, Fuzzy DM, or even scalar field DM \cite{1969PhRv..187.1767R, 1983PhLB..122..221B, 1994PhRvD..50.3650S, 2000PhRvL..85.1158H, 2001PhRvD..63f3506M, 2012MNRAS.422..282R, 2017PhRvD..95f3515S, 2017PhRvD..95d3541H, 2018PhRvD..98h3027B, 2020MNRAS.494.2027M, 2021PhRvD.103l3551C}.

(ii) The case of fully-degenerate fermions (i.e., in the zero temperature approximation under the Thomas-Fermi approach) with masses $m_{\rm df}c^2\sim$ few $\times 10^2$ eV \cite{2013NewA...22...39D, 2015JCAP...01..002D, 2017MNRAS.467.1515R, 2022PhRvD.106d3538C}. Or the case of self-gravitating fermions but distributed in the opposite limit, that is, in the dilute regime (i.e., in Boltzmannian-like fashion) which, however, \textit{do not} imply an explicit particle mass dependence when contrasted with halo observables (see, e.g., \cite{2014MNRAS.442.2717D}).

(iii) The more general case of self-gravitating fermions in a semi-degenerate regime (i.e., at finite temperature), which can include both regimes in the same system: that is, to be highly degenerate in the center and more diluted in the outer region (see \cite{1990A&A...235....1G,1998MNRAS.300..981C,2002PrPNP..48..291B,2015PhRvD..92l3527C,2022PhRvD.106d3538C} for a list of generic works). Recently, the phenomenology of this theory to the study of DM in real galaxies (using specific boundary conditions from observations), was developed in full general relativity either including for escape of particles \cite{2018PDU....21...82A, 2019PDU....24100278, 2020A&A...641A..34B, 2021MNRAS.505L..64B, 2022MNRAS.511L..35A, Krut_2023} or not (\cite{2015MNRAS.451..622R}), and leading to particle masses in the range $m_fc^2\sim$ few $10$--$100$ keV. The latter model is usually referred to in the literature as the Ruffini-Arg\"uelles-Rueda (RAR) model (it has been sometimes called the relativistic fermionic-King model).

A relevant aspect of the above models is the DM particle mass dependence on the density profiles, which differs from  phenomenological profiles used to fit results from classical N-body numerical simulations. Moreover, these type of self-gravitating systems of DM opens the possibility of having access to the very nature, mass, and explicit dependence on their phase-space distributions at the onset of DM halo formation in real galaxies (see, e.g., ~\cite{2017PhRvD..95d3541H} and refs. therein, for bosons and, e.g.~\cite{2019PDU....24100278, 2021MNRAS.502.4227A, 2022PhRvD.106d3538C} and refs. therein, for fermions).

In this review, we will focus on fermionic candidates as described in (iii), describe the huge progress in the last decade in their theory and phenomenology, and comment on future perspectives. This work is organized as follows. In Section \ref{sec:2}, we recall the main features of the RAR model, the theoretical framework including the fermion equation of state (EOS), the general relativistic equations of equilibrium, and the properties of the solutions such as the DM density profile and galactic rotation curves. Section \ref{sec:3} is devoted to applying the RAR model to the description of the MW, so the constraint that the Galactic data of the rotation curves impose on the fermion mass. Special emphasis is also given to the high-quality data of the motion of the S-cluster stars in the proximity of Sgr A*. We outline in Section \ref{sec:4} the application of the RAR model to other galaxy types, specifically to dwarf spheroidal, spiral, and elliptical galaxies, as well as galaxy clusters. The extension of the model to various galaxies allows us to verify its agreement with universal observational correlations. Section \ref{sec:5} discusses this topic. Section \ref{sec:6} discusses the possibility of the sterile neutrinos being the massive fermions of the RAR model, including the constraints on possible fermion self-interactions. Section \ref{sec:7} discusses the formation and stability of the fermionic DM from a cosmological evolution viewpoint, e.g., the non-linear structure and the relation with warm DM cosmologies. Section \ref{sec:8} outlines some additional astrophysical systems and observations that can help to constrain DM models, e.g., the gravitational lensing, the DM dynamical friction on the motion of binaries, and the link between DM halos and SMBHs at the centers of active galaxies. Finally, Section \ref{sec:9} concludes.

%%%%%%%%%%%%%%%%%%%%%%%%%%%%%%%%%%%%%%%%%%%%%%%%%%%%%%%%%%%%%%%%%%%
%%%%%%%%%%%%%%%%%%%%%%%%%%%%%%%%%%%%%%%%%%%%%%%%%%%%%%%%%%%%%%%%%%%
\section{The RAR model: theoretical framework}\label{sec:2}
%%%%%%%%%%%%%%%%%%%%%%%%%%%%%%%%%%%%%%%%%%%%%%%%%%%%%%%%%%%%%%%%%%%
%%%%%%%%%%%%%%%%%%%%%%%%%%%%%%%%%%%%%%%%%%%%%%%%%%%%%%%%%%%%%%%%%%%

The RAR model was proposed to evaluate the possible manifestation of the \textit{dense quantum core - classical halo} distribution in the astrophysics of real galaxies, i.e., as a viable possibility to establish a link between the dark central cores to DM halos within a unified approach in terms of DM fermions \cite{2015MNRAS.451..622R}. The RAR model equilibrium equations consist of the Einstein equations in spherical symmetry for a perfect fluid energy-momentum tensor. The Fermi-Dirac statistics gives the pressure and density, while the closure relations are determined by the Klein and Tolman thermodynamic equilibrium conditions \cite{2015MNRAS.451..622R}. The solution to this system of equations leads to continuous and novel profiles for galactic dark matter halos, whose global morphology depends on the fermionic particle mass. Such morphology has a universal behavior of the type \emph{dense~core~-~dilute~halo} that extends from the center to the galactic halo, which allows providing solutions to various tensions faced by standard cosmological paradigms on galactic scales. The outermost part of such distributions makes it possible to explain the galactic rotation curves (in a similar way as traditional dark matter profiles do), while their central morphology is characterized by high concentrations of semi-degenerate fermions (due to the Pauli exclusion principle) with important astrophysical consequences for galactic nuclei (see \cite{2015ARep...59..656S,2016JCAP...04..038A,2017IJMPD..2630007M} for its applications). Similar \emph{core-halo} profiles with applications to fermionic DM were obtained in \cite{2002PrPNP..48..291B} and more recently in \cite{2015PhRvD..92l3527C} from a statistical approach within Newtonian gravity. 

The above corresponds to the original version of the RAR model, with a unique family of density profile solutions that behaves as $\rho(r) \propto r^{-2}$ at large radial distances from the center. This treatment was extended in \cite{2018PDU....21...82A} by introducing a cutoff in momentum space in the distribution function (DF)  (i.e., accounting for particle-escape effects) that allows defining the galaxy border. The extended RAR model conceives the DM in galaxies as a general relativistic self-gravitating system of massive fermions (spin $1/2$) in hydrostatic and thermodynamic equilibrium. Following \cite{2018PDU....21...82A}, we solve the Tolman–Oppenheimer–Volkoff (TOV) equations using an equation of state (EOS) that takes into account (i) relativistic effects of the fermionic constituents, (ii) finite temperature effects, and (iii) particle escape effects at large momentum ($p$) through a cutoff in the Fermi-Dirac distribution $f_c$, 

\begin{equation}
f_c(\epsilon\leq\epsilon_c) = \frac{1-e^{(\epsilon-\epsilon_c)/k_B T}}{e^{(\epsilon-\mu)/k_B T}+1}, \qquad f_c(\epsilon>\epsilon_c)=
0\, ,
\label{fcDF}
\end{equation}

\noindent where $\epsilon=\sqrt{c^2 p^2+m^2 c^4}-mc^2$ is the particle kinetic energy, $\epsilon_c$ is the cutoff particle energy, $\mu$ is the chemical potential from which the particle rest-energy is subtracted, $T$ is the temperature, $k_B$ is the Boltzmann constant, $c$ is the speed of light, and $m$ is the fermion mass. 

{This extended version of the RAR model differs from the original one presented in \cite{2015MNRAS.451..622R} only in condition (iii). The inclusion of the cutoff parameter allows a more realistic description of galaxies since it allows us (1) to model their finite size and (2) to apply relaxation mechanisms that realize in nature the Fermi-Dirac distribution in Eq. (\ref{fcDF}), where the entropy can reach a maximum, unlike the original RAR model with no particle escape condition, i.e., $\epsilon_c \rightarrow \infty$ (see \cite{2015PhRvD..92l3527C} for a discussion). In fact,} it is important to highlight that quantum phase-space distribution (Eq.~\ref{fcDF}) can be obtained from a maximum entropy production principle \cite{1998MNRAS.300..981C}. It has been shown that there is a stationary solution of a generalized Fokker-Planck equation for fermions that includes the physics of violent relaxation and evaporation, appropriate within the non-linear stages of galactic DM halo structure formation \cite{1998MNRAS.300..981C}. The full set of dimensionless parameters of the model are 

\begin{equation}
    \beta = k_B T/(mc^2) , \qquad \theta = \mu/(k_BT) , \qquad W = \epsilon_c/(k_BT),
\end{equation}

\noindent where $\beta$, $\theta$, and $W$ are the temperature, degeneracy, and cutoff parameters, respectively. We do not consider anti-fermions because the temperature fulfills $T \ll mc^2/k_B$.

The stress-energy tensor is that of fermionic gas modeled as a perfect fluid, 

\begin{equation}
    T^{\alpha\nu} = \mathrm{diag}\left[c^2\rho(r),P(r),P(r),P(r)\right],
\end{equation}\

\noindent whose density and pressure are associated with the distribution function $f_c$, 

\begin{equation}\label{rho}
    \rho = m\frac{2}{h^3}\int_{0}^{\epsilon_c}f_c(p)\left(1+\frac{\epsilon(p)}{mc^2}\right)d^3p\ ,
\end{equation}

\begin{equation}\label{P}
    P = \frac{1}{3}\frac{4}{h^3}\int_{0}^{\epsilon_c}f_c(p)\,\epsilon\,\frac{1+\epsilon(p)/2mc^2}{1+\epsilon(p)/mc^2}d^3p,
\end{equation}

\noindent where the integration is carried out over the momentum space bounded by $\epsilon\leq\epsilon_c$.

The system is considered to be spherically symmetric, so we adopt the metric

\begin{equation}
        \label{eqn:metric}
        ds^2 = e^{\nu(r)}{\rm d}t^2 - e^{\lambda(r)}{\rm d}r^2 - r^2 {\rm d}\theta^2 + r^2 \sin^2{\theta} {\rm d}\phi^2,
\end{equation}

\noindent where ($r$,$\theta$,$\phi$) are the spherical coordinates, the metric functions $e^{\nu(r)}$ and $e^{-\lambda(r)} = 1- 2 M(r)/r$ are functions of $r$, and $M(r)$ is the mass function. The \citet{1930PhRv...35..904T} and \citet{1949RvMP...21..531K} thermodynamic equilibrium conditions, as well as the cutoff \cite{1989A&A...221....4M} condition obtained from energy conservation along a geodesic, can be written as 

\begin{subequations}
\begin{align}
    e^{\nu/2} T &= {\rm constant},\label{eq:cond1}\\
    e^{\nu/2}(\mu+m c^2) &= {\rm constant},\label{eq:cond2}\\
    e^{\nu/2}(\epsilon+m c^2) &=  {\rm constant}.\label{eq:cond3}
\end{align}
\end{subequations}

We set the constants on the right-hand side of Eqs. (\ref{eq:cond1})--(\ref{eq:cond3}) by evaluating the equations at the boundary radius, say $r_b$. For instance, the escape energy condition (\ref{eq:cond3}) becomes

\begin{equation}\label{eq:cond4}
    \left(1+W\beta\right) = e^{(\nu_b - \nu)/2},
\end{equation}

\noindent being $\nu_b \equiv \nu(r_b)$, $W(r_b) = \epsilon_c (r_b) = 0$ \cite{1989A&A...221....4M}. {This condition approaches the Newtonian escape velocity condition, $v^2_e = - 2 \Phi$, in the weak-field, non-relativistic limit, $c\to \infty$ and $e^{\nu/2} \approx 1 + \Phi/c^2$, where $\Phi$ is the Newtonian gravitational potential, and  setting $\Phi(r_b) = 0$.}

Therefore, the Einstein equations together with the equilibrium conditions (\ref{eq:cond1})--(\ref{eq:cond4}) form the following system of coupled non-linear ordinary integrodifferential equations

\begin{subequations}\label{RAR-eqtns}
\begin{align}
        \frac{d\hat M}{d\hat r}&=4\pi\hat r^2\hat\rho, \label{eq:eqs1}\\ 
        \frac{d\theta}{d\hat r}&=-\frac{1-\beta_0(\theta-\theta_0)}{\beta_0}
    \frac{\hat M+4\pi\hat P\hat r^3}{\hat r^2(1-2\hat M/\hat r)},\label{eq:eqs2}\\
    \frac{d\nu}{d\hat r}&=\frac{2(\hat M+4\pi\hat P\hat r^3)}{\hat r^2(1-2\hat M/\hat r)},\label{eq:eqs3} \\
    \beta(\hat r)&=\beta_0 e^{\frac{\nu_0-\nu(\hat r)}{2}}, \\
    W(\hat r)&=W_0+\theta(\hat r)-\theta_0\, ,\label{eq:Cutoff}
\end{align}
\end{subequations}

\noindent where the subscript `$0$' stands for variable evaluated at $r=0$. We have introduced dimensionless quantities: $\hat r=r/\chi$, $\hat M=G M/(c^2\chi)$, $\hat\rho=G \chi^2\rho/c^2$, $\hat P=G \chi^2 P/c^4$, with $\chi=2\pi^{3/2}(m_{\rm Pl}/m)\hbar/(mc)$, being $m_{\rm Pl}=\sqrt{\hbar c/G}$ the Planck mass. Equations~(\ref{eq:eqs1}) and (\ref{eq:eqs2}) are the relevant Einstein equations, Eq.~(\ref{eq:eqs3}) is a convenient combination of the Klein and Tolman relations for the gradient of $\theta = \mu/(k_BT)$, and Eq.~(\ref{eq:Cutoff}) is a direct combination of the Klein and cutoff energy conditions. Note that in the limit $W \rightarrow \infty$ (no particle escape: $\epsilon_c \rightarrow \infty$), the above equation system reduces to the equations obtained in the original RAR model \cite{2015MNRAS.451..622R}.

{
\citet{2018PDU....21...82A} have shown that including the cutoff parameter, besides its theoretical relevance, allows handling more stringent outer halo constraints and leads to central
cores with higher compactness compared to the ones produced by the unbounded solutions of the original RAR model \cite{2015MNRAS.451..622R}.}

%%%%%%%%%%%%%%%%%%%%%%%%%%%%%%%%%%%%%%%%%%%%%%%%%%%%%%%%%%%%%%%%%%%
%%%%%%%%%%%%%%%%%%%%%%%%%%%%%%%%%%%%%%%%%%%%%%%%%%%%%%%%%%%%%%%%%%%
\section{Constraints on fermionic DM from the Milky Way}
\label{sec:3}
%%%%%%%%%%%%%%%%%%%%%%%%%%%%%%%%%%%%%%%%%%%%%%%%%%%%%%%%%%%%%%%%%%%
%%%%%%%%%%%%%%%%%%%%%%%%%%%%%%%%%%%%%%%%%%%%%%%%%%%%%%%%%%%%%%%%%%%

Therefore, in the RAR model, the distribution of DM in the galaxy is calculated self-consistently by solving the Einstein equations, subjected to thermodynamic equilibrium conditions. The system of equations must be subjected to boundary conditions to satisfy galactic observables. In this section, we summarize the most relevant example, the case of the Milky Way (MW).

%%%%%%%%%%%%%%%%%%%%%%%%%%%%%%%%%%%%%%%%%%%%%%%%%%%%%%%%%%%%%%%%%%%
\subsection{The Milky Way rotation curves}
%%%%%%%%%%%%%%%%%%%%%%%%%%%%%%%%%%%%%%%%%%%%%%%%%%%%%%%%%%%%%%%%%%%

Thanks to the vast amount of rotation curve data (from the inner bulge to outer halo) \cite{2013PASJ...65..118S}, our Galaxy is the ideal scenario to test the RAR model. The observational data used in \cite{2013PASJ...65..118S} varies from a few pc to a few $100$ kpc, covering various orders of magnitude of the radial extent with different baryonic and dark mass structures, and can go down to the $\sim 10^{-4}$ pc scale when including the S-cluster stars \cite{2017ApJ...837...30G}. To fit the observed rotation curve, and according to \cite{2013PASJ...65..118S}, the following matter components of the MW must be assumed:

\renewcommand{\theenumi}{\roman{enumi})}
\begin{enumerate}
    \item A central region ($r \sim 10^{-4}$--$2$ pc) of young stars and molecular gas whose dynamic is dictated by a dark and compact object centered at Sgr~A*. 
    
    \item An intermediate spheroidal bulge structure ($r \sim 2$--$10^3$ pc) composed mostly of older stars, with inner and main mass distributions explained by the exponential spheroid model.
    
    \item An extended flat disk ($r \sim 10^3$--$10^{4}$ pc) including star-forming regions, dust, and gas, whose surface mass density is described by an exponential law.
    
    \item A spherical halo ($r \sim 10^{4}$--$10^5$ pc) dominated by DM, followed by a decreasing density tail with a slope steeper than $r^{-2}$.
\end{enumerate}
 
\begin{figure}[H]
    \centering
        \includegraphics[width=0.49\hsize,clip]{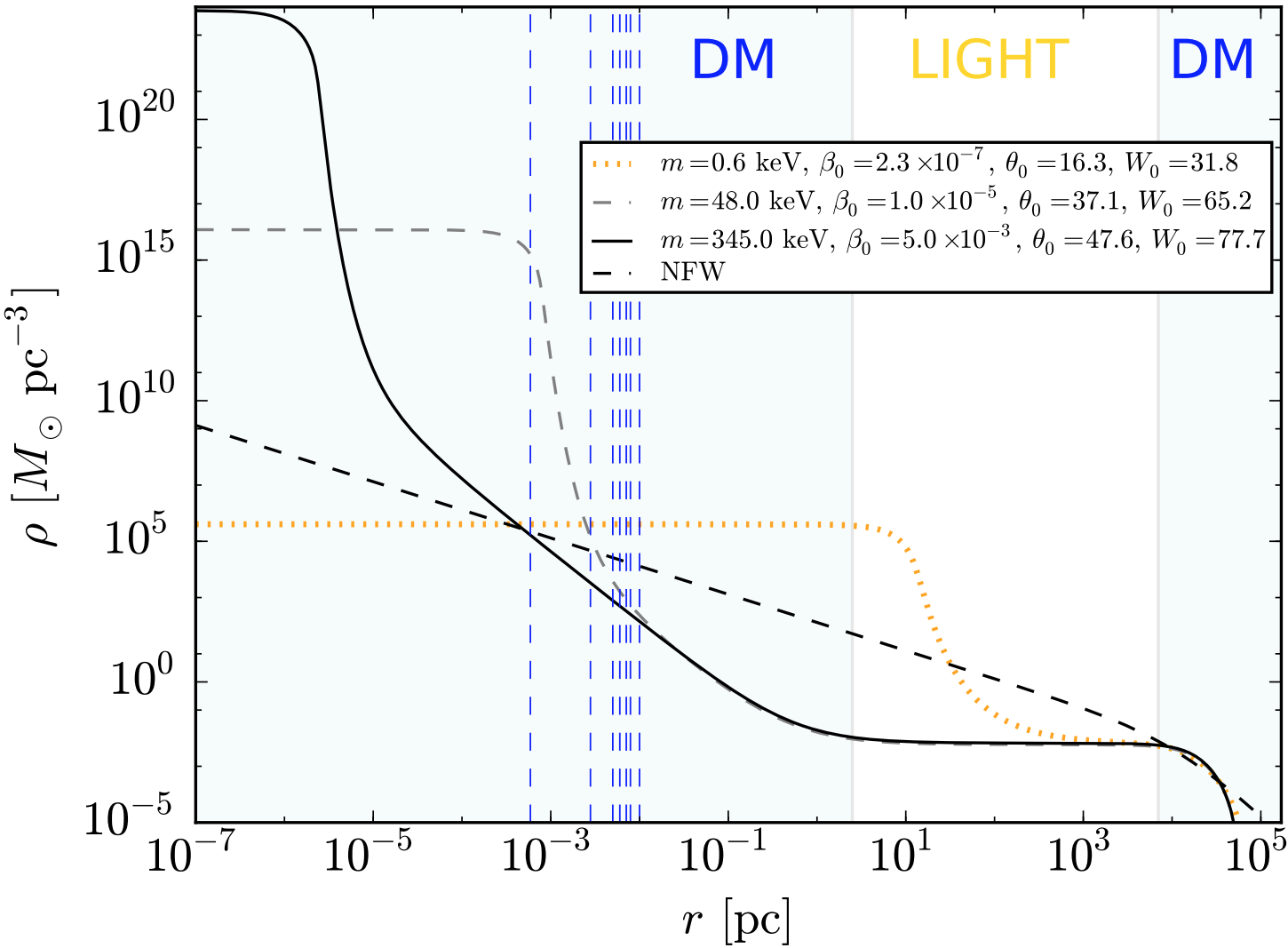}
        \includegraphics[width=0.49\hsize,clip]{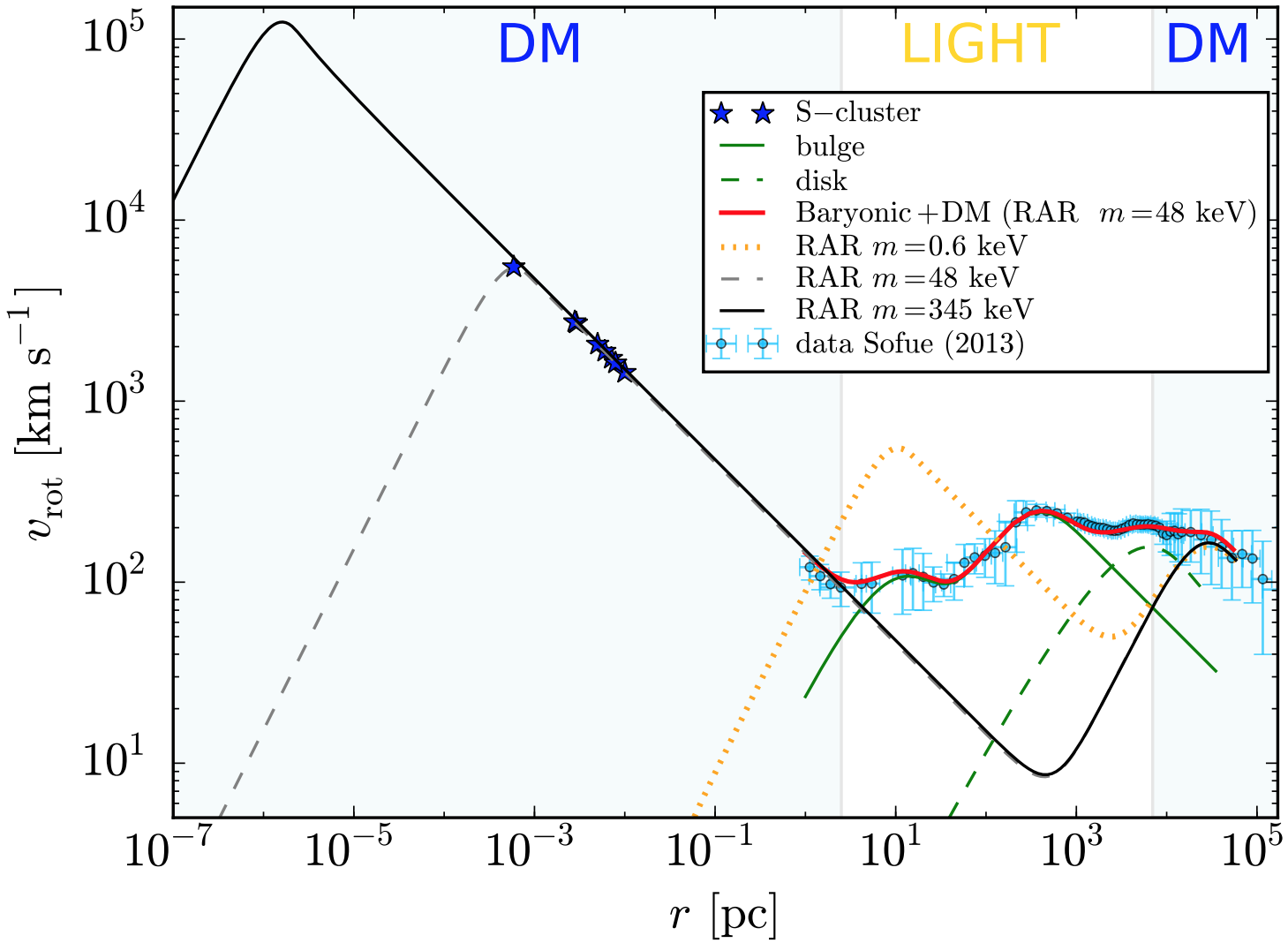}
        \caption{DM density profiles (left) and rotation curves (right), from $10^{-7}$ pc to $10^{5}$ pc, for three representative fermion masses $m c^2 = 0.6$, $48$, and $345$ keV. The vertical dashed blue lines in the left plot indicate the position of the S-cluster stars \cite{2009ApJ...707L.114G}, and the dashed black curve represents the NFW density profile obtained in \citet{2013PASJ...65..118S}. {Reprinted from \cite{2018PDU....21...82A}, Copyright (2018), with permission from Elsevier}.}\label{fig:density_vs_r}
\end{figure}

\begin{figure}[H]
    \centering
        \includegraphics[width=0.6\hsize,clip]{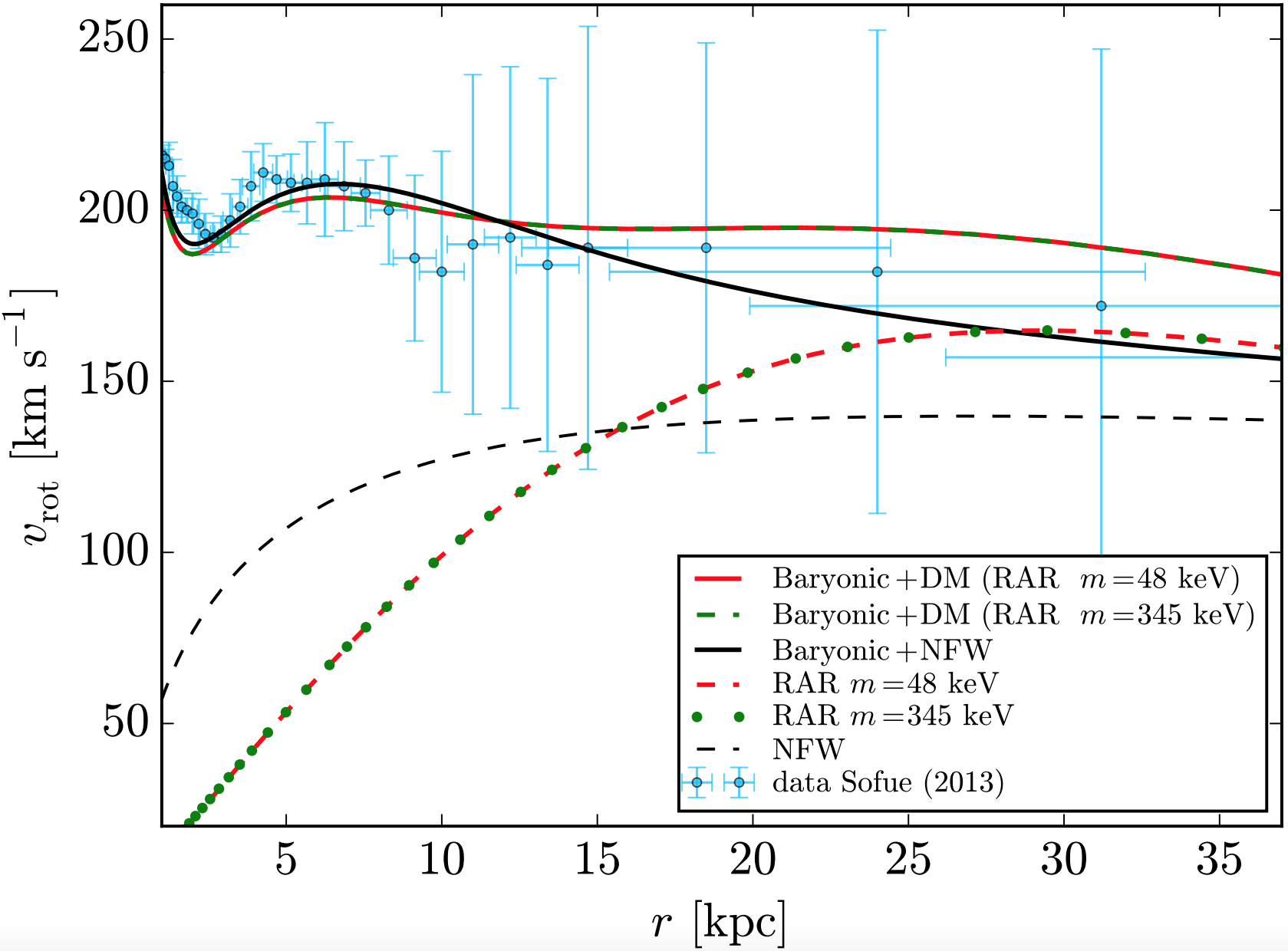}
        \caption{Zoom of the most relevant curves in the $1$--$35$ kpc region and in the linear scale of the plot in the right panel of Fig. \ref{fig:density_vs_r}. The graphic allows us to appreciate better the difference between the diverse DM models in the radial window where the Rotation Curve is most relevant. {Reprinted from \cite{2018PDU....21...82A}, Copyright (2018), with permission from Elsevier}.}\label{fig:zoom_vrot_vs_r}
\end{figure}

Considering the previous mass distributions for the Galaxy, the extended RAR model was successfully applied to explain the MW rotation curve as shown in Figs. \ref{fig:density_vs_r} and \ref{fig:zoom_vrot_vs_r}, implying a more general \emph{dense~core~--~diluted~halo} behavior for the DM distribution:
\begin{itemize}
\item
A DM core with radius $r_c$ (defined at the first maximum of the twice-peaked rotation curve), whose value is shown to be inversely proportional to the particle mass $m$, in which the density is nearly uniform. This central core is supported against gravity by the fermion degeneracy pressure, and general relativistic effects are appreciable.
\item
Then, there is an intermediate region characterized by a sharply decreasing density where quantum corrections are still important, followed by an extended and diluted plateau. This region extends until the halo scale-length $r_h$ is achieved (defined at the second maximum of the rotation curve).
\item
Finally, the DM density reaches a Boltzmann regime supported by thermal pressure with negligible general relativistic effects. It shows a behavior $\rho\propto r^{-n}$ with $n>2$ that is due to the phase-space distribution cutoff\footnote{It was recently shown the full possible range of density tail slopes within the RAR model when applied to typical rotation-supported galaxies: they can go from polytropic-like ($n=5/2$) to power law-like ($n=3$), see right panel of Fig. \ref{fig:profile-illustration-mep} and \cite{Krut_2023}.}. This leads to a DM halo bounded in radius (i.e., $\rho\approx 0$ occurs when the particle escape energy approaches zero).
\end{itemize}

The different regimes in the density profiles are also revealed in the DM rotation curve showing (see right panel of Fig. \ref{fig:density_vs_r}): 

\begin{itemize} 
\item A linearly increasing circular velocity $v \propto r$ reaching a first maximum at the quantum core radius $r_c$.  

\item a Keplerian power law, $v \propto r^{-1/2}$, with decreasing behavior representing the transition from quantum degeneracy to the dilute regime. After a minimum, highlighting the plateau, the circular velocity follows a linear trend until reaching the second maximum, which is adopted as the one-halo scale length in the fermionic DM model.

\item A decreasing behavior consistent with the power-law density tail $\rho \propto r^{-n}$ due to the cutoff constraint. 
\end{itemize}\

The Galaxy data analysis allowed us to rule out the fermion mass range $m c^2 < 10$ keV because the corresponding rotation curve starts to exceed the total velocity observed in the baryon-dominated region. On the other hand, by focusing only on the quantum core, it was possible to derive constraints that further limit the allowed fermion mass to $m c^2\approx 48$--$345$ keV. The dynamics of the S-cluster give the mass a lower bound. The S-stars analysis made through a simplified circular velocity analysis in general relativity showed that the fermion mass should be  $m c^2 \geq 48$ keV. Namely, the quantum core radius of the solutions for $m c^2 < 48$ keV are always greater than the radius of the S2 star pericenter, i.e.,  $r_c > r_{p_{(S2)}} = 6\times 10^{-4}$ pc, which rules out fermions lighter than $48$ keV. The mass upper bound of $m c^2=345$ keV corresponds to the last stable configuration before reaching the critical mass for gravitational collapse, $M^{\rm cr}_c \propto m^3_{\rm Pl}/m^{2}$  \cite{2018PDU....21...82A, 2014IJMPD..2342020A, 2014JKPS...65..809A}.

As was explicitly shown in \cite{2019IJMPD..2843003A, 2019PDU....24100278, 2018PDU....21...82A}, this new type of \emph{dense~core~--~diluted~halo} density profile suggests that the DM might explain the mass of the dark compact object in Sgr~A* as well as the halo mass. It applies not only to the MW but also to other galactic structures, from dwarfs and ellipticals to galaxy clusters \cite{2019PDU....24100278}. Specifically, an MW analysis \cite{2018PDU....21...82A} has shown that this DM profile can indeed explain the dynamics of the closest S-cluster stars (including S2) around Sgr~A*and to the halo rotation curve without changing the baryonic bulge-disk components (see Figs. \ref{fig:density_vs_r} and \ref{fig:zoom_vrot_vs_r}). Therefore, for a fermion mass between $48$--$345$ keV, the RAR solutions explain the Galactic DM halo and, at the same time, provide a natural alternative to the central BH scenario.

%

%%%%%%%%%%%%%%%%%%%%%%%%%%%%%%%%%%%%%%%%%%%%%%%%%%%%%%%%%%%%%%%%%%
\subsection{The orbits of S2 and G2}
%%%%%%%%%%%%%%%%%%%%%%%%%%%%%%%%%%%%%%%%%%%%%%%%%%%%%%%%%%%%%%%%%%

The extensive and continuous monitoring of the closest stars to the Galactic center has produced, over decades, a large amount of high-quality data on their positions and velocities. The explanation of these data, especially the S2 star motion, reveals a compact source, Sagittarius A* (Sgr~A*), whose mass must be about $4\times 10^6~M_\odot$. This result has been the protagonist of the Nobel Prize in Physics 2020 to Reinhard Genzel and Andrea Ghez \textit{for the discovery of a supermassive compact object at the centre of our galaxy}. Traditionally, the nature of Sgr A* has been attributed to a supermassive black hole (SMBH), even though direct proof of its existence is absent. Further, recent data on the motion of the G2 cloud show that its post-peripassage velocity is lower than expected from a Keplerian orbit around the hypothesized SMBH. An attempt to overcome this difficulty has used a friction force, produced (arguably) by an accretion flow whose presence is also observationally unconfirmed. 

We have advanced in \citet{2018PDU....21...82A} and in  \citet{2020A&A...641A..34B,2021MNRAS.505L..64B,2021MNRAS.tmpL.114A} an alternative scenario that identifies the nature of the supermassive compact object in the MW center, with a highly concentrated core of DM made of fermions (referred from now on as \textit{darkinos}). The existence of a high-density core of DM at the center of galaxies had been demonstrated in \citet{2015MNRAS.451..622R}, where it was shown that \textit{core-halo} profiles are obtained from the RAR fermionic DM model. The DM galactic structure is calculated in the RAR model treating the \textit{darkinos} as a self-gravitating system at finite temperatures, in thermodynamic equilibrium, and in general relativity. It has already been shown that this model, for \textit{darkinos} of $48$--$345$~keV, successfully explains the observed halo rotation curves of the MW \citep{2018PDU....21...82A} and other galaxy types \citep{2019PDU....24100278}.

Therefore, since 2020 we move forward by performing first in \citet{2020A&A...641A..34B} and then in \citet{2021MNRAS.505L..64B,2021MNRAS.tmpL.114A,2022MNRAS.511L..35A}, observational tests of the theoretically predicted dense quantum core at the Galactic center within the DM-RAR model. Namely, to test whether the quantum core of \textit{darkinos} could work as an alternative to the SMBH scenario for SgrA*. For this task, the explanation of the multiyear accurate astrometric data of the S2 star around Sgr~A*, including the relativistic redshift that has recently been verified, is particularly important. Another relevant object is G2, whose most recent observational data, as we have recalled, challenge the scenario of an SMBH.

\begin{figure}[hbtp!]
\centering 
\includegraphics[width=0.8\hsize,height=0.8\hsize]{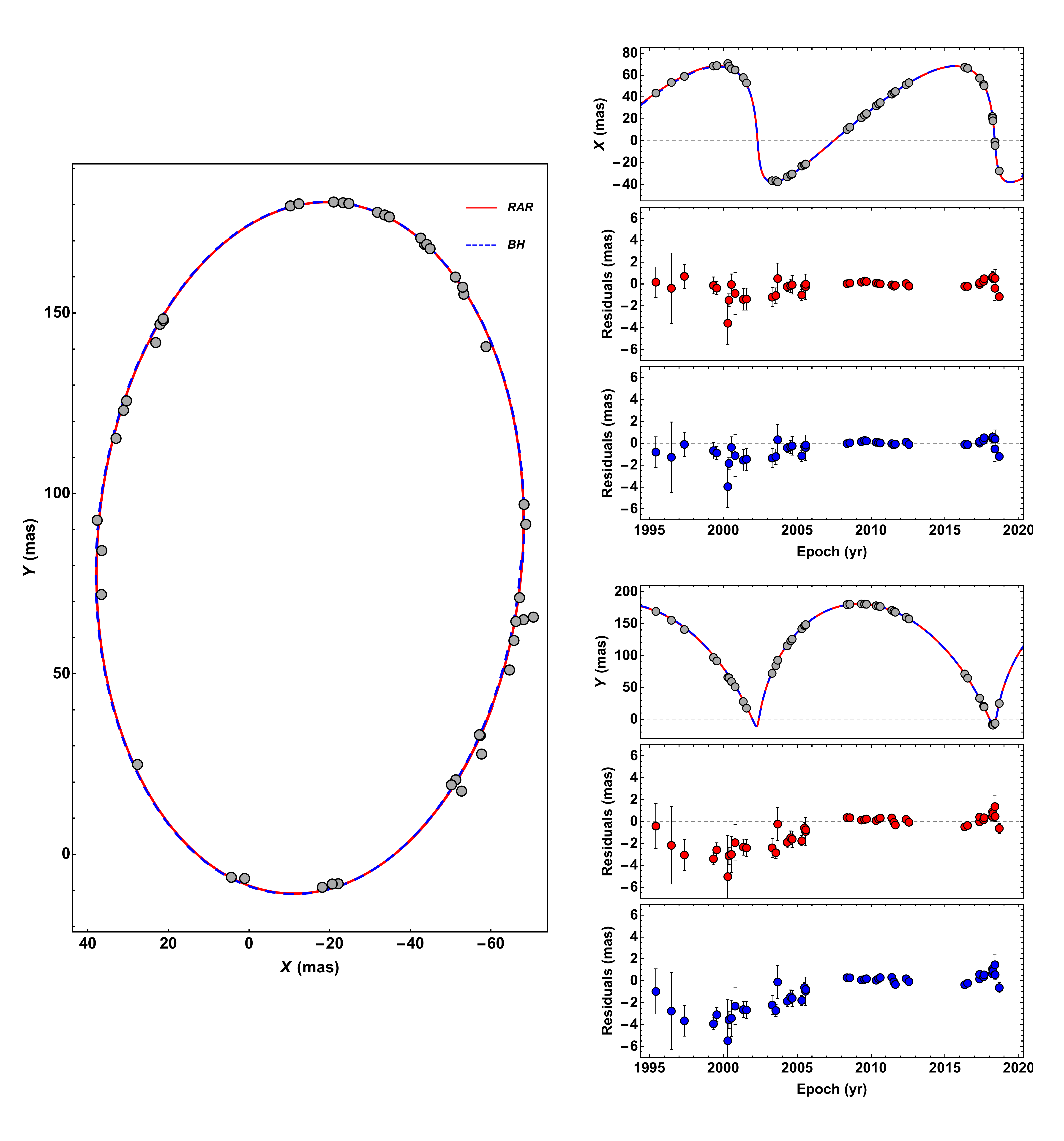} \caption{Theoretical (central BH and RAR models) and observed orbit of S2 around Sgr A*. The left panel shows the orbit, i.e., the right ascension ($X$) vs. declination ($Y$), while the right panel shows $X$ and $Y$ vs. observation time, with the residuals of the best-fit for the BH (blue) and the RAR (red) models. The theoretical models are calculated by solving the equations of motion of a test particle in the gravitational field of 1) a Schwarzschild BH of $4.075\times 10^6~M_\odot$, and 2) the DM-RAR model for $56$ keV-fermions (leading to a quantum core mass $3.5\times 10^6~M_\odot$). {Reproduced from \cite{2020A&A...641A..34B} with permission from Astronomy \& Astrophysics, Copyright ESO}.}% 
\label{fig:S2-orbit}% 
\end{figure} 

\begin{figure}[hbtp!]
\centering 
\includegraphics[width=0.49\hsize,clip]{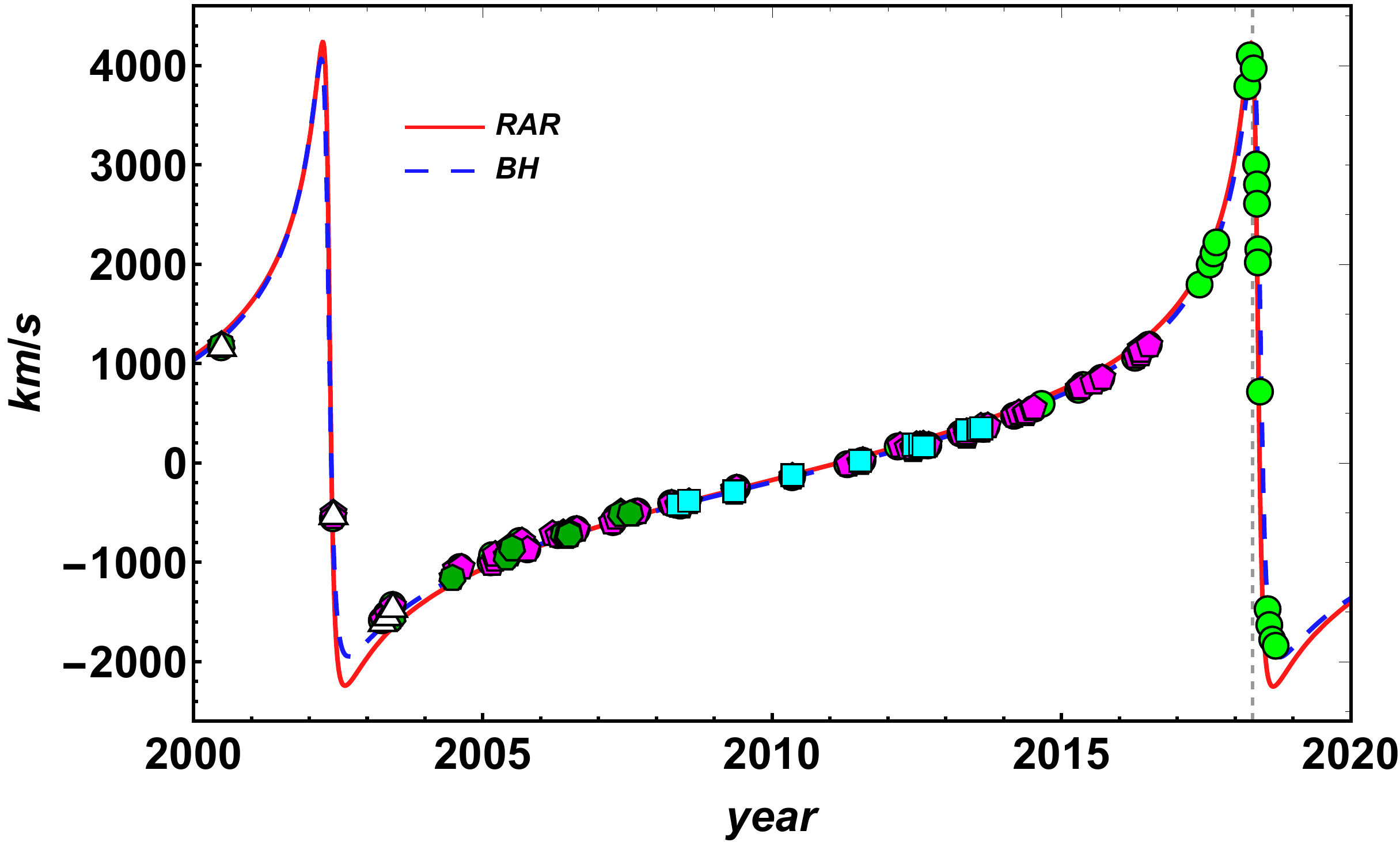} \includegraphics[width=0.49\hsize,clip]{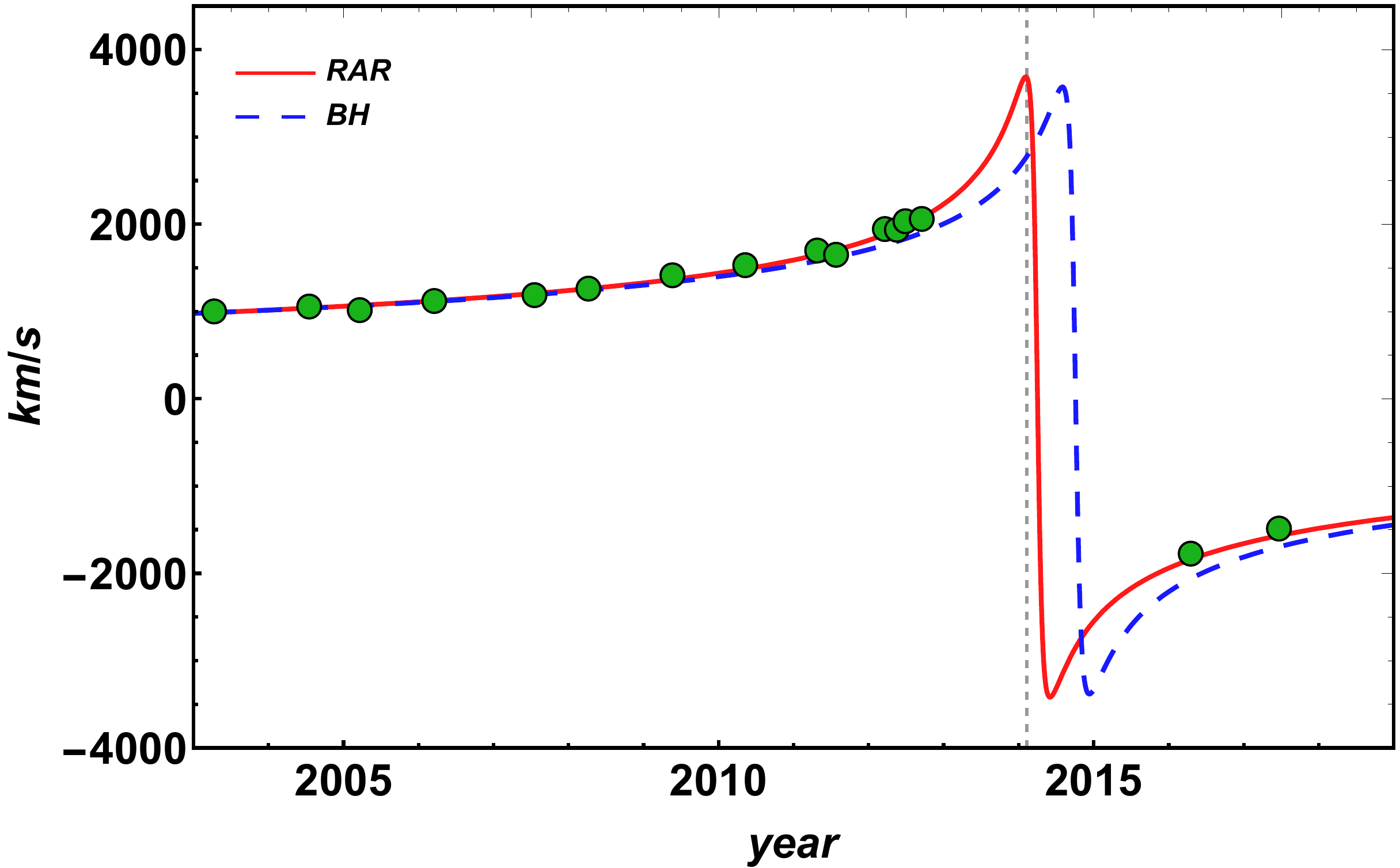}\caption{Theoretical and observed line-of-sight radial velocity for S2 (left) and G2 (right), calculated for the same models as in Fig.~\ref{fig:S2-orbit}. Both models can fit the data with similar precision S2, and for G2, the RAR model fits the data with noticeably better precision than the BH. {Reproduced from \cite{2020A&A...641A..34B} with permission from Astronomy \& Astrophysics, Copyright ESO}.} 
\label{fig:S2-redshift}% 
\end{figure}

We show in \citet{2020A&A...641A..34B} that the solely gravitational potential of such a DM profile for a fermion mass of $56$~keV explains (see Figs.~\ref{fig:S2-orbit} and \ref{fig:S2-redshift}):
\begin{enumerate}
    \item 
    all the available time-dependent data of the position (orbit) and {line-of-sight} radial velocity (redshift function $z$) of S2,
    \item 
    the combination of the special and general relativistic redshift measured for S2,
    \item 
    the currently available data on the orbit and $z$ of G2,
    \item 
    its post-pericenter passage deceleration without introducing a drag force.
\end{enumerate}

For both objects, it was found that the RAR model fits the data better than the BH scenario. The reduced chi-squares of the time-dependent orbit and $z$ data are $\langle\bar{\chi}^2\rangle_{\mathrm{S2, RAR}}\approx 3.1$ and $\langle\bar{\chi}^2\rangle_{\mathrm{S2, BH}}\approx 3.3$ for S2 and $\langle\bar{\chi}^2\rangle_{\mathrm{G2, RAR}}\approx 20$ and $\langle\bar{\chi}^2\rangle_{\mathrm{G2, BH}}\approx 41$ for G2. The fit of the $z$ data shows that, while for S2 the fits are comparable, i.e. $\bar{\chi}^2_{z,\mathrm{RAR}}\approx 1.28$ and $\bar{\chi}^2_{z,\mathrm{BH}}\approx 1.04$, for G2 only the RAR model fits the data: $\bar{\chi}^2_{z,\mathrm{RAR}}\approx 1.0$ and $\bar{\chi}^2_{z,\mathrm{BH}}\approx 26$. Therefore, the sole DM core, for $56$~keV fermions, explains the orbits of S2 and G2. No drag force nor other external agents are needed, i.e., their motion is purely geodesic.

The above robust analysis (and detailed in \cite{2020A&A...641A..34B}) has tested the extended RAR model \cite{2018PDU....21...82A} with the precise astrometric data of the S2 star (and also of G2), showing an excellent agreement for a particle mass of $m c^2=56$ keV. Thus, it is confirmed that the elliptic orbits of these two objects moving in the surroundings of the DM-core made of $56$ keV fermions are compact enough so that the core radius $r_c$ is always smaller than the pericenter of the stars and have their foci coinciding with the Galactic center as obtained by observations (see, e.g., \cite{2018A&A...615L..15G}). The above results invalidate recently claimed drawbacks of the RAR model raised in \cite{2022MNRAS.513L...6Z} since none of the working hypotheses therein apply to the RAR model.

%%%%%%%%%%%%%%%%%%%%%%%%%%%%%%%%%%%%%%%%%%%%%%%%%%%%%%%%%%%
\subsection{The orbits of all S-cluster stars}
%%%%%%%%%%%%%%%%%%%%%%%%%%%%%%%%%%%%%%%%%%%%%%%%%%%%%%%%%%%

\begin{figure}%
	\centering%
	\includegraphics[width=0.7\hsize,height=0.8\hsize]{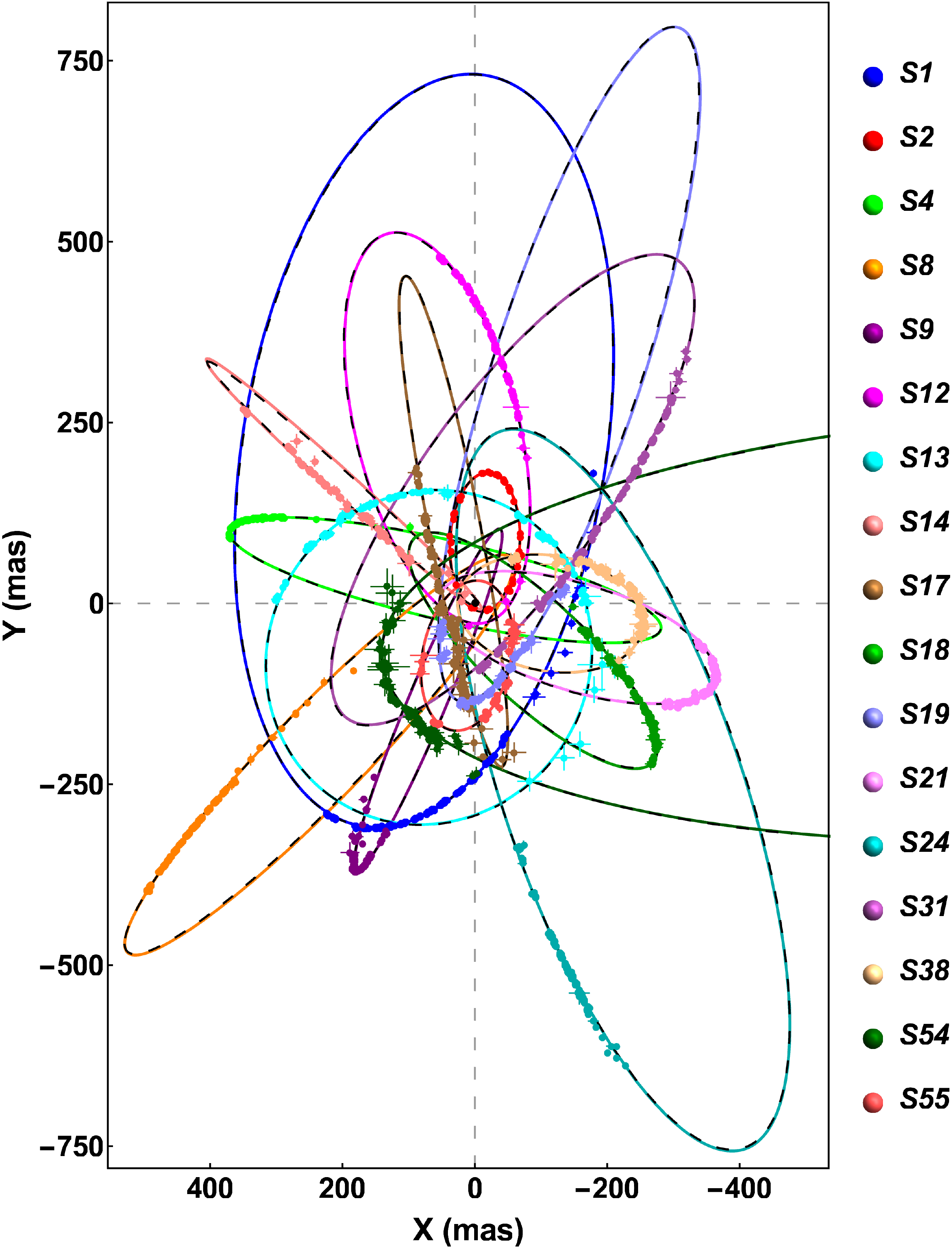}
	\caption{Best-fit orbits for the $17$ best-resolved S-star orbiting Sgr~A*. It shows the projected orbit in the sky, $X$ vs. $Y$, where $X$ is right ascension and $Y$ is declination. The \textit{black dashed curves} correspond to the BH model and the \textit{colored curves} to the RAR model for $m c^2\approx 56$ keV fermions. {Reproduced from \cite{2021MNRAS.505L..64B} with the authors' permission}.}
	\label{fig:Orbits}%
\end{figure}  

More recently, in \citet{2021MNRAS.505L..64B}, we have extended the above results to all the best observationally resolved S-cluster stars, namely to the up-to-date astrometry data of the $17$ S-stars orbiting Sgr~A*, achieving to explain the dynamics of the S-stars with similar (and some cases better) accuracy compared to a central BH model (see Fig.~\ref{fig:Orbits}).  

Table 1 in \citet{2021MNRAS.505L..64B} summarizes the best-fit model parameters and reduced-$\chi^2$ for the position and the line-of-sight radial velocity of the $17$ S-cluster stars for the central BH and the RAR model. The average reduced-$\chi^2$ of the RAR model was $1.5741$ and the corresponding value of the central BH model was $1.6273$. 

Therefore, a core of fermionic DM at the Galactic center explains the orbits of the S-stars with similar accuracy compared to a central BH model. The same \textit{core -- halo} distribution of $56$ keV fermions also explains the MW rotation curves \citep{2018PDU....21...82A, 2020A&A...641A..34B}. Data of the motion of objects near Sgr A*, if accurate enough, could put additional constraints on the fermion mass. The recently detected \textit{hot-spots} apparently in a circular motion at $7$--$23~G M_{\rm BH}/c^2$ radius \citep{2018A&A...618L..10G, 2020MNRAS.497.2385M}, however, fail in this task because the highly-model dependence of the objects real orbit inference, including the lack of information on the object's nature. In addition, the quality of their astrometry data is low relative to the S-stars data. We hope the data quality of these spots or similar objects can increase so that they can put relevant constraints on Sgr A* models. We keep our eyes open to the data of newly observed S-stars (S62, S4711--S4714). Their motion models based on a central BH predict they have pericenter distances $\sim 400~G M_{\rm BH}/c^2$ \citep{2020ApJ...889...61P, 2020ApJ...899...50P}. If confirmed, these new S-stars might constrain the DM core size, so the fermion mass.

%%%%%%%%%%%%%%%%%%%%%%%%%%%%%%%%%%%%%%%%%%%%%%%%%%%%%%%%%%%%%%%%%%%
\subsection{The precession of the S2 orbit}
%%%%%%%%%%%%%%%%%%%%%%%%%%%%%%%%%%%%%%%%%%%%%%%%%%%%%%%%%%%%%%%%%%%

\begin{figure}
  \centering
  \includegraphics[width=0.65\textwidth,clip]{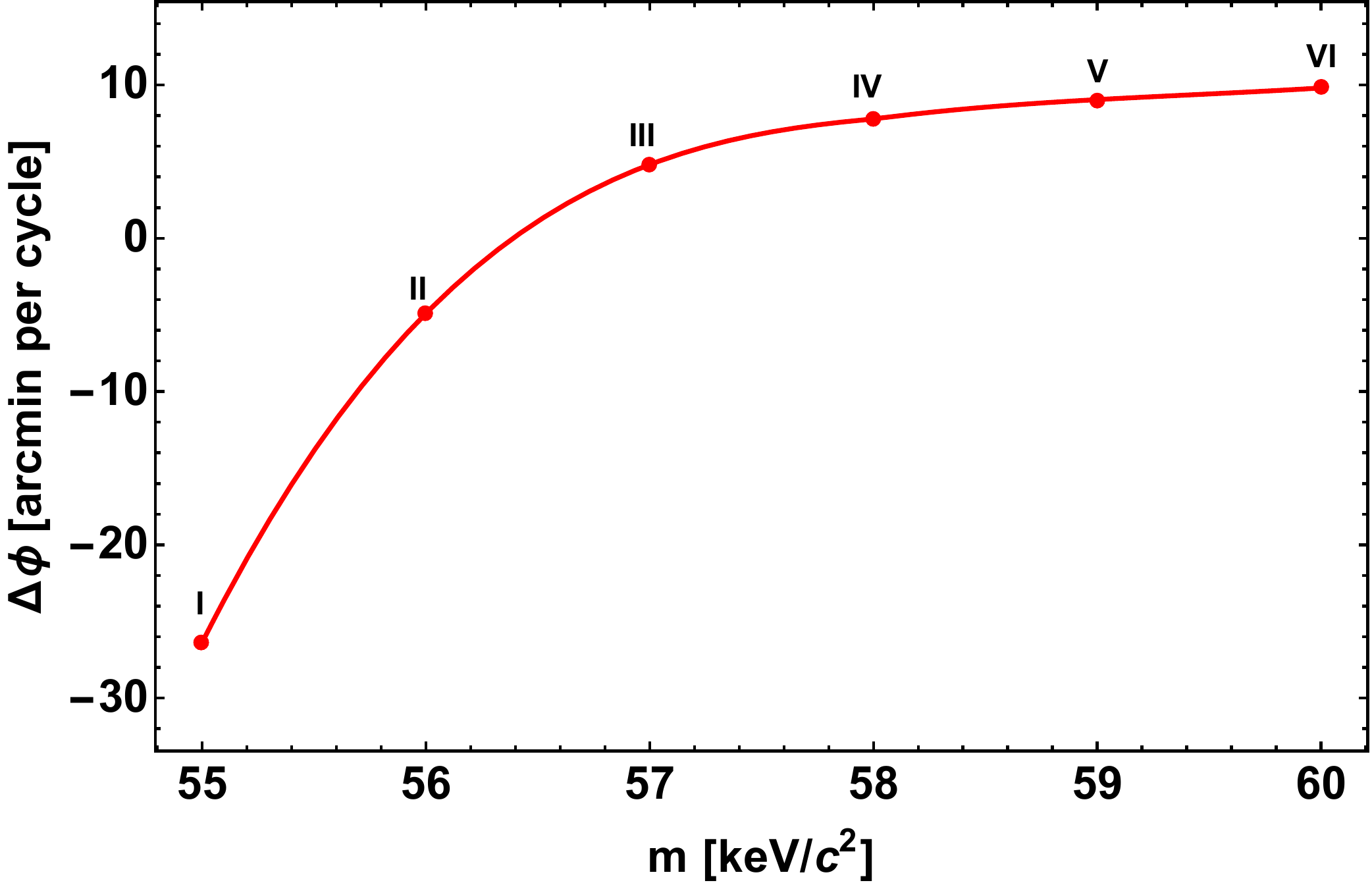}
  \caption{Relativistic periapsis precession $\Delta\phi$ per orbit as a function of the \textit{darkino} mass as predicted by the RAR DM models for the S2-star. The precession is retrograde for $m c^2<56.4$~keV while it becomes prograde for $m c^2>56.4$~keV (see also Table below). {Reproduced from \cite{2022MNRAS.511L..35A} with the authors' permission}.}
  \label{fig:precession}
\end{figure}

Furthermore, \cite{2022MNRAS.511L..35A} focused on the periapsis precession of the S2 orbit. We have there quantified, for the first time within the RAR-DM model, the effects on the S2-star periapsis (precession) shift due to an extended DM mass filling the S2 orbit, in contrast with the vacuum solution of the traditional Schwarzschild BH. The main result is as follows. While the central BH scenario predicts a unique prograde precession, in the DM scenario, it can be either retrograde or prograde, depending on the amount of DM mass enclosed within the S2 orbit, which in turn is a function of the fermion mass (see Fig. \ref{fig:precession} and Table \ref{fig:precession}).

\begin{table*}
\centering
\caption{Comparison of the BH and RAR DM models that best fit all the publicly available data of the S2 orbit. The $2nd$ column shows the central object mass, $M_{\rm CO}$. For the Schwarzschild BH model, $M_{\rm CO} = M_{\rm BH}$, while for the RAR model,  $M_{\rm CO} = M_c$, with $M_c$ the DM core mass. The $3rd$ column shows the radius of the central object, $r_c$. For the Schwarzschild BH model, $r_c$ is given by the event horizon radius, $R_{\rm Sch} = 2 G M_{\rm BH}/c^2$. The $4th$ column shows the DM mass enclosed within the S2 orbit, $\Delta M_{\rm DM}/M_{\rm CO}$. The best-fitting pericenter and apocenter radii of the S2 orbit are given in the $5th$ and $6th$ columns. The values of the average reduced-$\chi^2$ of the best fits, defined as in \citealp{2020A&A...641A..34B}, are given in the $7th$ column. The last two columns show the model predictions of the periapsis precession of the real orbit, $\Delta \phi$, and of the sky-projected orbit, $\Delta\phi_{\rm sky}$. {Reproduced from \cite{2022MNRAS.511L..35A} with the authors' permission}.}
\label{tab:parameters}

\resizebox{\textwidth}{!}{%
\begin{tabular}{clcccccccccccccccc}
\hline
\multicolumn{2}{c}{\textbf{Model}} & & \textbf{\begin{tabular}[c]{@{}c@{}}$M_{CO}$\\ $\left[10^6 M_\odot\right]$\end{tabular}} & & \textbf{\begin{tabular}[c]{@{}c@{}}$r_c$\\ $\left[mpc\right]$ \end{tabular}} & & $\Delta M_{DM}/M_{CO}$ & & \textbf{\begin{tabular}[c]{@{}c@{}}$r_p$\\ $\left[as\right]$\end{tabular}} &  & \textbf{\begin{tabular}[c]{@{}c@{}}$r_a$\\ $\left[as\right]$\end{tabular}} &  &
\textbf{\begin{tabular}[c]{@{}c@{}}$\langle \bar{\chi}^2\rangle$ \end{tabular}} &  &
\textbf{\begin{tabular}[c]{@{}c@{}}$\Delta \phi$\\ $\left[arcmin\right]$\end{tabular}} &  & \textbf{\begin{tabular}[c]{@{}c@{}}$\Delta\phi_{sky}$\\ $\left[ arcmin\right]$\end{tabular}}\\
\cline{1-2} \cline{4-4} \cline{6-6} \cline{8-8} \cline{10-10} \cline{12-12} \cline{14-14} \cline{16-16} \cline{18-18} 
I   & RAR ($m c^2 = 55$ keV) &  & $3.55$ &  & $0.446$ &  & $1.39\times 10^{-2}$ &  & $0.01417$ &  & $0.23723$ &  & $2.9719$ &  & $-26.3845$ &  & $-32.1116$ \\
II  & RAR ($m c^2= 56$ keV) &  & $3.50$ &  & $0.427$ &  & $5.99\times 10^{-3}$ &  & $0.01418$ &  & $0.23618$ &  & $3.0725$ &  & $-4.9064$ &  & $-5.9421$ \\
III & RAR ($m c^2 = 57$ keV) &  & $3.50$  &  & $0.407$ &  & $2.21\times 10^{-3}$ &  & $0.01417$ &  & $0.23617$ & & $3.2766$ &  & $4.8063$ &  & $5.8236$\\
IV  & RAR ($m c^2 = 58$ keV) &  & $3.50$  &  & $0.389$ &  & $7.13\times 10^{-4}$ &  & $0.01424$ &  & $0.23609$ & & $3.2814$ &  & $7.7800$ &  & $9.4243$\\
V   & RAR ($m c^2 = 59$ keV) &  & $3.50$  &  & $0.371$ &  & $2.93\times 10^{-4}$ &  & $0.01418$ &  & $0.23613$ & & $3.3356$ &  & $9.0456$ &  & $10.9613$\\
VI  & RAR ($m c^2 = 60$ keV) &  & $3.50$  &  & $0.355$ &  & $1.08\times 10^{-4}$ &  & $0.01423$ &  & $0.23610$ & & $3.3343$ &  & $9.8052$ &  & $11.8764$\\
      & BH &  & $4.07$ &  & $3.89 \times 10^{-4}$ &  & $0$ &  & $0.01427$ &  & $0.23623$ &  & $3.3586$ &  & $11.9501$ &  & $14.4947$ \\\hline
\end{tabular}%
}
\end{table*}

{Therefore, for larger and larger particle masses, the behavior of the RAR model tends to be that of the BH. As Fig. \ref{fig:precession} shows, the precession tends nearly asymptotically to $12$ arcmin ($\approx 0.2$ deg), the predicted precession of a central Schwarzschild BH (see, e.g., \cite{2020A&A...636L...5G}). The RAR model produces the same prograde precession of a central BH for $345$ keV fermions, corresponding to the unstable DM core for gravitational collapse into a BH. For $56.4$ keV fermions, the prograde and retrograde contributions balance each other leading to a zero net precession. For lower masses, the net precession is retrograde. It has been shown in \cite{2022MNRAS.511L..35A} that currently, available data constrains the amount of retrograde precession, imposing a lower limit to the fermion mass of $\approx 57$ keV, hence an upper limit to the amount of mass enclosed in the orbit of $\approx 0.1\%$ of the core's mass (see Table \ref{tab:parameters}). Indeed, the latter limit agrees with the one obtained by \cite{2020A&A...636L...5G}. For fermion masses above $57$ keV, the prograde precession of the RAR model and that of a central BH agree within the experimental uncertainties. The reason has been explained in \cite{2022MNRAS.511L..35A}: the most accurate S2 data correspond to its pericenter passage \cite{2019Sci...365..664D, 2020A&A...636L...5G}, while the best place to analyze orbital precession is around the apocenter.}

\begin{figure*}
  \centering
  \includegraphics[width=\hsize,clip]{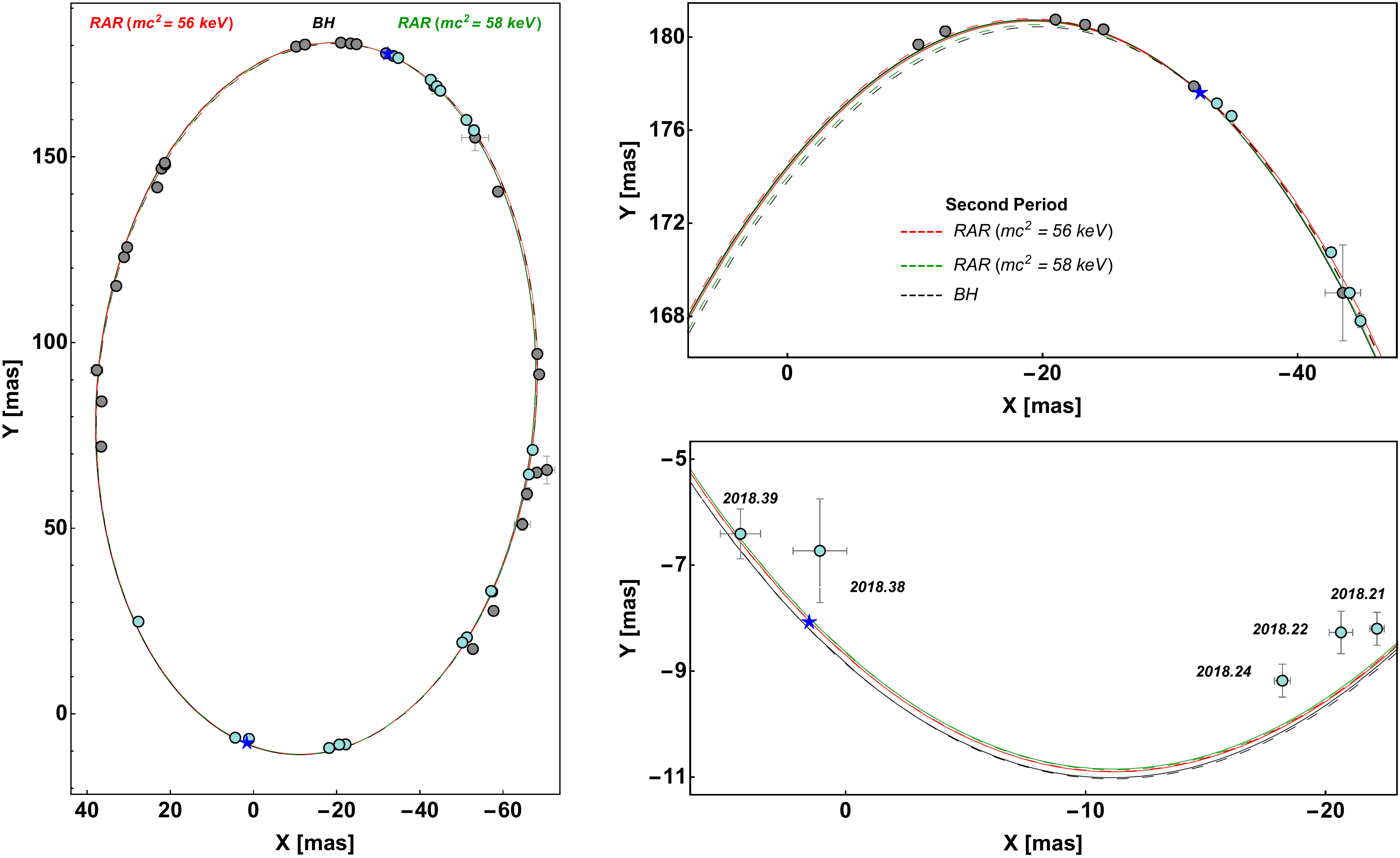}
  \caption{Relativistic precession of S2 in the projected orbit on the plane of the sky, predicted in the BH and RAR DM models. While it is prograde for the BH and RAR ($m c^2=58$ keV) (in dashed black and green respectively), it is retrograde for the RAR DM model ($m c^2=56$ keV) (in dashed red). The solid (theoretical) curves and gray (data) points correspond to the first period ($\approx 1994$--$2010$), while the dashed (theoretical) curves and cyan (data) points to the second period ($\approx 2010$--$2026$). \textit{Right panels}: zoom of the region around apocenter \textit{(top panel)} and pericenter \textit{(bottom panel)}. {Reproduced from \cite{2022MNRAS.511L..35A} with the authors' permission}.}
  \label{fig:precession_data}
\end{figure*}

{The above is shown in Fig.~\ref{fig:precession_data} that plots} the relativistic precession of S2 projected orbit in a \textit{right ascension - declination} plane. It can be seen that while the positions in the plane of the sky nearly coincide about the last pericenter passage in the three models, they can be differentiated close to the next apocenter. Specifically, the upper right panel evidences the difference at the apocenter between the prograde case (as for the BH and RAR model with $m c^2=58$~keV) and the retrograde case (i.e., RAR model with $m c^2=56$~keV).

The bad news is that, as evidenced by Fig.~\ref{fig:precession_data}, all the current and publicly available data of S2 can not discriminate between the two models. The good news is that the upcoming S2 astrometry data close to the next apocentre passage could potentially establish if a classical BH or a quantum DM system governs Sgr~A*.

A further interesting consequence of this scenario is that a core made of \textit{darkinos} becomes unstable against gravitational collapse into a BH for a threshold mass of $\sim 10^8~M_\odot$. Collapsing DM cores can provide the BH seeds for forming SMBHs in active galaxies (such as M87) without the need for prior star formation or other BH seed mechanisms involving super-Eddington accretion rates as demonstrated in \cite{2021MNRAS.502.4227A} from thermodynamic arguments. This topic is of major interest, and further consequences and ramifications are currently being studied, including: 

\begin{itemize}
    \item to propose a new paradigm for SMBH formation and growth in a cosmological framework, which is neither based on the baryonic matter nor early Universe physics;
    \item to study the problem of disk accretion around such DM-cores starting with the generalization of the Shakura \& Sunyaev disk equations in the presence of a high concentration of regular matter (i.e., instead of a singularity);
    \item to use fully relativistic ray-tracing techniques to predict the corresponding shadow-like images around these fermion cores and compare them with the shape and sizes of the ones obtained by the EHT.
\end{itemize}

%%%%%%%%%%%%%%%%%%%%%%%%%%%%%%%%%%%%%%%%%%%%%%%%%%%%%%%%%%%%%%%%%%%
%%%%%%%%%%%%%%%%%%%%%%%%%%%%%%%%%%%%%%%%%%%%%%%%%%%%%%%%%%%%%%%%%%%
\section{Fermionic DM in other galaxy types}\label{sec:4}
%%%%%%%%%%%%%%%%%%%%%%%%%%%%%%%%%%%%%%%%%%%%%%%%%%%%%%%%%%%%%%%%%%%
%%%%%%%%%%%%%%%%%%%%%%%%%%%%%%%%%%%%%%%%%%%%%%%%%%%%%%%%%%%%%%%%%%%

%%%%%%%%%%%%%%%%%%%%%%%%%%%%%%%%%%%%%%%%%%%%%%%%%%%%%%%%%%%%%%%%%%%
\subsection{The RAR model in dwarf, spiral, elliptical galaxies, and galaxy clusters}
%%%%%%%%%%%%%%%%%%%%%%%%%%%%%%%%%%%%%%%%%%%%%%%%%%%%%%%%%%%%%%%%%%%

\begin{figure*}%
	\centering%
	\includegraphics[width=\hsize]{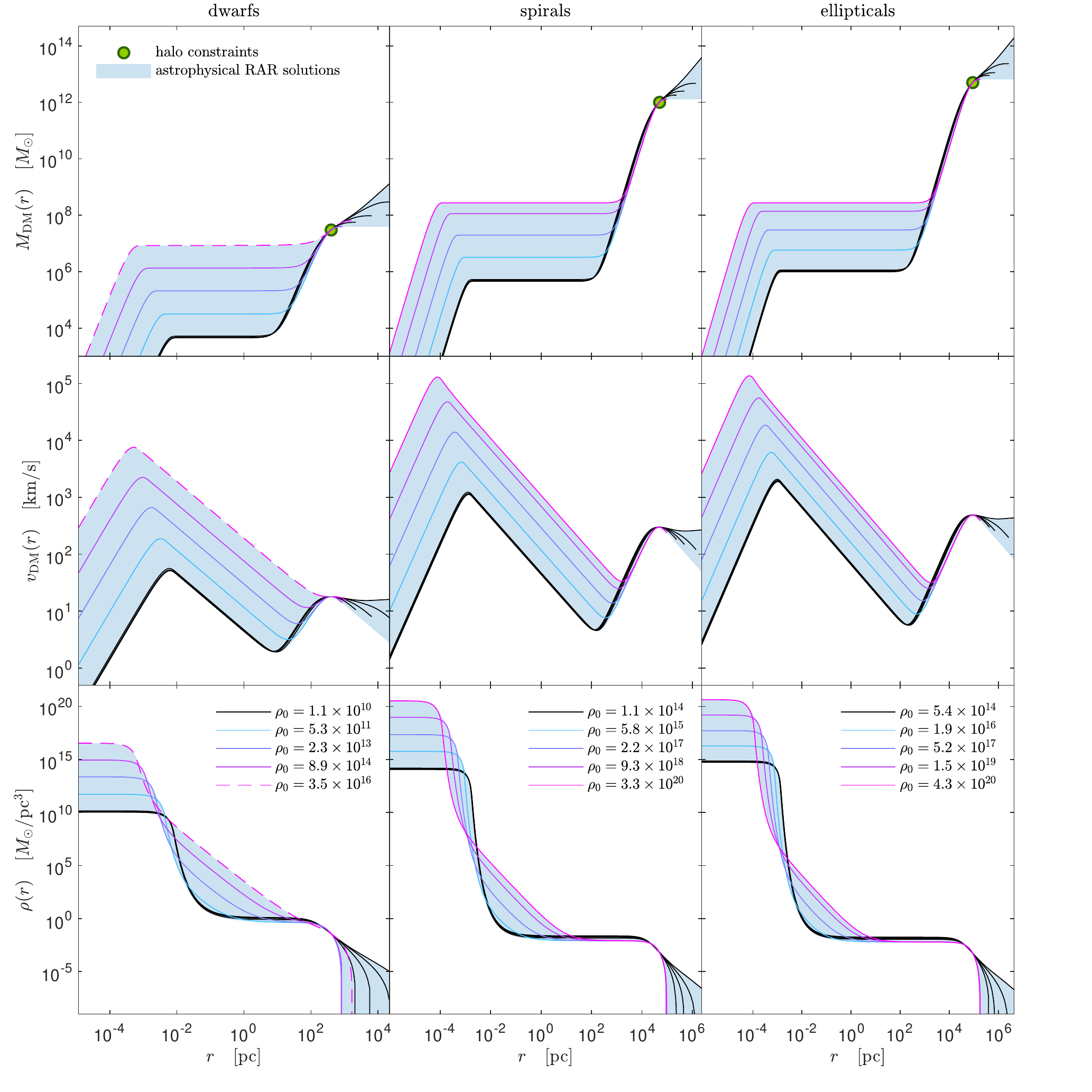}
	\caption{Astrophysical RAR solutions, for the relevant case of $mc^2=48$ keV, observationally fulfill given DM halo restrictions $(r_h, M_h)$ for typical dwarf (left), spiral (middle) and elliptical galaxies (right). Shown are density profiles (bottom), rotation curves (middle), and DM mass distributions (top). The full window for each galaxy type is illustrated by a blue shaded region and enveloped approx. by $5$ benchmark solutions inside. Each solution is labeled with the central density in units of $M_\odot$ pc$^{-3}$. The continuous-magenta curves, occurring only for spiral and elliptical galaxies, indicate the critical solutions which develop compact, critical cores (before collapsing to a BH) of $M_c^{\rm crit} = 2.2\times 10^8 M_\odot$. The dashed-magenta curves for dwarfs are limited (instead) by the astrophysical necessity of a maximum in the halo rotation curve. The bounding black solutions correspond to the ones having the minimum core mass (or minimum $\rho_0$), which in turn imply larger cutoff parameters (implying $\rho\propto r^{-2}$ when $W_0 \to\infty$. {Reprinted from \cite{2019PDU....24100278}, Copyright (2019), with permission from Elsevier}.}
	\label{fig:profiles}
\end{figure*}

In the preceding section, we have shown that the RAR model can perfectly explain the MW rotation curve while providing an excellent alternative to the BH scenario in SgrA* when constrained through the dynamics of the S-cluster stars, including its relativistic effects. In this section, we summarize the main results of the RAR model when applied to different galaxy types, such as dwarfs, spirals, ellipticals, and galaxy clusters, as detailed in \cite{2019PDU....24100278}. Thus, for the relevant case of a fermion mass of $m c^2\approx 50$ keV (motivated by the MW phenomenology as shown in \ref{sec:3}), we make full coverage of the remaining free RAR model parameters ($\beta_0$,$\theta_0$,$W_0$) and present the complete family of density profiles which satisfy realistic halo boundary conditions as inferred from observations (see Fig. \ref{fig:profiles}). 

For galaxy types located far away from us, the observational inferences of their DM distributions are limited to a narrow window of galaxy radii, typically from a few up to several half-light radii. Generally, we have no access to observations for possible detection of a central dark compact object (as for SgrA* in our Galaxy) nor to constrain the system's total (or virial) mass. Thus, we adopt here (see \cite{2019PDU....24100278} for further details) a similar methodology as applied to the MW (as shown in Fig. \ref{sec:3} and detailed in \citep{2018PDU....21...82A}), but only limited to halo-scales where observational data is available (i.e., we do not set any boundary mass conditions in the central core). In particular, we select as the boundary conditions a characteristic halo radius $r_h$ with the corresponding halo $M_h \equiv M(r_h)$. The halo radius is defined as the location of the maximum in the halo rotation curve, which is defined as the one-halo scale length in the RAR model. We list below the parameters ($r_h$, $M_h$) adopted for the different DM halos as constrained from observations in typical dwarf spheroidal (dSph), spiral, elliptical galaxies, and bright galaxy clusters (BCG). We only exemplified the first three cases; see \cite{2019PDU....24100278} for the BCGs.

%%%%%%%%%%%%%%%%%%%%%%%%%%%%%%%%%%%%%%%%%%%%%%%%%%%%%%%%%%%%%%
\subsection{Typical dSph galaxies}
%%%%%%%%%%%%%%%%%%%%%%%%%%%%%%%%%%%%%%%%%%%%%%%%%%%%%%%%%%%%%%

\begin{table*}
\centering
\caption{Typical halo radius and mass for dsPh, spirals, elliptical galaxies, and galaxy cluster, adopted in \cite{2018PDU....21...82A} for testing the RAR model.}
\label{tab:typicalgalaxies}
\resizebox{\textwidth}{!}{%
\begin{tabular}{ccccc}
\hline
 & typical dSph & typical spiral & typical elliptical & typical galaxy cluster \\ 
 \hline
 $r_{h(d)}$ (kpc) & $0.4$ & $50$ & $90$ & $600$\\
 $M_{h(d)}$ $(M_\odot)$ & $3\times 10^{7}$ & $1\times 10^{12}$ & $5\times 10^{12}$ & $3\times 10^{14}$ \\ 
 \hline
\end{tabular}%
}
\end{table*}

The halo parameters in Table \ref{tab:typicalgalaxies} are an average of the eight best-resolved dSphs of the MW studied in \cite{2009ApJ...704.1274W} by solving the Jeans equations and adopting a cored-Hernquist DM profile, which is similar to the mid-outer region of the RAR profiles \cite{2019PDU....24100278}.  

%%%%%%%%%%%%%%%%%%%%%%%%%%%%%%%%%%%%%%%%%%%%%%%%%%%%%%%%%%%%%%
\subsection{Typical spiral galaxies}
%%%%%%%%%%%%%%%%%%%%%%%%%%%%%%%%%%%%%%%%%%%%%%%%%%%%%%%%%%%%%%

The halo parameters shown in Table \ref{tab:typicalgalaxies} are obtained from a group of nearby disk galaxies taken from the THINGS data sample \cite{2008AJ....136.2648D}, where it was possible to obtain a DM model-independent evidence for a maximum in the rotation curves. This was obtained by accounting for baryonic (stars and gas) components in addition to the (full) observed rotation curve from HI tracers (see \cite{2019PDU....24100278}, and refs. therein for details).

%%%%%%%%%%%%%%%%%%%%%%%%%%%%%%%%%%%%%%%%%%%%%%%%%%%%%%%%%%%%%%
\subsection{Typical elliptical galaxies}
%%%%%%%%%%%%%%%%%%%%%%%%%%%%%%%%%%%%%%%%%%%%%%%%%%%%%%%%%%%%%%

The halo parameters in Table \ref{tab:typicalgalaxies} are obtained from (i) a sample of elliptical galaxies from \cite{2005ApJ...635...73H}, studied via weak lensing, from which in \cite{2009MNRAS.397.1169D} it was provided the corresponding halo mass models for the tangential shear of the distorted images; and (ii) the largest and closest elliptical M87 as studied in \cite{2001ApJ...553..722R}, accounting for combined halo mass tracers like stars, globular clusters, and X-ray sources (see \cite{2019PDU....24100278}, and refs. therein for further details).

%%%%%%%%%%%%%%%%%%%%%%%%%%%%%%%%%%%%%%%%%%%%%%%%%%%%%%%%%%%%%%
\subsection{Typical galaxy clusters}
%%%%%%%%%%%%%%%%%%%%%%%%%%%%%%%%%%%%%%%%%%%%%%%%%%%%%%%%%%%%%%

The halo parameters in Table \ref{tab:typicalgalaxies} are obtained from a sample of $7$ BCGs from \cite{2013ApJ...765...25N}. There, the luminous and dark components were disentangled to obtain DM distributions which were reproduced by a generalized NFW (gNFW) model \cite{1996MNRAS.278..488Z}, developing a maximal velocity at the one-halo length scale $r_{\rm max(bcg)}$. In \cite{2013ApJ...765...25N}, the data among all these $7$ cases (i.e., weak lensing and stellar kinematics) support such a maximum within a radial extent from $10$ kpc up to $3$ Mpc; whose corresponding averages in $r_h$ and $M_h$ are given in Table \ref{tab:typicalgalaxies}.\\

The main conclusions of having applied the RAR model for $m c^2\approx 50$ keV to different galaxy types are summarized as follows (see \cite{2019PDU....24100278} for a more detailed explanation): 

\begin{enumerate}
    \item [(I)] Typical dwarf galaxies can harbor dense and compact DM cores with masses from $M_c\sim 10^3 M_\odot$ up to $M_c\sim 10^6 M_\odot$ (see Fig. \ref{fig:profiles}), offering a natural explanation for the so-called intermediate-mass BHs (IMBH). Since the total mass of the typical dSphs here analyzed is below the critical mass of core-collapse (i.e., $M_{\rm tot(d)}\sim 10^7 M_\odot < M_c^{\rm cr}\approx 2\times 10^8 M_\odot$), the core can never become critical and thus will never collapse to a BH. Therefore, the RAR model predicts (for a particle mass of $m c^2\approx 50$ keV) that dSph galaxies can never develop a BH at their center, a result that may explain why these galaxies never become active.
    \item [(II)] Typical spiral and elliptical galaxies can harbor denser and more compact DM cores (w.r.t dSphs) with masses from $M_c\sim 10^5 M_\odot$ up to $M_c^{\rm cr}\approx 2\times10^8 M_\odot$ (see Fig. \ref{fig:profiles}). Thus, they offer a natural alternative to the supermassive BHs hypothesis (see Fig. \ref{sec:3} for the MW). Since the total mass in spirals and ellipticals is much larger than $M_c^{\rm cr}$, the core mass can become critical and eventually collapse towards an SMBH of $\sim 10^8 M_\odot$ which may then grow even larger by accretion.
    \item [(III)] Typical bright cluster of galaxies (BCGs) can harbor dense and compact DM cores with masses from $M_c\sim 10^6 M_\odot$ up to $M_c^{\rm cr}\approx 2\times10^8 M_\odot$ (see Fig. \ref{fig:profiles}). The implications of this prediction for BCGs are still unclear, mainly given the limited spatial resolution achieved by actual observational capabilities below the central kpc. More work is needed (for example, using strong lensing observations) to evaluate if galaxy clusters show an enhancement in DM density similar to the one predicted by the RAR model.
    \item [(IV)] By combining the range of DM core masses, inner halo densities, and total halo masses as predicted by the RAR model across all of these systems, it is possible to test whether or not this model can answer different Universal scaling relations. A point which is studied in the next section (and further detailed in \cite{2019PDU....24100278} and \cite{Krut_2023}).
\end{enumerate}

%%%%%%%%%%%%%%%%%%%%%%%%%%%%%%%%%%%%%%%%%%%%%%%%%%%%%%%%%%%%%%%%%%%%%
%%%%%%%%%%%%%%%%%%%%%%%%%%%%%%%%%%%%%%%%%%%%%%%%%%%%%%%%%%%%%%%%%%%%%
\section{Universal galaxy scaling relations}\label{sec:5}
%%%%%%%%%%%%%%%%%%%%%%%%%%%%%%%%%%%%%%%%%%%%%%%%%%%%%%%%%%%%%%%%%%%%%
%%%%%%%%%%%%%%%%%%%%%%%%%%%%%%%%%%%%%%%%%%%%%%%%%%%%%%%%%%%%%%%%%%%%%

Galaxies follow different scaling relations (or Universal scaling relations) such as the DM surface density relation (SDR) \cite{2009MNRAS.397.1169D}, the radial acceleration relation \cite{2016PhRvL.117t1101M}, the mass discrepancy acceleration relation (MDAR) \cite{2004ApJ...609..652M} and the Ferrarese relation \cite{2002ApJ...578...90F, 2015ApJ...800..124B} between the total halo mass and its supermassive central object mass. Thus, this section's main goal is to illustrate/summarize the ability of the RAR model to agree with all these relations, covering a really broad window of radial scales across very different galaxy types.  {The analyses are exemplified for a fermion of rest mass-energy of $\approx 50$ keV.}

For given halo parameters $(r_h, M_h)$ as inferred from observations in the smallest up to the largest galaxy types (see Table \ref{tab:typicalgalaxies}), the RAR model predicts a window of halo values on other radial-scales: (a) the \textit{plateau} density (or inner-halo constant density) $\rho_{\rm pl}$; (b) the total halo mass $M_{\rm tot}$; and (c) the fermion-core mass $M_c$. We will then use this predictive power of the RAR model (shown Fig. \ref{fig:profiles}) to test if it can answer for the above Universal scaling relations as reported in the literature. We will focus first on the Ferrarese relation between the total halo mass and its supermassive central object mass and in the SDR.

\begin{figure*}%[t!]%
	\centering%
    \includegraphics[width=0.48\hsize]{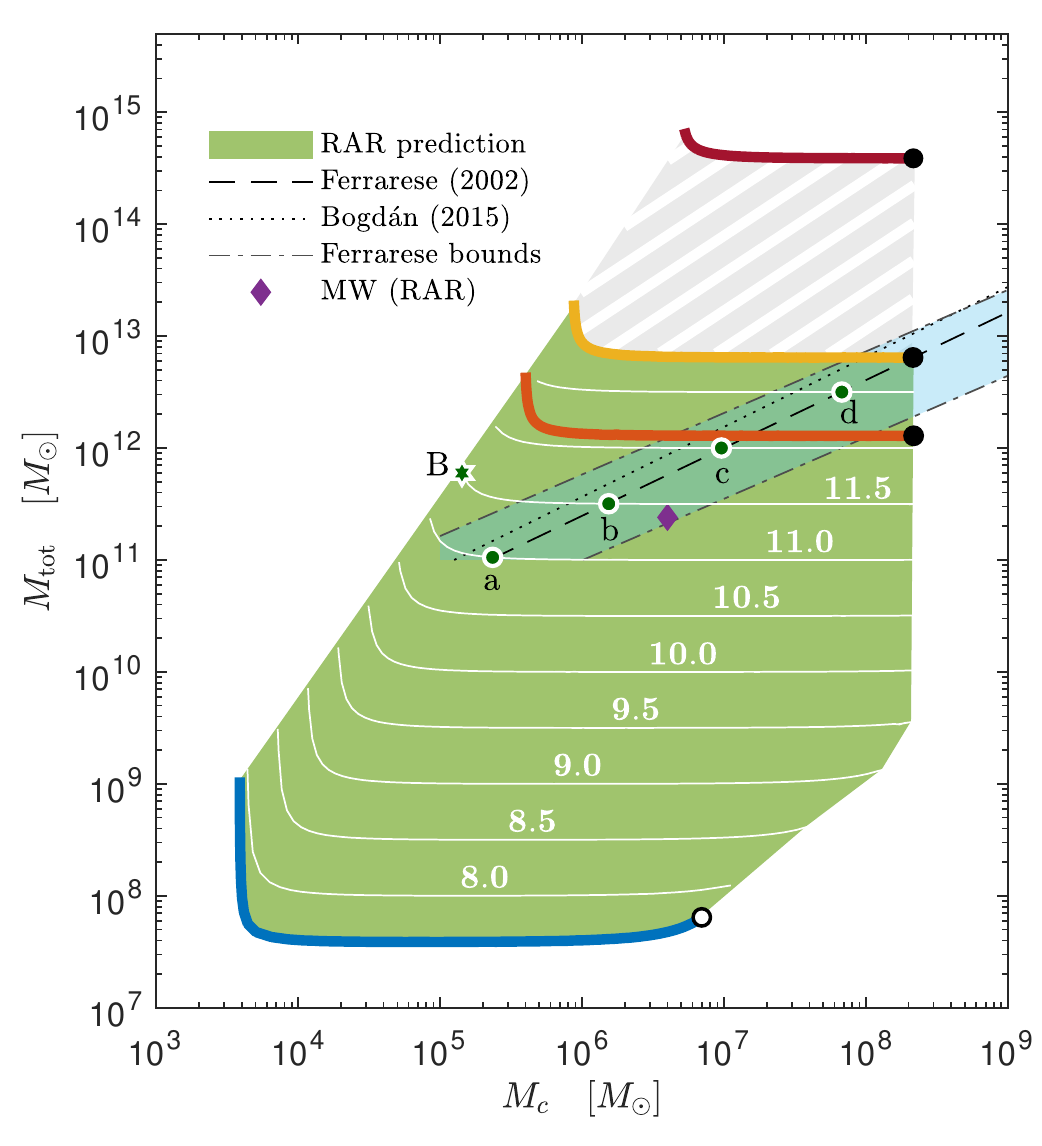}
	\includegraphics[width=0.48\hsize]{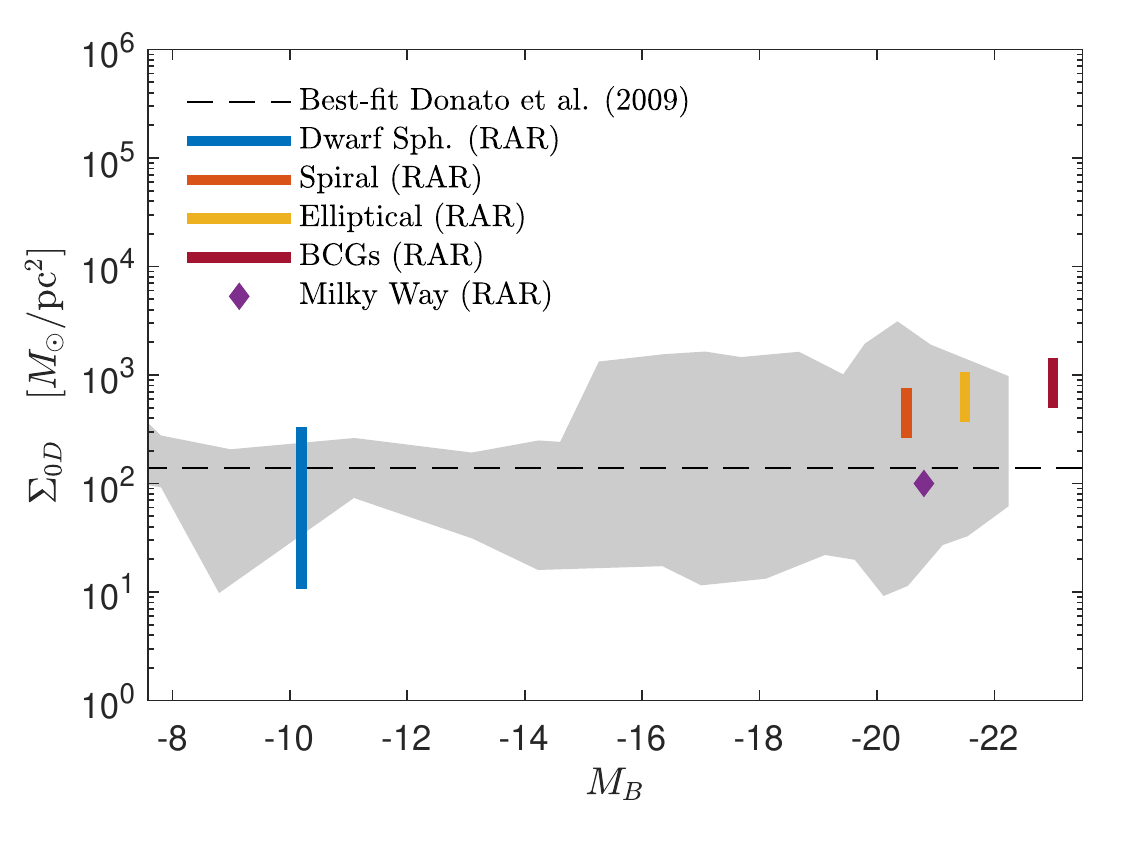}
    \caption{Left: theoretical $M_c$-$M_{\rm tot}$ relation within three parametric RAR model (for $m c^2 = 48$ keV). The colored lines read for each galaxy type in correspondence with the astrophysical RAR solutions in Fig. \ref{fig:profiles}. The green area covers the RAR predictions for a given halo mass in the range $M_h \approx 10^7$--$10^{12} M_\odot$ and fulfilling $\Sigma_{0D}\propto \rho_{\rm pl}\, r_h = 140 M_\odot$ pc$^{-2}$, as inferred from the Donato relation (see right panel). The white lines show a set of families with given halo mass $M_h$ (and labeled by the $M_{\rm tot}$ value in the horizontal regime). The results show the RAR model agrees with the different $M_{\rm BH}$-$ M_{\rm tot}$ relations, as considered in the literature and shown in the blue-ish stripe. The filled-black dots correspond to the critical core mass $M_c^{\rm crit}$, and the empty-black dot indicates the limiting maximum core mass $M_c^{\rm max}$ for dwarfs. Right: surface DM density (or Donato) relation as predicted by the RAR model (see vertical color lines) for each galactic structure in correspondence with the astrophysical solutions (i.e., blue-shaded regions in Fig. \ref{fig:profiles}). The dashed horizontal line represents the Universal relation from the best fit of the data as found by \cite{2009MNRAS.397.1169D}. The dark-gray region indicates the delimited area by the $3-\sigma$ error bars of all the data points. {Reprinted from \cite{2019PDU....24100278}, Copyright (2019), with permission from Elsevier}.}%
	\label{fig:scaling-relations}%
\end{figure*}

%%%%%%%%%%%%%%%%%%%%%%%%%%%%%%%%%%%%%%%%%%%%%%%%%%%%%%%%%%%%%%%%%%%%%%%%%
\subsection{The Ferrarese relation}
%%%%%%%%%%%%%%%%%%%%%%%%%%%%%%%%%%%%%%%%%%%%%%%%%%%%%%%%%%%%%%%%%%%%%%%%%

This relation establishes that, for large enough galaxies\footnote{This kind of correlation has been shown to brake for the case of small and bulgeless galaxies in \cite{2011Natur.469..377K}.}, the more massive the halo, the larger (in mass) is the supermassive compact object lying at its center. This relation is shown in dashed in the left panel Fig. \ref{fig:scaling-relations}, as taken from \cite{2002ApJ...578...90F}, together with other more updated versions of such relation such as the one found in \cite{2015ApJ...800..124B} (shown in dotted line in the same figure). Interestingly enough, the whole family of RAR solutions, with its corresponding window of predicted core and total DM masses $M_c, M_{\rm tot}$ (see left panel of Fig. \ref{fig:scaling-relations}), from typical dwarf to ellipticals galaxies (see green area), do overlap with the Ferrarese relation. That is, the RAR model contains the observational strip (in light blue), while at the same time, it extends out such a Ferrarese strip, indicating a yet-unseen or otherwise unphysical cases (see \cite{2019PDU....24100278} for further details).

Even if no observational data exist yet at the down-left corner of Fig. \ref{fig:scaling-relations} (left panel), special attention has to be given to the RAR model predictions for dwarf galaxies: recent observations towards the center of some ultra-compact dwarf galaxies of total mass few $\sim 10^7 M_\odot$ (e.g., \cite{2014Natur.513..398S, 2017ApJ...839...72A, 2018MNRAS.477.4856A}), indicate the existence of putative massive BH of few $\sim 10^6 M_\odot$. Interestingly, the RAR model naturally allows for a slight difference (less than an order of magnitude) between $M_c$ and $M_{\rm tot}$. However, more work is needed on this particular dwarf galaxy to make a more definite statement. 

%%%%%%%%%%%%%%%%%%%%%%%%%%%%%%%%%%%%%%%%%%%%%%%%%%%%%%%%%%%%%%%%%%%%%%%%%
\subsection{The DM surface density relation (DSR)}
%%%%%%%%%%%%%%%%%%%%%%%%%%%%%%%%%%%%%%%%%%%%%%%%%%%%%%%%%%%%%%%%%%%%%%%%%

The DSR relation establishes that the central surface DM density in galaxies is roughly constant, spanning more than 14 orders in absolute magnitude ($M_B$): $\rho_{0D} r_0 \approx 140 M_\odot$ pc$^{-2}$ (with $\rho_{0D}$ the inner-halo -or sometimes called central- DM halo density measured at the Burkert halo radius $r_0$) \citep{2009MNRAS.397.1169D}.

Since the Burkert central density corresponds to the plateau density of the RAR density profiles (see right panel of Fig. \ref{fig:profile-illustration-mep}), i.e., $\rho_{0D}\equiv \rho_{\rm pl}$, and $r_0 \approx 2/3 \,r_h$ (as shown in \cite{2019PDU....24100278} connecting the Burkert and RAR one-halo scale-lengths), it is possible to check the DSR by calculating the product $2/3 \,\rho_{\rm pl} r_h$ along the entire family of astrophysical RAR solutions of Section \ref{sec:4} (including the Milky Way). This is shown in the right panel of Fig. \ref{fig:scaling-relations}.

It can be seen that the RAR model predictions agree with the observed relation, the latter displayed in Fig. \ref{fig:scaling-relations} (right panel) in the dark-grey region delimited (or enveloped) within the $3-\sigma$ error bars along all the data points considered in \citet{2009MNRAS.397.1169D}. That figure shows that each typical galaxy type's predicted RAR surface density (vertical solid lines) is within the expected $3-\sigma$ data region. We further notice that typical bright clusters are beyond the observed window as reported by \citet{2009MNRAS.397.1169D}, who considered only up to elliptical structures.

%%%%%%%%%%%%%%%%%%%%%%%%%%%%%%%%%%%%%%%%%%%%%%%%%%%%%%%%%%%%%%%%%%%%%%%%%
\subsection{The radial acceleration and mass discrepancy acceleration relations}
%%%%%%%%%%%%%%%%%%%%%%%%%%%%%%%%%%%%%%%%%%%%%%%%%%%%%%%%%%%%%%%%%%%%%%%%%

\begin{figure*}[t!]
	\centering%
	\includegraphics[width=0.9\hsize]{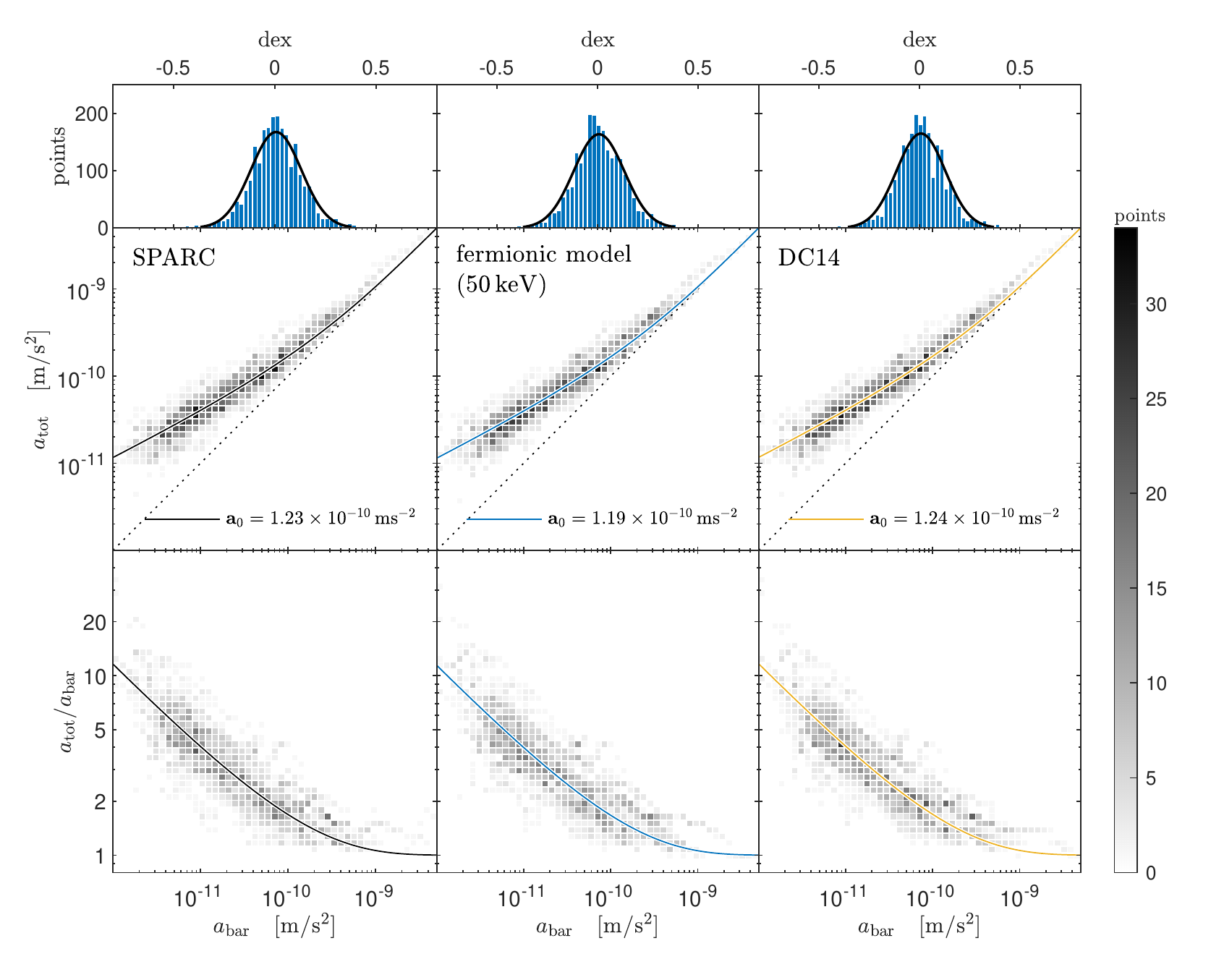}
	\caption{Radial acceleration relation (top) and mass discrepancy acceleration relation (bottom) for SPARC data and competing DM halo models. Each plot is divided into $50\times 50$ equal bins. The baryonic centripetal acceleration $\mathfrak{a}_{\rm bar}$ is inferred from luminosity observables while the total acceleration $\mathfrak{a}_{\rm tot}$ is inferred independently from velocity fields. For DM halo models, the total acceleration is composed of the predicted dark and inferred baryonic components, i.e., $\mathfrak{a}_{\rm tot} = \mathfrak{a}_{\rm DM}+ \mathfrak{a}_{\rm bar}$. The corresponding solid curves are the best fits characterized by a specific $\mathfrak{a}_0$. The histogram plots (upper row) show a Gaussian distribution of $\log_{10}(\mathfrak{a}_0/\mathfrak{a}_0)$. The grayscale legend shows the number of points per bin ($2396$ for the $120$ SPARC-galaxies used). {Reproduced from \cite{Krut_2023}}.}\label{fig:acceleration:grid}
\end{figure*}

We now confront the RAR model predictions with two closely related Universal relations, though this time not based on typical galaxies with given (averaged) parameters ($r_h, M_h$) inferred from observables (as done before), but from a sample of 120 rotationally supported galaxies taken from the SPARC data set \cite{2016AJ....152..157L}.

The radial acceleration relation is a non-linear correlation between the radial acceleration exerted on the total matter distribution and the one caused by the baryonic matter component only (see Eq. \ref{eq:mcgaugh-fit} below). Thus it offers an important restriction to any model that explains the DM halo in galaxies. 

The different mass components in a galaxy, like a bulge, disk, gas, etc., trace their own contribution to the centripetal or radial acceleration $\mathfrak{a} = v^2/r$. As originally shown in \cite{2016PhRvL.117t1101M}, it exists a relatively tight relation between the radial acceleration owing to the total mass, say $\mathfrak{a}_{\rm tot}$, and that produced by the baryonic mass component, say $\mathfrak{a}_{\rm bar}$, which the empirical formula can well describe
\begin{equation}
	\label{eq:mcgaugh-fit}
	\mathfrak{a}_{\rm tot} = \frac{\mathfrak{a}_{\rm bar}}{1 - e^{-\sqrt{\mathfrak{a}_{\rm bar}/{\mathfrak{a}_0}}}},
\end{equation}
with $\mathfrak{a}_0$ the only adjustable parameter. Notice that for $\mathfrak{a}_{\rm bar} \ll \mathfrak{a}_0$, the DM dominates, and the relation deviates from linear, while for $\mathfrak{a}_{\rm bar} \gg \mathfrak{a}_0$, the baryonic component dominates and the linear relation is recovered (see Fig. \ref{fig:acceleration:grid}). The relation has been shown to apply applies to different galaxy types, e.g., disk, elliptical, lenticular, dwarf spheroidal, and low-surface-brightness galaxies, so it looks like a real universal law \cite{2016PhRvL.117t1101M, 2017ApJ...836..152L, 2019ApJ...873..106D}. 

We also include in the analysis a close relation: the mass discrepancy acceleration relation (MDAR) relation between baryonic and total mass components. It is defined by $D= M_{\rm tot}/M_{\rm bar}$ with $M_{\rm bar}$ the total baryonic mass and $M_{tot}$ the total mass content of the galaxy including the DM component. If we define the radial acceleration through its definition in terms of the gravitational potential of a spherically symmetric mass distribution, then the MDAR can be written as $D = \mathfrak{a}_{tot}/\mathfrak{a}_{\rm bar}$. 

In Fig. \ref{fig:acceleration:grid}, we show the results of the best fits for the radial acceleration relation and MDAR for both the RAR model (central panels) and the DC14 DM model (right panels) accounting for baryonic-feedback \cite{2014MNRAS.441.2986D} (see also the full Figure 2 in \cite{Krut_2023} for best-fits to both relations by other DM typical halo models used in the literature). Since both models achieve very close fitted values of $\mathfrak{a}_0 \approx 1.2\times 10^{-10}$ m s$^{-2}$ as the one obtained from the SPARC data only (i.e., done in DM model-independent fashion, see left panels), then it is concluded that this Universal relation does not help to prefer one model over the other statistically (see \cite{Krut_2023} for details). Instead, as demonstrated in \cite{Krut_2023}, the individual fitting of each rotation curve (covering the full SPARC sample studied) allows to favor of the cored profiles over the cuspy Navarro-Frenk-White profile (not shown here).

\begin{figure}
	\centering%
	\includegraphics[width=0.48\hsize]{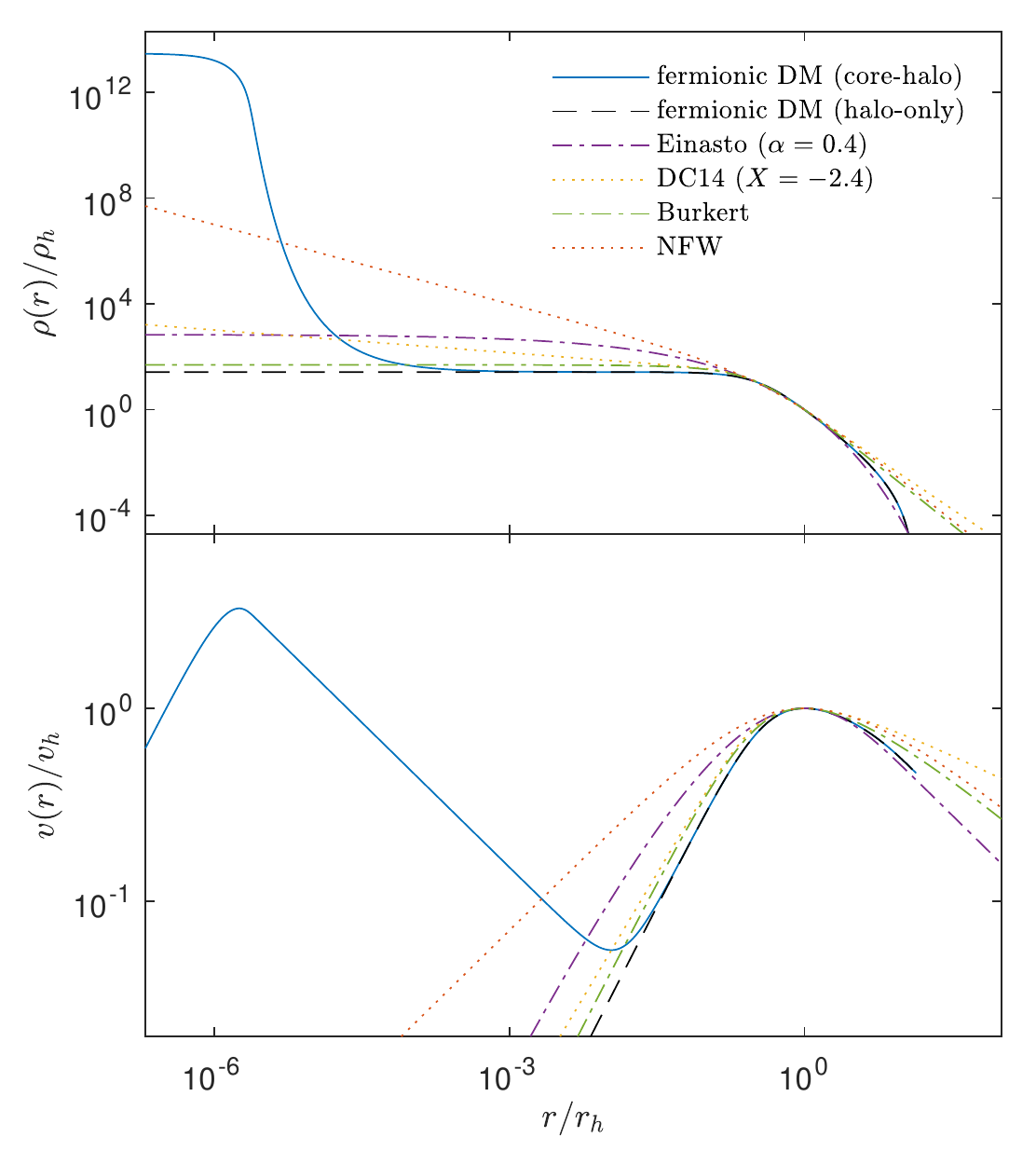}
    \includegraphics[width=0.48\hsize]{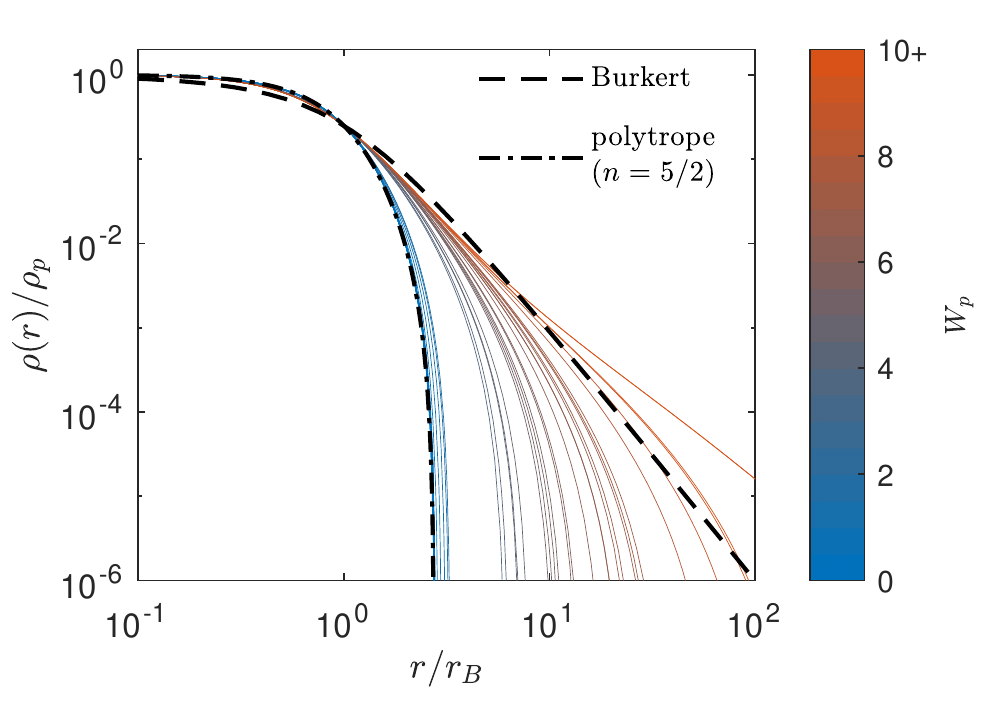}
	\caption{Left: density profile (top) and rotation curve (bottom) for selected DM models for the typical configuration parameters presented in \cite{Krut_2023}. The core-halo solution of the RAR model is illustrated by the configuration parameters $\beta_0 = 10^{-7}$, $\theta_0 = 30$ and $W_0 = 60$, in combination with a corresponding halo-only solution. Other DM models (NEW, DC14, Burkert, Einasto) are added for comparison. All profiles are normalized to the halo radius $r_h$, defined at the halo velocity maximum, with $\rho_h = r(r_h)$ and $v_h = v(r_h)$. Right: best-fit DM density RAR profiles (outer region) to the full SPARC sample considered in \cite{Krut_2023} (in solid) for different values of the cut-off parameter $W(r_{\rm pl})\equiv W_p$: the outer halo of the RAR model can go from polytropic-like (polytropic index $n=5/2$ in dot-dashed) to power law-like (e.g., $\rho \propto r^{-3}$) resembling the Burkert profile (in dashed). {Reproduced from \cite{Krut_2023}.}}\label{fig:profile-illustration-mep}
\end{figure}

Finally, we have also compared the DM halo models by their density profiles and rotation curves for typical SPARC galaxies in the left panel of Fig. \ref{fig:profile-illustration-mep}. Most models show similarities in the halo tails, but the RAR model is the only one showing a dense-fermion core at the galaxy's center, which acts as an alternative to the SMBH scenario. It is also relevant to mention that while many of the models (such as Einasto or DC14) need the baryonic feedback to produce the statistically favored cored density profiles, the RAR model naturally achieves a cored behavior (i.e., the inner-halo plateau) due to the quasi-thermodynamic equilibrium of the particles at formation (see \cite{Krut_2023} and references therein for further discussions).   

{Although we do not expect significant qualitative or quantitative differences in the galaxy structure parameters for fermion masses around the above-explored value of $50$ keV, studying the implications of different fermion masses for the universal relations will be interesting. We plan to have new results on this topic soon.}

%%%%%%%%%%%%%%%%%%%%%%%%%%%%%%%%%%%%%%%%%%%%%%%%%%%%%%%%%%%%%%%%%%%
%%%%%%%%%%%%%%%%%%%%%%%%%%%%%%%%%%%%%%%%%%%%%%%%%%%%%%%%%%%%%%%%%%%
\section{Fermionic DM and particle physics}\label{sec:6}
%%%%%%%%%%%%%%%%%%%%%%%%%%%%%%%%%%%%%%%%%%%%%%%%%%%%%%%%%%%%%%%%%%%
%%%%%%%%%%%%%%%%%%%%%%%%%%%%%%%%%%%%%%%%%%%%%%%%%%%%%%%%%%%%%%%%%%%

%%%%%%%%%%%%%%%%%%%%%%%%%%%%%%%%%%%%%%%%%%%%%%%%%%%%%%%%%%%%%%%%%%%
\subsection{Are the sterile neutrinos the fermions of the RAR model?}
%%%%%%%%%%%%%%%%%%%%%%%%%%%%%%%%%%%%%%%%%%%%%%%%%%%%%%%%%%%%%%%%%%%

In this section, we will discuss a possible connection to particle physics (i.e., beyond the standard model of particles), analyzing the possibility that the DM particles are self-interacting DM. Namely, DM particles interact among themselves via some unknown fundamental interaction besides gravity. It is a very active field of research within the DM community since self-interacting DM (SIDM) has been proposed as a possible solution to several challenges which is facing the standard cosmological model on galactic scales (see \cite{2017IJMPD..2630007M,2018PhR...730....1T} for different reviews on the subject).

In 2016, our group presented an extension of the RAR model to include fermion self-interactions, referred to as the Arg\"uelles, Mavromatos, Rueda and Ruffini (AMRR) model~\cite{amrr}. In the AMRR model, it has been advanced that the \textit{darkinos} might be the right-handed sterile neutrinos introduced in the minimum standard model extension ($\nu$MSM) paradigm~\cite{nuMSM}. The AMRR model adopts right-handed neutrinos self-interacting via dark-sector massive (axial) vector mediators.

In 2020, \cite{2020PDU....3000699Y} explored the radiative decay channel of such sterile neutrinos into X-rays due to the Higgs portal interactions of the $\nu$MSM. This work shows that such generalized RAR profiles, including fermion self-interactions, agree with the overall MW rotation curve. In addition, the window of self-interacting DM cross-sections that satisfy the known bullet cluster constraints has been identified.

\begin{figure}[hbtp!]
\centering 
\includegraphics[width=0.7\hsize,clip]{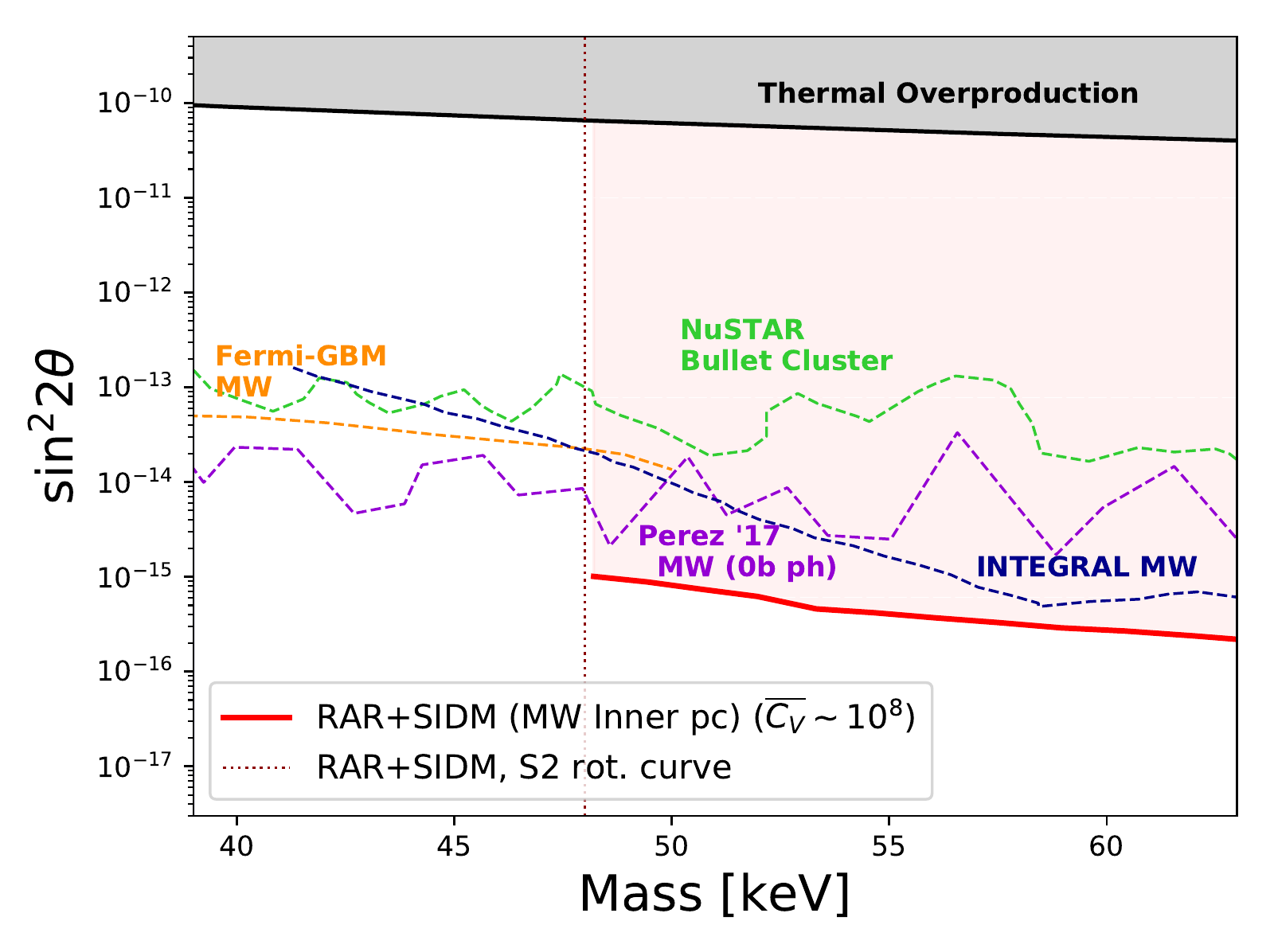}
\caption{Sterile neutrino parameter space limits obtained for Galactic Center observations using the AMRR profiles (continuous red line) when assuming DM production due to self-interactions through a massive vector field mediator. The light-red shaded region above the continuous red line corresponds to the AMRR limits given by X-ray bounds (i.e.~indirect detection analysis), while the vertical shaded region below $48$~keV indicates the smallest DM mass compatible with S-cluster stars' rotation curve data. The upper shaded region corresponds to production mechanism bounds: non-resonant production under no lepton asymmetry. Other dotted lines refer to several X-ray bounds for different DM halo profiles, including 0-bounce photon analysis. {Reprinted from \cite{2020PDU....3000699Y}, Copyright (2020), with permission from Elsevier}.}%
\label{fig:X-ray}%
\end{figure}

To further constrain the AMRR-$\nu$MSM model, an indirect detection analysis has been performed using X-ray observations from the Galactic center by the Nustar mission \cite{2020PDU....3000699Y}. Figure~\ref{fig:X-ray} summarizes all the observational constraints.

It has also advanced a new generation mechanism based on vector-meson decay, able to produce these sterile neutrinos in the early Universe.

Summarizing, by considering a DM profile that self-consistently accounts for the particle-physics model, the analysis of NuSTAR X-ray data shows how sterile-neutrino self-interactions affect the $\nu$MSM parameter-space constraints. The decay of the massive vector field that mediates the self-interactions affects standard production mechanisms in the early Universe. This mechanism might broaden the allowed parameter space compared to the standard $\nu$MSM scenario.

%%%%%%%%%%%%%%%%%%%%%%%%%%%%%%%%%%%%%%%%%%%%%%%%%%%%%%%%%%%%%%%%%%%
%%%%%%%%%%%%%%%%%%%%%%%%%%%%%%%%%%%%%%%%%%%%%%%%%%%%%%%%%%%%%%%%%%%
\section{Fermionic DM and cosmology}\label{sec:7}
%%%%%%%%%%%%%%%%%%%%%%%%%%%%%%%%%%%%%%%%%%%%%%%%%%%%%%%%%%%%%%%%%%%
%%%%%%%%%%%%%%%%%%%%%%%%%%%%%%%%%%%%%%%%%%%%%%%%%%%%%%%%%%%%%%%%%%%

%%%%%%%%%%%%%%%%%%%%%%%%%%%%%%%%%%%%%%%%%%%%%%%%%%%%%%%%%%%%%%%%%%%
\subsection{Formation and stability of fermionic DM halos in a cosmological framework}
%%%%%%%%%%%%%%%%%%%%%%%%%%%%%%%%%%%%%%%%%%%%%%%%%%%%%%%%%%%%%%%%%%%

The formation and stability of collisionless self-gravitating systems are long-standing problems that date back to the work of D. Lynden-Bell in 1967 on violent relaxation and extend to the virialization of DM halos. Such a relaxation process predicts that a Fermi-Dirac phase-space distribution can describe spherical equilibrium states when the extremization of a coarse-grained entropy is reached. In the case of DM fermions, the most general solution develops a degenerate compact core surrounded by a diluted halo. As we have recently shown \cite{2018PDU....21...82A, 2019PDU....24100278, 2020A&A...641A..34B}, the \textit{core-halo} profiles obtained within the fermionic DM-RAR model explain the galaxy rotation curves, and the DM core can mimic the effects of a central BH. A yet open problem is whether these astrophysical \textit{core-halo} configurations can form in nature and if they remain stable within cosmological timescales. These issues have been recently assessed in \cite{2021MNRAS.502.4227A}.

\begin{figure}[hbtp!]
\centering 
\includegraphics[width=0.5\hsize,clip]{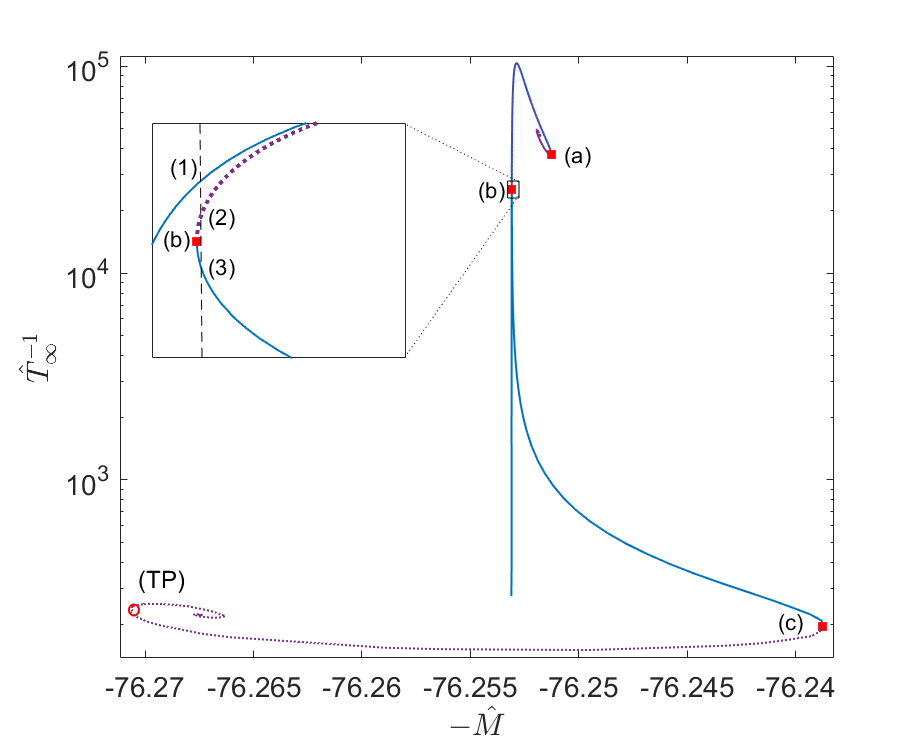}\includegraphics[width=0.5\hsize,clip]{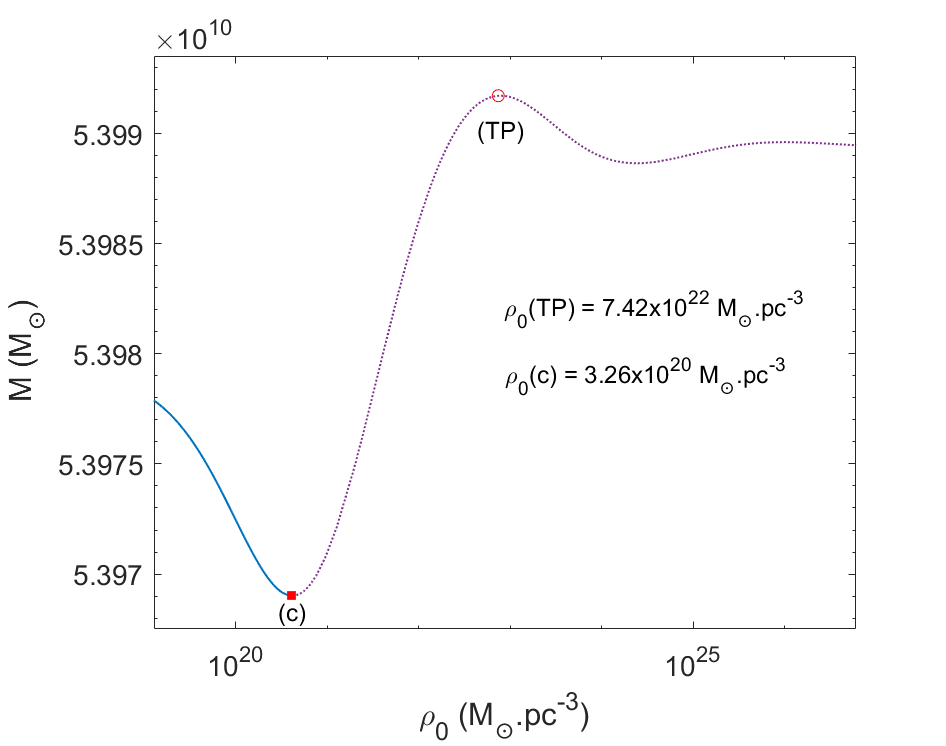}
\caption{Left: equilibrium solutions along the caloric curve for tidally-truncated configurations of $m c^2=48$~keV fermions with fixed $N$. The states within the continuous-blue branches are thermodynamically (and dynamically) stable (i.e., either local or global entropy maxima), while the dotted-violet branches - between (a) and (b) and after (c) - are unstable (i.e., either minimum or saddle point of entropy). Solution (3) is stable and fulfills the virialization conditions. The second spiral of relativistic origin for high $T_\infty$ is characteristic of caloric curves at fixed $N$ within general relativity. It implies the existence of a turning point in a mass-central density curve (see right panel). Right: equilibrium states with $N > N_{\rm OV}$ along the $M$ vs. $\rho_0$ curve, in correspondence with the caloric curve of the left panel. The last stable configuration at the onset of the core collapse occurs at the minimum of this curve and \textit{prior} to the turning point instability (corresponding with point (c) in the left panel). Such a critical solution has a core mass $M_c^{\rm cr}\approx 2\times 10^8 M_\odot$, thus forming an SMBH from DM core-collapse. {Reproduced from \cite{2021MNRAS.502.4227A} with the authors' permission}.}%
\label{fig:thermo1}%
\end{figure}

Specifically, we performed a thermodynamic stability analysis in the microcanonical ensemble for solutions with given particle numbers at halo virialization in a cosmological framework. For the first time, we demonstrate that the above \textit{core-halo} DM profiles are stable (i.e., maxima of entropy) and extremely long-lived. We find a critical point at the onset of instability of the \textit{core-halo} solutions, where the fermion-core collapses towards a supermassive black hole. For particle masses in the keV range, the core-collapse can only occur for  $M_{\rm vir} \gtrsim 10^9 M_\odot$ starting at $z_{\rm vir}\approx 10$ in the given cosmological framework. This is a key result since the fermionic DM system can provide a novel mechanism for SMBH formation in the early Universe, offering a possible solution to the yet-open problem of how SMBHs grow so massive so fast. We recall the result of Fig. \ref{fig:thermo1}, which evidences the existence of a last stable configuration before the core collapse into an SMBH. We refer the reader to \cite{2021MNRAS.502.4227A} for details on this relevant result.

This thermodynamic approach allows a detailed description of the relaxed halos from the very center to the periphery, which $N$-body simulations do not allow due to finite inner-halo resolution. In addition, it includes richer physical ingredients such as (i) general relativity --- necessary for a proper gravitational DM core-collapse to an SMBH; (ii) the quantum nature of the particles --- allowing for an explicit fermion mass dependence in the profiles; (iii) the Pauli principle self-consistently included in the phase-space distribution function --- giving place to novel \textit{core-halo} profiles at (violent) relaxation.

Such treatment allows linking the behavior and evolution of the DM particles from the early Universe to the late stages of non-linear structure formation. We obtain the virial halo mass, $M_{\rm vir}$, with associated redshift $z_{\rm vir}$. The fermionic halos are assumed to be formed by fulfilling a maximum entropy production principle at virialization. It allows obtaining a most likely distribution function of Fermi-Dirac type, as first shown in \citet{1998MNRAS.300..981C} (generalizing Lynden-Bell results), here applied to explain DM halos. Finally, the stability and typical lifetime of such equilibrium states and their possible astrophysical applications are studied with a thermodynamic approach.
   
\begin{figure}[hbtp!]
\centering 
\includegraphics[width=0.6\hsize,clip]{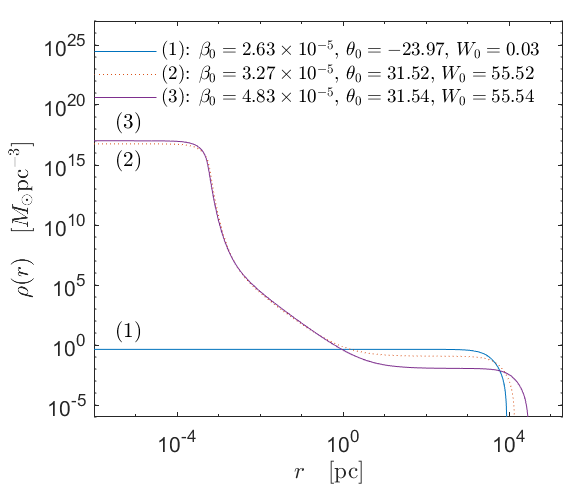} 
\caption{Density profiles for $m c^2=48$~keV corresponding with the equilibrium states of the caloric curve in Fig.~\ref{fig:thermo1} with corresponding fixed total halo mass. Only profiles (1) (resembling a King distribution) and the \textit{core-halo} one (3) are stable, while profile (2) is thermodynamically unstable. Interestingly, solutions like (3) were successfully applied to explain the DM halo in the MW in \cite{2018PDU....21...82A}. They are stable, extremely long-lived, and fulfill the observed surface DM density relation and the expected value of the DM dispersion velocity. {Reproduced from \cite{2021MNRAS.502.4227A} with the authors' permission}.}% 
\label{fig:thermo2}%
\end{figure}

For the first time, we have calculated the caloric curves for self-gravitating, tidally-truncated matter distributions of $\mathcal{O} (10)$ keV fermions at finite temperatures within general relativity. We applied this framework to realistic DM halos (i.e., sizes and masses). With the precise shape of the caloric curve, we establish the families of stable as well as astrophysical DM profiles (see Figs.~\ref{fig:thermo1} and \ref{fig:thermo2}). They are either King-like or develop a \textit{core-halo} morphology that fits the rotation curve in galaxies \citep{2018PDU....21...82A,2019PDU....24100278}. In the first case, the fermions are in the dilute regime and correspond to a global entropy maximum. In the second case, the degeneracy pressure (i.e., Pauli principle) holds the quantum core against gravity and corresponds to a local entropy maximum. Those metastable states are extremely long-lived, and, as such, they are the more likely to arise in nature. Thus, these results prove that DM halos with a \textit{core-halo} morphology are a very plausible outcome within nonlinear stages of structure formation.

%%%%%%%%%%%%%%%%%%%%%%%%%%%%%%%%%%%%%%%%%%%%%%%%%%%%%%%%%%%%%%%%%%%
\subsection{Interactions in warm DM: a view from cosmological perturbation theory}
%%%%%%%%%%%%%%%%%%%%%%%%%%%%%%%%%%%%%%%%%%%%%%%%%%%%%%%%%%%%%%%%%%%

The traditional $\Lambda$CDM paradigm of cosmology is in remarkable agreement with large-scale cosmological observations and galaxy properties. However, there are increasing tensions of the $\Lambda$CDM with observations on smaller scales, such as the so-called \textit{missing} DM sub-halo problem and the \textit{core-cusp} discrepancy.
 
High-resolution cosmological simulations of average-sized halos in $\Lambda$CDM predict an overproduction of small-scale structures significantly larger than the observed number of small satellite galaxies in the Local Group. Moreover, $N$-body simulations of CDM predict a cuspy density profile for virialized halos, while observations show dSphs having flattened smooth density profiles in their central regions. 

A possible alternative to alleviate or try to resolve such tensions is to consider \textit{warm} dark matter (WDM) particles, meaning that they are semi-relativistic during the earliest stages of structure formation with non-negligible free-streaming particle length. 

WDM models feature an intermediate velocity dispersion between HDM and CDM that results in a suppression of structures at small scales due to free-streaming. If this free streaming scale today is smaller than the size of galaxy clusters, it can solve the missing satellites problem. However, thermally produced WDM suffers from the so-called \textit{catch-22} problem when studied within $N$-body simulations \cite{Shao_2013}. Such WDM-only simulations either show unrealistic core sizes for particle masses above the keV range or acquire the right halo sizes for sub-keV masses in direct conflict with phase-space and Lyman-$\alpha$ constraints.

Another compelling alternative to collisionless CDM, apart from WDM, is to consider interactions in CDM. This consideration relaxes the assumption that CDM interacts only gravitationally after early decoupling and includes interactions between DM and SM particles, additional hidden particles, or among DM particles. These later models are denominated as \textit{self-interacting} DM models (SIDM).

To shed light on this matter, we have recently provided a general framework for self-interacting WDM in cosmological perturbations by deriving from first principles a Boltzmann hierarchy that retains certain independence from an interaction Lagrangian \cite{2020JCAP...09..041Y}. Elastic interactions among the massive particles were considered to obtain a more general hierarchy than those usually obtained for non-relativistic (cold DM) or ultra-relativistic (neutrinos) approximations. The more general momentum-dependent kernel integrals in the Boltzmann collision terms are explicitly calculated for different field-mediator models, including a scalar or massive vector field. 

In particular, if the self-interactions maintain the DM fluid in kinetic equilibrium until the fluid becomes non-relativistic, the background distribution function at that moment will switch into a non-relativistic form. This constitutes the scenario known as non-relativistic self-decoupling (a.k.a. late kinetic decoupling). The consequences of this scenario are poorly explored in the literature, and only some preliminary results have been recently obtained within simplified DM fluid approximations \cite{2021JCAP...05..013E}. However, more recently \cite{2021ARep...65.1068Y}, this late kinetic decoupling physics was fully explored with self-interactions treated from first principles using interaction Lagrangian (i.e., superseding the fluid approximation), following the formalism developed in \cite{2020JCAP...09..041Y}. There, it was found that if one imposes continuity of the limiting expressions for the energy density, the non-relativistic distribution function can be found in an analytic expression (see \cite{2021ARep...65.1068Y} for details).

\begin{figure}[htb]
\centering
\includegraphics[width=1.0\hsize,clip]{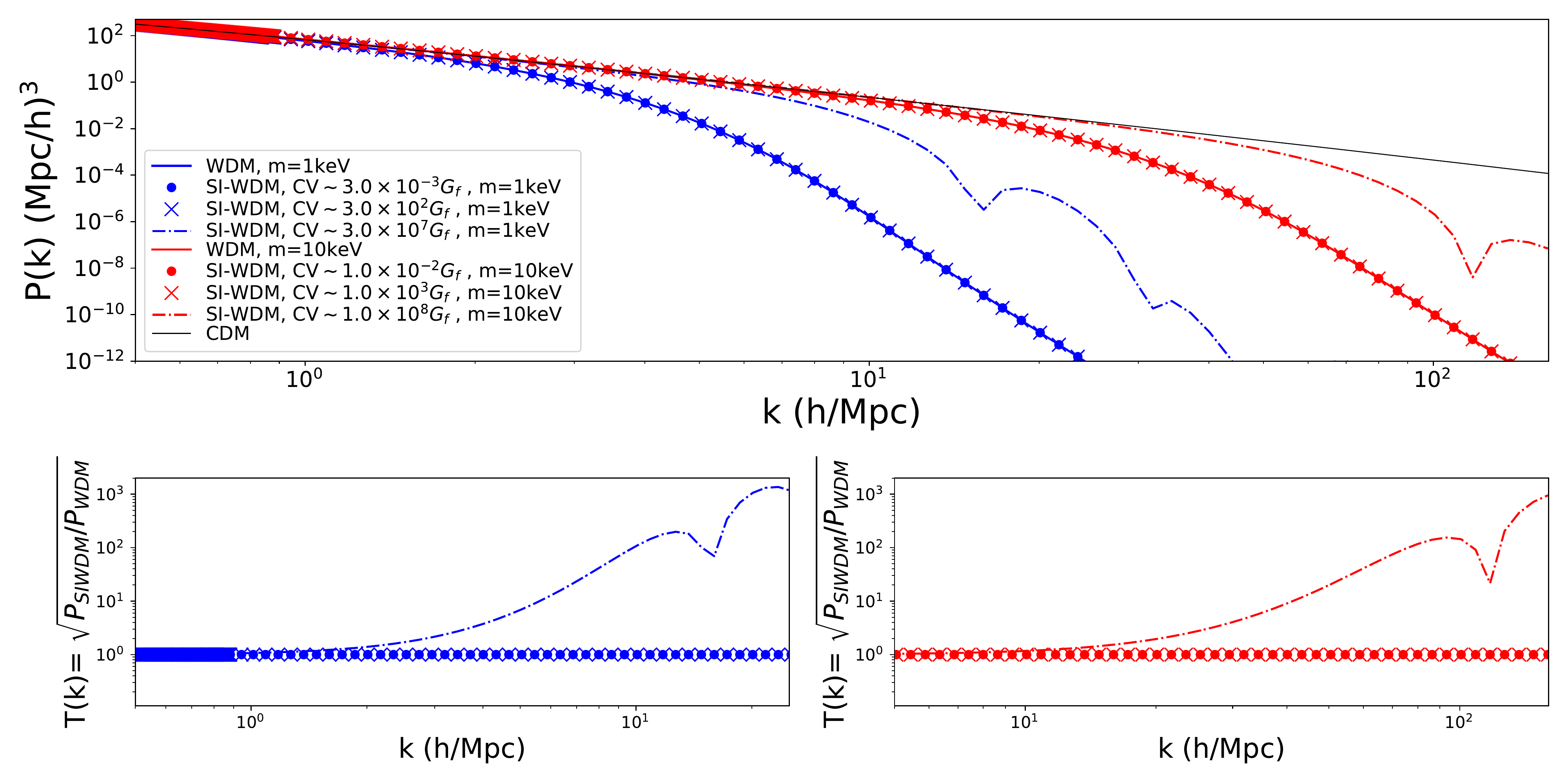}
\caption{
Power Spectrum (\textit{top panel}) and transfer functions compared to standard WDM (\textit{bottom panels}) for a vector field SI-WDM model, using a modification to CLASS. We assume the relaxation time approximation and consider two values of the DM particle mass: $1$ and $10$ keV. Also plotted are the power spectra of CDM and WDM models with DM mass of $1$ and $10$ keV. All WDM and SI-WDM models consider a nonresonant production scenario (Dodelson-Widrow mechanism, \cite{1993sndm.rept.....D}) with $T \sim (4/11)^{1/3} T_\gamma$. Importantly, the effect of large enough self-interactions increases the amount of small structure formed for these models (see dot-dashed curves), implying that the few keV (traditional) WDM models which were excluded in the past, can now be back to an agreement with observations in this new SI-WDM scenarios. {Reproduced from \cite{2021ARep...65.1068Y}, with permission from SNCSC}.}
\label{fig:SIWDM}
\end{figure}

Figure~\ref{fig:SIWDM} illustrates the effects of self-interactions in the matter power spectrum for the case of a massive scalar field-mediator while including the late kinetic decoupling case. We have used an extended version of the CLASS code incorporating our results for SI-WDM with particle masses in the $\sim$ keV range. There, we see some of the particular features of the models. With the inclusion of self-interactions, for models with non-relativistic self-decoupling (i.e., late kinetic decoupling), the resulting power spectra may differ significantly from their relativistic counterparts. Indeed, we find that in this regime, the models are ``colder'' (i.e., as if they correspond to a higher particle mass) and show, even at smaller $k$ values, a distinctive oscillatory pattern (see, e.g., dot-dashed curves in Fig. \ref{fig:SIWDM}. This increases the small structures for these models, implying that the few keV (traditional) WDM models excluded from phase-space arguments now agree with observations in this new SI-WDM scenario. 

Most tensions inherent to $\nu$MSM WDM models arise from structure formation, i.e., the MW satellite counts and Lyman-$\alpha$ observations: the preferred parameter ranges may underproduce small structures and almost rule out the available parameter space. So, including self-interactions can significantly relax the existing bounds on this family of models. The prediction of these models for the number of MW satellites and the observations of the Lyman-$\alpha$ forest was presented in \cite{2022JCAP...02..024Y} (see Fig. \ref{fig:Obs_Results_SInuMSM}). 

\begin{figure}
	\centering
	\includegraphics[width=0.48\hsize]{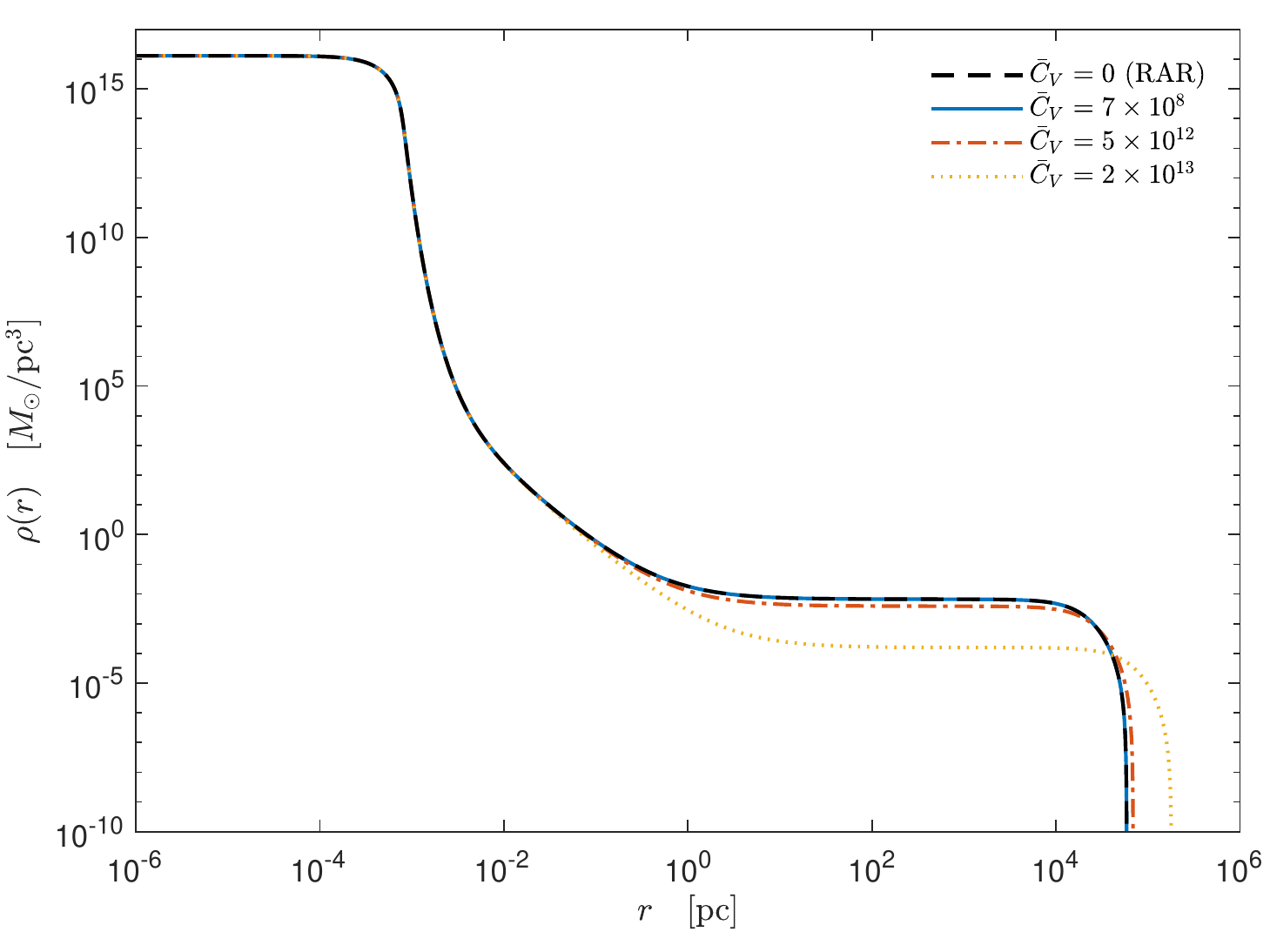}
	\includegraphics[width=0.48\hsize]{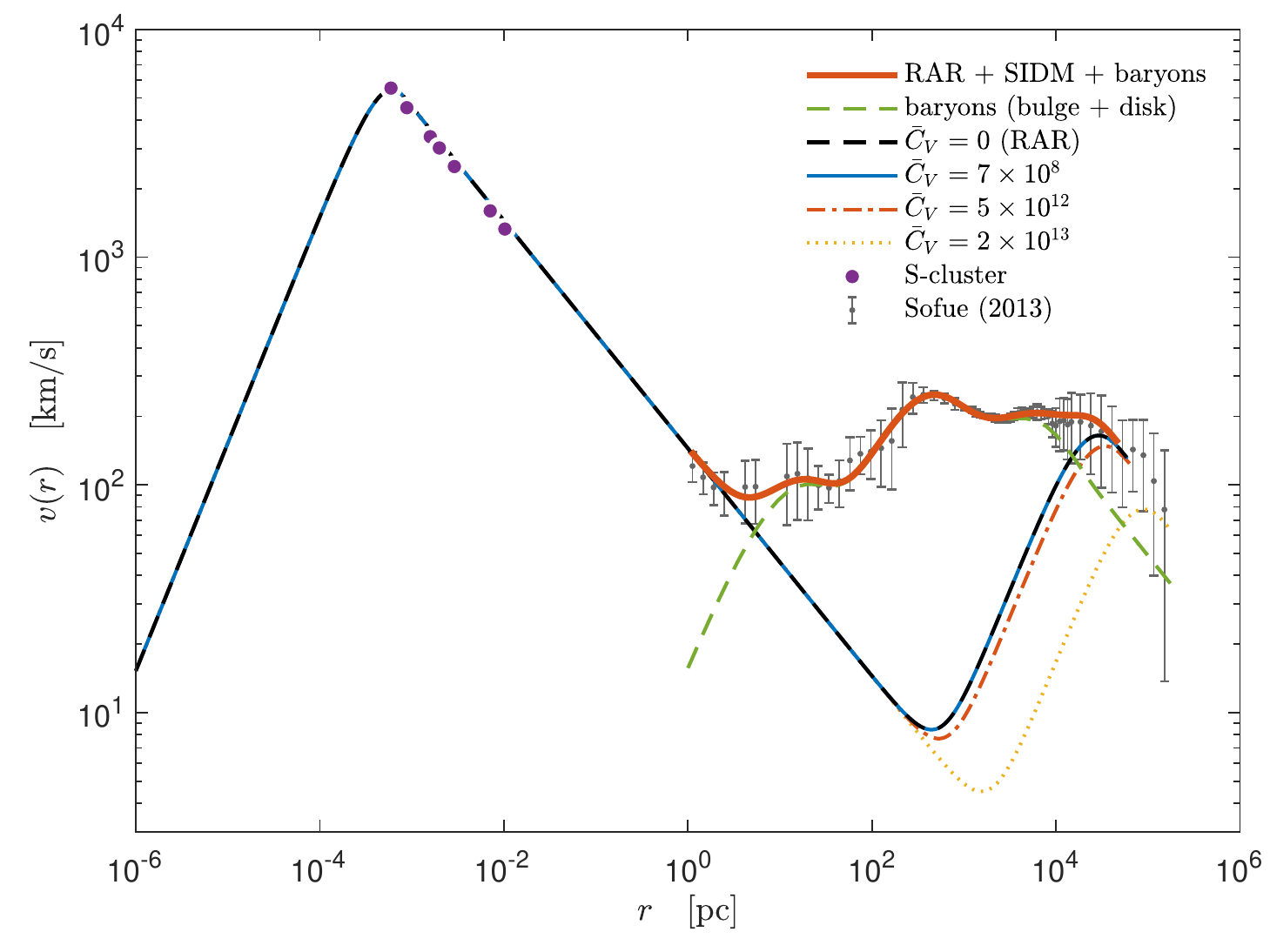}
    \caption{Density profiles (left panel) and rotation curves (right panel) of the extended RAR model with both self-interactions and escape of particle effects. Solutions are shown with four different self-interaction constant values $\bar C_V$, including the case of $\bar C_V = 0$ in black dashed, corresponding to the original Milky Way fit presented in \cite{2018PDU....21...82A}. The model fits the observed total rotation curve data from \cite{2013PASJ...65..118S} (with error bars). The predicted rotation curve (in solid red) shows the excellent fit to the data for $\bar C_V= 7\times10^8$ (in solid blue) and baryonic components (in green dashed). Notice that the DM halo becomes more extended and less dominant, implying an underfit to the data for large enough $\bar C_V > 10^{13}$, due to the additional pressure term contribution of the boson mediator field. {Reprinted from \cite{2020PDU....3000699Y}, Copyright (2020), with permission from Elsevier}.}.
	\label{fig:rho-vcirc-SIDM}
\end{figure}

\begin{figure}[htbp]
\centering
\includegraphics[width=0.8\hsize,clip]{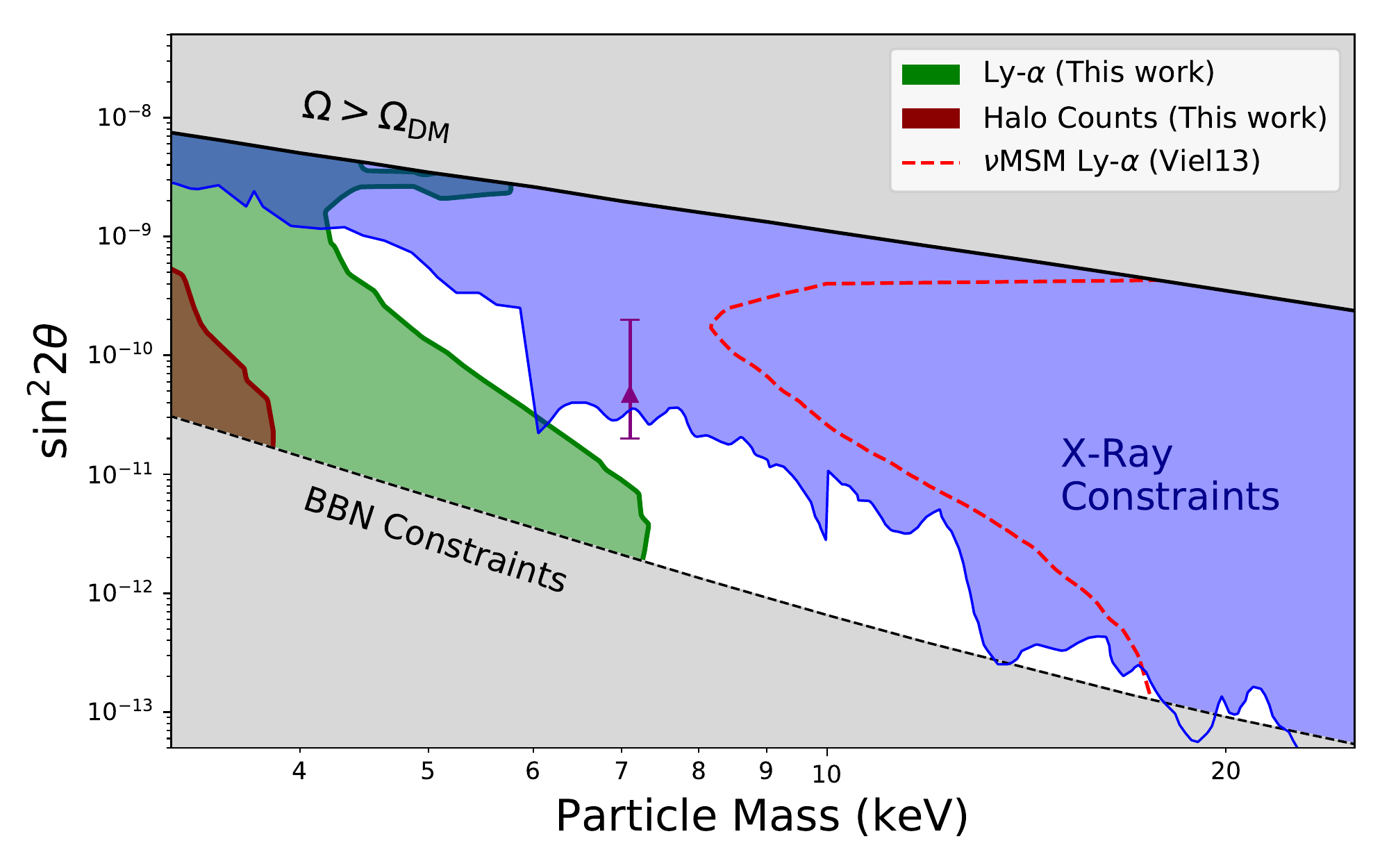}
\caption{\small 
Parameter space constraints for $\nu$MSM, where MW satellite halo counts and Lyman-$\alpha$ forest bounds are analyzed under a self-interacting model as outlined above. For each point $(\theta,m)$ in the parameter space, we consider a self-interacting model under a vector field mediator, with its interaction constant given by $\sigma/m \sim 0.144 C_v^2 / m^3 = 0.1$ cm$^2$ g$^{-1}$, the upper limit given by Bullet Cluster constraints (see \cite{2022JCAP...02..024Y} for details). For comparison, we plot the Lyman-alpha bounds for the non-interacting case for a comparable analysis, plus other bounds to the $\nu$MSM parameter space for informative purposes, namely X-Ray indirect detection bounds (in blue) and sterile neutrino production bounds (in grey). We also plot the sterile neutrino model compatible with a tentative $3.5$ keV DM signal, the subject of debate in recent years, as a purple triangle. The complete list of references of all such bounds can be found in \cite{2022JCAP...02..024Y}. {Reprinted from \cite{2022JCAP...02..024Y}. Copyright IOP Publishing. Reproduced with permission}.}
\label{fig:Obs_Results_SInuMSM}
\end{figure}

{While in the above paragraphs, we have discussed the effects of self-interacting $\mathcal{O}(1-10)$~keV light fermionic candidates on the linear-structure formation (i.e., its suppression effects on the linear matter power spectrum and its consequences on Lyman-$\alpha$ forest and satellite counts \cite{2022JCAP...02..024Y}), we will discuss in next its effects on DM halo profiles. For this, we will follow the specific self-interacting model where the \textit{darkinos} are right-handed neutrinos self-interacting via massive (axial) vector-boson mediators (also considered as a possible case in \cite{2022JCAP...02..024Y}). The main consequences of this model for DM halo structures were studied in \cite{2020PDU....3000699Y}, for a particle mass $m c^2\approx 50$ keV. In particular, it mainly investigated the consequences of the Milky Way DM halo and the bullet cluster. The massive boson mediator adds a pressure term in the equilibrium equations (see, e.g., \cite{2016JCAP...04..038A, 2020PDU....3000699Y}) which, for normalized
interaction constants up to $\bar C_V\sim 10^{12}$ (in Fermi constant units), causes no appreciable effects in the Milky Way rotation curve. However, values larger than $\bar C_V\sim 10^{13}$ are ruled out since the additional pressure term is enough to push forward the halo, spoiling the fit to the data, see Fig. \ref{fig:rho-vcirc-SIDM}. Interestingly, values of interaction strengths between our light fermionic candidates of $\bar C_V\sim 10^8$ agree with the bullet cluster measurements. Indeed, on cluster scales, it was demonstrated in \cite{2020PDU....3000699Y} (see section 3.2 therein) that such interaction constants resolve the tensions between predictions of $\Lambda$CDM-based numerical simulations and observations since the corresponding self-interacting DM (SIDM) cross section\footnote{The connection between the self-interaction constant ($\bar C_V$) and the cross section is given by $\sigma_{SIDM}\approx C^2_V 29 m^2/(4^3\pi)$ as calculated in \cite{2016JCAP...04..038A} within an electroweak-like formalism for an elastic scattering process.} ($\sigma_{SIDM}$) lies in the expected range~\cite{2015Sci...347.1462H}
\begin{equation}
0.1 \,\,{\rm cm}^2\,{\rm g}^{-1} \le \frac{\sigma_{\rm SIDM}}{m} \, \le 0.47\,\,{\rm cm}^2\,{\rm g}^{-1}.
\label{eq:vector_newconstr}
\end{equation}
}

%%%%%%%%%%%%%%%%%%%%%%%%%%%%%%%%%%%%%%%%%%%%%%%%%%%%%%%%%%%%%%%%%%%
%%%%%%%%%%%%%%%%%%%%%%%%%%%%%%%%%%%%%%%%%%%%%%%%%%%%%%%%%%%%%%%%%%%
\section{Additional fermionic DM probes}\label{sec:8}
%%%%%%%%%%%%%%%%%%%%%%%%%%%%%%%%%%%%%%%%%%%%%%%%%%%%%%%%%%%%%%%%%%%
%%%%%%%%%%%%%%%%%%%%%%%%%%%%%%%%%%%%%%%%%%%%%%%%%%%%%%%%%%%%%%%%%%%

%%%%%%%%%%%%%%%%%%%%%%%%%%%%%%%%%%%%%%%%%%%%%%%%%%%%%%%%%%%%%%%%%%%
\subsection{Gravitational lensing}
%%%%%%%%%%%%%%%%%%%%%%%%%%%%%%%%%%%%%%%%%%%%%%%%%%%%%%%%%%%%%%%%%%%

Over the last decade, gravitational lensing has become a very effective tool to probe the DM distribution of systems from galaxy to cluster scales. The availability of high-quality HST imaging and integral field spectroscopy with the VLT has allowed high-precision strong lensing models to be developed in recent years, with tight constraints on the inner mass distribution of galaxy clusters (e.g., \cite{2019A&A...631A.130B}). By comparing high-resolution mass maps of galaxy cluster cores obtained by such lens models with cosmological hydrodynamical simulations in the LCDM framework, a tension has emerged in the mass profile of the sub-halo population associated with cluster galaxies \cite{2020Sci...369.1347M}. Namely, the observed sub-halos appear more compact than those in simulations, with circular velocities inferred from the lens models, which are higher, for a given sub-halo halo mass, than those derived from simulations. It is still unclear whether this tension is due to some limitations of the numerical simulations in the modeling of baryonic physics (back-reaction effects on DM) or rather a more fundamental issue related to the CDM paradigm. 

In \cite{2016PhRvD..94l3004G}, the lensing effects of the core-halo DM distribution were computed. The NFW and nonsingular isothermal sphere (NSIS) DM models and the central Schwarzschild BH model were compared at distances near the Galaxy center. The DM density profiles lead to small deviations of light (of $0.1$ arcsec) in the halo part ($\sim 8$ kpc). However, the RAR profile produces strong lensing effects when the dense core becomes highly compact. For instance, the density distribution of the RAR model for a fermion mass of $\sim 100$ keV generates strong lensing features at distances $\sim 0.1$ mpc (see \cite{2016PhRvD..94l3004G} for additional details). The DM quantum core has no photon sphere inside or outside but generates multiple images and Einstein rings. Thus, it will be interesting to compare and contrast lensing images produced by the RAR model with the ones produced by phenomenological profiles.

It is also interesting to construct the shadow-like images produced by the fermion DM cores of the RAR model and analyze them in light of the EHT data from Sgr A*. {Indeed, horizonless objects can be compact, lack a hard surface, and cast a shadow surrounded by a ring-like feature of lensed photons \cite{2019LRR....22....4C}. This has been shown for boson stars for Sgr A* in \cite{2016CQGra..33j5015V} and further studied by \cite{2020MNRAS.497..521O} within Numerical Relativity simulations. Using full general relativistic ray tracing techniques \cite{2022MNRAS.515.1316P}, we have recently computed relativistic images of the fermionic DM cores of the RAR model under the assumption that photons are emitted from a surrounding accretions disc (self-consistently computed in the given metric) and have shown that there exists a particle mass range in which the shadow feature acquires the typical sizes as resolved for the EHT in Sgr A* (Pelle, Argüelles, et al., to be submitted).}

%%%%%%%%%%%%%%%%%%%%%%%%%%%%%%%%%%%%%%%%%%%%%%%%%%%%%%%%%%%%%%%%%%%
\subsection{Dynamical friction}
%%%%%%%%%%%%%%%%%%%%%%%%%%%%%%%%%%%%%%%%%%%%%%%%%%%%%%%%%%%%%%%%%%%

The measured orbital period decay of relativistic compact-star binaries (e.g., the famous Hulse-Taylor binary pulsar) has been explained with very-high precision by the gravitational-wave emission predicted by general relativity of an inspiraling binary, assuming the binary is surrounded by empty space and within the point-like approximation of the two bodies. However, the presence of DM around the binary might alter the orbital dynamics because of a traditionally neglected phenomenon: DM dynamical friction (DMDF). The binary components interact with their own gravitational wakes produced by the surrounded DM, leading to an orbital evolution that can be very different from the orbital evolution solely driven by the emission of gravitational waves. 

In \cite{2015PhRvD..92l3530P, 2017PhRvD..96f3001G}, the effect of the DMDF on the motion of NS-NS, NS-WD, and WD-WD was evaluated. Quantitatively, the crucial parameters are the orbital period and the value of the DM density at the binary location in the galaxy. A comparison among the DMDF produced by the NFW, the NSIS, and the RAR model was presented. 

For NS-NS, NS-WD, and WD-WD with measured orbital decay rates, the energy loss by gravitational waves dominates over the DMDF effect. However, there are astrophysically viable conditions for which the two effects become comparable. For example, $1.3$--$0.2$ $M_\odot$ NS-WD, $1.3$--$1.3$~$M_\odot$ NS-NS, and $0.25$--$0.50$~$M_\odot$ WD-WD, located at $0.1$~kpc, the two effects compete with each other for a critical orbital period in the range $20$--$30$ days for the NFW model. For the RAR model, it occurs at an orbital period $\sim 100$ days (see \cite{2017PhRvD..96f3001G} for details). Closer to the Galactic center, the DMDF effect keeps increasing, so the above values for the critical orbital period become shorter. For certain system parameters, the DMDF can lead to an orbital widening rather than a shrinking. 

The DMDF depends on the density profile, velocity distribution function, and velocity dispersion profile. Therefore, measuring with high accuracy the orbital decay rate of compact-star binaries at different Galactic locations and as close as possible to the Galactic center might reveal a powerful tool to test DM models.

{Indeed, during the peer-reviewing process of this review, \citet{2023ApJ...943L..11C} published an analysis of the DMDF contribution to the orbital decay rate of the X-ray binaries formed by stellar-mass BHs and stellar companions. Specifically, they analyzed the case of A0620--00 and XTE J1118+480, finding that the DMDF can explain their fast orbital decay. The analysis in \cite{2023ApJ...943L..11C} uses a phenomenological DM profile, so it will be interesting to apply the DMDF with the RAR model, as done in \cite{2017PhRvD..96f3001G}, to check whether or not X-ray binaries can further constrain the fermion mass.}

%%%%%%%%%%%%%%%%%%%%%%%%%%%%%%%%%%%%%%%%%%%%%%%%%%%%%%%%%%%%%%%%%%%
\subsection{Gravitational collapse of DM cores and SMBH formation}
%%%%%%%%%%%%%%%%%%%%%%%%%%%%%%%%%%%%%%%%%%%%%%%%%%%%%%%%%%%%%%%%%%%

The explanation of the formation, growth, and nature of SMBHs observed at galaxy centers are amongst the most relevant problems in astrophysics and cosmology. Relevant questions yet waiting for an answer are: how can they increase their size in a relatively short time to explain the farthest quasars \cite{2012Sci...337..544V, 2019PASA...36...27W}?; where do the BH seeds come from and how large they must be to form SMBHs of $\sim 10^8$--$10^9 M_\odot$ at high cosmological redshift \cite{2022MNRAS.514.5583Z}?; and what is the connection between the host galaxy and central SMBH masses \cite{2021NatRP...3..732V}?

Scenarios on the origin of SMBHs can be divided according to the formation channel (see \cite{2020ARA&A..58...27I, 2021NatRP...3..732V} for recent reviews):
\begin{enumerate}
    \item[(I)] Channels that advocate a baryonic matter role (gas and stars). (Ia) \textit{Population III stars} and (Ib) \textit{direct collapse to a BH (DCBH)}. Scenarios (Ia) produce BH seeds $\sim 10^2 M_\odot$ \cite{2001ApJ...551L..27M, 2016ApJ...824..119H}, so very-high accretion rates are needed to grow them to $10^9 M_\odot$ in a few billion years. Simulations show that BHs of $<10^3 M_\odot$ cannot grow to $10^8 M_\odot$ at cosmological redshift $z \sim 6$ due to radiative feedback \cite{2022MNRAS.514.5583Z}. DCBH scenarios (Ib) produce BH seeds in the range $10^4$--$10^5 M_\odot$ \cite{2006MNRAS.370..289B, 2008MNRAS.387.1649B, 2017ApJ...842L...6W}. Hydrodynamic N-body simulations show some preference for DCBH scenarios \cite{2022MNRAS.514.5583Z, 2022Natur.607...48L}, although numerical and ad-hoc assumptions limit the results generality \cite{2022MNRAS.514.5583Z}.
    \item[(II)] Early Universe channels where BH seeds form before galaxy formation. They include primordial BHs \cite{2020ARNPS..70..355C} and exotic candidates like topological defects \cite{2015JCAP...06..007B}. However, these scenarios are difficult to prove or disprove since those processes are hypothesized to occur in early cosmological epochs not accessed by observations.
\end{enumerate}

Recently, we have proposed in Arguelles et al. (submitted) an SMBH formation channel conceptually different from cases (I) and (II). The new scenario is based on the gravitational collapse and subsequent growth of dense fermionic DM cores. The cores originate at the center of the halos and start to grow as they form. As we have recalled in this work, those \textit{dense core}-\textit{diluted halo} DM density distributions are predicted by maximum entropy production principle models of halo formation \cite{2021MNRAS.502.4227A, 2022IJMPD..3130002A}. We have recalled how the core made of fermions of $\sim 50$ keV rest mass-energy becomes unstable against gravitational collapse into a BH for a threshold mass of $\sim 10^8~M_\odot$. Therefore, the new scenario produces larger BH seeds that comfortably grow to SMBH mass values of $10^9 M_\odot$ in a relatively short time without super-Eddington accretion (Arguelles et al., submitted).

%%%%%%%%%%%%%%%%%%%%%%%%%%%%%%%%%%%%%%%%%%%%%%%%%%%%%%%%%%%%%%%
%%%%%%%%%%%%%%%%%%%%%%%%%%%%%%%%%%%%%%%%%%%%%%%%%%%%%%%%%%%%%%%
\section{Conclusions}\label{sec:9}
%%%%%%%%%%%%%%%%%%%%%%%%%%%%%%%%%%%%%%%%%%%%%%%%%%%%%%%%%%%%%%%
%%%%%%%%%%%%%%%%%%%%%%%%%%%%%%%%%%%%%%%%%%%%%%%%%%%%%%%%%%%%%%%

Possibly one of the most relevant steps has been to develop a framework that allows obtaining the DM density (and other related physical properties) profiles from first principles. The distribution of DM in the galaxy is given by the solution of the Einstein equations subjected to appropriate boundary conditions to fulfill observational data. The nature of the particle candidate sets the equation of state, so the energy-momentum tensor. It turns out that neutral massive fermion DM particles distribute throughout the galaxy following a dense core-dilute halo density profile. See Section \ref{sec:2} for details.

Thanks to the above, the nature of the DM particle, like the rest mass, can be constrained using stellar dynamics data, e.g., the MW rotational curves data and the most accurate data of the S-cluster stellar orbits around Sgr A*. See Section \ref{sec:3} for details.

A remarkable and intriguing result is that the orbits of the S-stars are correctly described by the presence of the dense core of fermions without the need for the presence of a massive BH at the MW center. We have shown that even the most accurate available data of the S2 star can not distinguish between the two models since both models describe the data with comparable accuracy. However, we have shown that the data of the periapsis precession of the S2 star in the forthcoming three years (i.e., by 2026) could help to discriminate the two scenarios. See Section \ref{sec:3} for details.

In addition to the MW, the RAR model describes the DM component observed in other galaxy types, namely dSphs, ellipticals, and galaxy clusters. This point includes the explanation of existing observational Universal relations and the comparison with other DM models regarding the same observables. See Section \ref{sec:4} for details.

All the above constrains the DM particle nature limiting the fermion mass to the $\sim 50$--$350$ keV range, constituting a relevant starting point for complementary particle physics analyses. Among the plethora of particle-physics candidates, the beyond-standard model sterile neutrinos (also including self-interactions) remain an interesting candidate for being the fermions of the RAR model. See Section \ref{sec:5} for details.

None DM discussion is complete without examining the cosmological implications. In this order of ideas, it has been first shown that the core-halo configurations can indeed form in the Universe in appropriate cosmological timescales thanks to the mechanism of violent relaxation that predicts equilibrium states described by the Fermi-Dirac phase-space distribution. Second, it has been shown that those equilibrium states are long-lasting, with a lifetime of several cosmological timescales, so well over the Universe's lifetime. Third, it has been shown that those small-scale DM core-halo substructures can form from non-linear cosmological density perturbations during the cosmological evolution. See Section \ref{sec:6} for details.

Last but not least, we have discussed some additional theoretical and observational scenarios that can help to probe DM models. Specifically, if DM permeates galaxies, stellar objects do not sit in a perfect vacuum. Thus, the distribution of DM can cause dynamical friction, a purely gravitational effect that can alter a purely Keplerian motion of binaries, and which, under certain conditions, can become as large as gravitational wave emission losses. Strong gravitational lensing is potentially sensitive to the density profile of DM at sufficiently small scales where DM model profiles differ. We have also outlined how the SMBHs observed at the center of active galaxies can be formed from BH seeds from the gravitational collapse of dense cores of fermionic DM. See Section \ref{sec:6} for details.

In summary, we express that only theoretical models that join microphysics and macrophysics, such as the RAR model, can lead to a comprehensive set of predictions ranging from participle physics to galactic stellar dynamics and to cosmology, which can be put to direct observational scrutiny. We hope that future observations and a further refined analysis of DM models, including the quantum nature of the particles, will lead to a breakthrough in revealing the DM nature.

%%%%%%%%%%%%%%%%%%%%%%%%%%%%%%%%%%%%%%%%%%
\vspace{6pt} 

%%%%%%%%%%%%%%%%%%%%%%%%%%%%%%%%%%%%%%%%%%
%% optional
%\supplementary{The following supporting information can be downloaded at:  \linksupplementary{s1}, Figure S1: title; Table S1: title; Video S1: title.}


%%%%%%%%%%%%%%%%%%%%%%%%%%%%%%%%%%%%%%%%%%
% \authorcontributions{For research articles with several authors, a short paragraph specifying their individual contributions must be provided. The following statements should be used ``Conceptualization, X.X. and Y.Y.; methodology, X.X.; software, X.X.; validation, X.X., Y.Y. and Z.Z.; formal analysis, X.X.; investigation, X.X.; resources, X.X.; data curation, X.X.; writing---original draft preparation, X.X.; writing---review and editing, X.X.; visualization, X.X.; supervision, X.X.; project administration, X.X.; funding acquisition, Y.Y. All authors have read and agreed to the published version of the manuscript.'', please turn to the  \href{http://img.mdpi.org/data/contributor-role-instruction.pdf}{CRediT taxonomy} for the term explanation. Authorship must be limited to those who have contributed substantially to the work~reported.}

\funding{This research received no external funding.}

\dataavailability{No new data were created or analyzed in this study. Data sharing is not applicable to this article.} 

%\acknowledgments{In this section you can acknowledge any support given which is not covered by the author contribution or funding sections. This may include administrative and technical support, or donations in kind (e.g., materials used for experiments).}

\conflictsofinterest{The authors declare no conflict of interest.} 

%%%%%%%%%%%%%%%%%%%%%%%%%%%%%%%%%%%%%%%%%%
%% Optional
% \sampleavailability{Samples of the compounds ... are available from the authors.}

%% Only for journal Encyclopedia
%\entrylink{The Link to this entry published on the encyclopedia platform.}

% \abbreviations{Abbreviations}{
% The following abbreviations are used in this manuscript:\\

% \noindent 
% \begin{tabular}{@{}ll}
% MDPI & Multidisciplinary Digital Publishing Institute\\
% DOAJ & Directory of open access journals\\
% TLA & Three letter acronym\\
% LD & Linear dichroism
% \end{tabular}
% }

%%%%%%%%%%%%%%%%%%%%%%%%%%%%%%%%%%%%%%%%%%
%% Optional
\appendixtitles{yes} % Leave argument "no" if all appendix headings stay EMPTY (then no dot is printed after "Appendix A"). If the appendix sections contain a heading then change the argument to "yes".

\appendixstart
\appendix

%%%%%%%%%%%%%%%%%%%%%%%%%%%%%%%%%%%%%%%%%%%%%%%%%%%%%%%%%%%%%%%
%%%%%%%%%%%%%%%%%%%%%%%%%%%%%%%%%%%%%%%%%%%%%%%%%%%%%%%%%%%%%%%
\section{Equations of motion and effective potential}
%%%%%%%%%%%%%%%%%%%%%%%%%%%%%%%%%%%%%%%%%%%%%%%%%%%%%%%%%%%%%%%
%%%%%%%%%%%%%%%%%%%%%%%%%%%%%%%%%%%%%%%%%%%%%%%%%%%%%%%%%%%%%%%

In general relativity, the Lagrangian for a free particle in a gravitational field can be expressed as
   
   \begin{equation}
      \displaystyle{ \mathcal{L}\left(x^\alpha, \dfrac{dx^\alpha}{d\lambda}\right) = \dfrac{1}{2} g_{\mu\nu}(x^\alpha)\dfrac{dx^\mu}{d\lambda}\dfrac{dx^\nu}{d\lambda} \equiv \dfrac{1}{2} g_{\mu\nu}(x^\alpha) \dot{x}^\mu \dot{x}^\nu },
   \end{equation}\
   
   \noindent where $g_{\mu\nu}$ are the covariant components of the metric tensor, $x^\beta$ are the space-time coordinates, and $\lambda$ is an affine parameter. In the space of the curves describe by $\{x^\mu(\lambda),\lambda\in[\lambda_i,\lambda_f]\}$, the action is defined as
   
    \begin{equation}
        \mathcal{S}=\int \mathcal{L}(x^\alpha,\dot{x}^\alpha) d\lambda = \dfrac{1}{2} \int g_{\mu\nu}(x^\alpha) \dot{x}^\mu \dot{x}^\nu d\lambda.
   \end{equation}\
   
   The Euler-Lagrange equations are obtained, as usual, by varying the action with respect to the coordinates, and by setting the variation equal to zero. For massive particles, we can change the parameter $\lambda$ by the proper time $\tau$, so by varying a curve

   \begin{equation}
       x^\mu(\tau)\longrightarrow x^\mu(\tau) + \delta x^\mu(\tau)\ ,
   \end{equation}
   \noindent with 
   \begin{equation}
   \delta x^\mu(\tau_i) = \delta x^\mu(\tau_f) = 0\ , 
   \end{equation}

   \noindent the action variation is

    \begin{equation}\label{eq:action}
      \displaystyle{  \delta \mathcal{S}=\int \left[\dfrac{\partial\mathcal{L}}{\partial x^\mu}\delta x^\mu + \dfrac{\partial\mathcal{L}}{\partial \dot{x}^\mu}\delta( \dot{x}^\mu)\right] d\tau }.
   \end{equation}\\

   Since \ $\displaystyle{\delta( \dot{x}^\mu) = \delta(dx^\mu/d\tau) = d\delta x^\mu/d\tau}$, \ the last term in Eq.~(\ref{eq:action}) can be written as 
   
    \begin{equation}
        \dfrac{\partial\mathcal{L}}{\partial \dot{x}^\mu}\delta( \dot{x}^\mu) = \dfrac{\partial\mathcal{L}}{\partial \dot{x}^\mu} \dfrac{d\delta x^\mu}{d\tau} = \dfrac{d}{d\tau}\left(\dfrac{\partial\mathcal{L}}{\partial \dot{x}^\mu}\delta x^\mu\right) - \dfrac{d}{d\tau}\left(\dfrac{\partial\mathcal{L}}{\partial \dot{x}^\mu}\right) \delta x^\mu.
   \end{equation}\

   When integrated between $\tau_i$ and $\tau_f$ the first term on the right-hand side in the above equation vanishes because \ $\delta\dot{x}^\mu(\tau_i)=\delta\dot{x}^\mu(\tau_f)=0$, \, therefore, Eq.~(\ref{eq:action}) become 

    \begin{equation}
      \displaystyle{  \delta \mathcal{S}=\int \left[\dfrac{\partial\mathcal{L}}{\partial x^\mu}\delta x^\mu - \dfrac{d}{d\tau}\left(\dfrac{\partial\mathcal{L}}{\partial \dot{x}^\mu}\right) \delta x^\mu\right] d\tau },
   \end{equation}\   
   
   \noindent so the action variation vanish for all $\delta x^\mu$ if, and only if, it is satisfied that
   
   \begin{equation}\label{eq:Euler-Lagrange}
     \dfrac{\partial\mathcal{L}}{\partial x^\mu} - \dfrac{d}{d\tau}\left(\dfrac{\partial\mathcal{L}}{\partial \dot{x}^\mu}\right) = 0.
   \end{equation}\  
   
  For the spherically symmetric metric (\ref{eqn:metric}), the Lagrangian of a free particle is
   
   \begin{equation}\label{eq:lagrangian}
        \mathcal{L}=\dfrac{1}{2} \left[ g_{00}(r) \ \dot{t}^2 - g_{11}(r) \ \dot{r}^2 - r^2 \ \dot{\theta}^2 - r^2 \sin^2\theta \ \dot{\phi}^2\right],
   \end{equation}\
   
   \noindent where the dot indicates differentiation with respect to $\tau$, i.e., $\dot{x}^\mu=dx^\mu/d\tau$. Replacing the Lagrangian, Eq.~(\ref{eq:lagrangian}), in the Euler-Lagrange expression, Eq.~(\ref{eq:Euler-Lagrange}), the equations of motion for $\dot{t}$, $\dot{r}$, $\dot{\theta}$ and $\dot{\phi}$ are\\
   
   \noindent for $\dot{t}$:
   \begin{equation*}
        \dfrac{\partial\mathcal{L}}{\partial t} - \dfrac{d}{d\tau}\left(\dfrac{\partial\mathcal{L}}{\partial \dot{t}}\right) = 0 \quad \longrightarrow \quad \dfrac{d}{d\tau}\left(g_{00}\dot{t}\right)=0 \quad \longrightarrow \quad g_{00} \ \dot{t}=\rm{const}\equiv E, 
   \end{equation*}\
   
   \noindent for $\dot{r}$:
   \begin{eqnarray*}
        \dfrac{\partial\mathcal{L}}{\partial r} - \dfrac{d}{d\tau}\left(\dfrac{\partial\mathcal{L}}{\partial \dot{r}}\right) = 0 & \quad   \rightarrow  \quad & \dfrac{d}{d\tau}\left(-g_{11}\dot{r}\right) - \dfrac{1}{2}(\dot{g}_{00}\dot{t}^2 - \dot{g}_{11}\dot{r}^2 - 2r\dot{\theta}^2 - 2r\sin^2\theta\dot{\phi}^2 )=0 \\
        & \quad   \rightarrow  \quad & g_{11} \ \ddot{r} + \dfrac{1}{2} \ \dot{g}_{00} \ \dot{t}^2 + \dfrac{1}{2} \ \dot{g}_{11} \ \dot{r}^2 - r \ \dot{\theta}^2 - r\sin^2\theta \ \dot{\phi}^2 = 0,
   \end{eqnarray*}
   
   \noindent for $\dot{\theta}$:
   \begin{eqnarray*}
        \dfrac{\partial\mathcal{L}}{\partial \theta} - \dfrac{d}{d\tau}\left(\dfrac{\partial\mathcal{L}}{\partial \dot{\theta}}\right) = 0 & \quad   \rightarrow  \quad & \dfrac{d}{d\tau}\left(-r^2\dot{\theta}\right) - ( -r^2 \sin\theta\cos\theta\dot{\phi}^2 )=0 \\
        & \quad   \rightarrow  \quad & r^2 \ \ddot{\theta} + 2r \ \dot{\theta} \ \dot{r} - r^2\sin\theta\cos\theta \ \dot{\phi}^2 = 0 ,
   \end{eqnarray*}
   
    \noindent for $\dot{\phi}$:
   \begin{equation*}
        \dfrac{\partial\mathcal{L}}{\partial\phi} - \dfrac{d}{d\tau}\left(\dfrac{\partial\mathcal{L}}{\partial \dot{\phi}}\right) = 0 \quad \rightarrow \quad \dfrac{d}{d\tau}\left(-r^2\sin^2\theta\dot{\phi}\right)=0  \ \rightarrow \ r^2\sin^2\theta \ \dot{\phi}=\rm{const}\equiv L .
   \end{equation*}\
   
   Due to the spherical symmetry, the metric is invariant under rotations of the polar coordinate. Therefore, we can assume without loss of generality $\theta=\pi/2$. In this case, our set of EOM is the following
   \begin{eqnarray}
       & & g_{00} \ \dot{t} = E,\label{eq:EOM1}\\
       & & g_{11} \ \ddot{r} + \dfrac{1}{2} \ \dot{g}_{00} \ \dot{t}^2 + \dfrac{1}{2} \ \dot{g}_{11} \ \dot{r}^2 - r \dot{\phi}^2 = 0,\\
       & & r^2 \dot{\phi} = L ,\label{eq:EOM3}
   \end{eqnarray}
   
   \noindent where $E$ and $L$ are the conserved energy and the angular momentum of particle per unit mass.
   
    From the condition for mass particle geodesics, $g_{\mu\nu}\dot{x}^\mu\dot{x}^\nu=1$, and the equation of motion for $t(\tau)$ and $\phi(\tau)$, we can obtain the effective potential. For massive particles must be fulfilled that
    \begin{equation}\label{eq:geodesic}
        g_{\mu\nu}\dot{x}^\mu\dot{x}^\nu = g_{00}(r) \ \dot{t}^2 - g_{11}(r) \ \dot{r}^2 - r^2 \ \dot{\theta}^2 - r^2 \sin^2\theta \ \dot{\phi}^2 = 1,
    \end{equation}\
    
    \noindent so for $\theta=\frac{\pi}{2}$ and replacing Eqs.~(\ref{eq:EOM1}) and (\ref{eq:EOM3}) in Eq.~(\ref{eq:geodesic}) we obtain
    \begin{equation}
        g_{00}(r) \  g_{11}(r) \ \dot{r}^2 = E^2 - g_{00}(r)\left[1 + \left(\dfrac{L}{r}\right)^2\right]
    \end{equation}\
    
    \noindent where the second term of the right-hand side of the previous equation is the well-known effective potential
    \begin{equation}
        U^2_{eff} (r) \equiv  g_{00}(r)\left(1 + \dfrac{L^2}{r^2}\right).
    \end{equation}

%%%%%%%%%%%%%%%%%%%%%%%%%%%%%%%%%%%%%%%%%%%%%%%%%%%%%%%%%%%%%%%
%%%%%%%%%%%%%%%%%%%%%%%%%%%%%%%%%%%%%%%%%%%%%%%%%%%%%%%%%%%%%%%
\section{Projection of orbit onto the plane of sky}
%%%%%%%%%%%%%%%%%%%%%%%%%%%%%%%%%%%%%%%%%%%%%%%%%%%%%%%%%%%%%%%
%%%%%%%%%%%%%%%%%%%%%%%%%%%%%%%%%%%%%%%%%%%%%%%%%%%%%%%%%%%%%%%

\begin{figure}[H]%
        \centering%
        \includegraphics[width=0.8\hsize,clip]{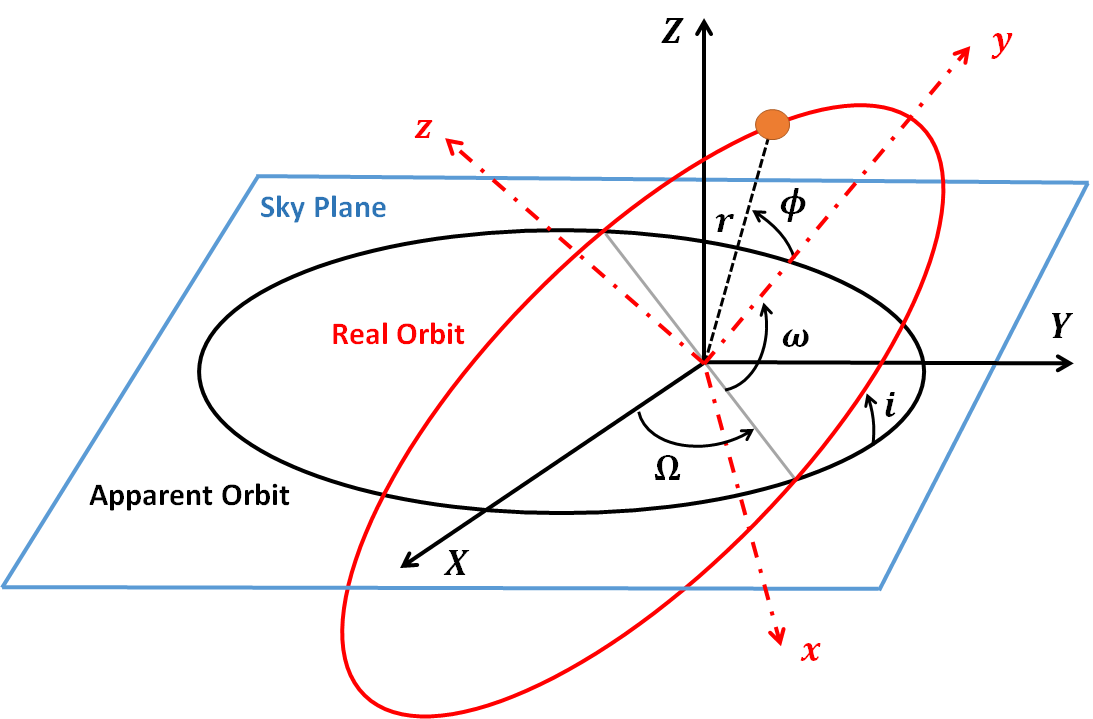}
        \caption{Projection of the real orbit onto the plane of the sky. The axes originate at Sgr A* (the focus of the ellipse). The picture illustrates the orbital parameters: $\phi$ {is the azimuth angle of the spherical system of coordinates associated with the $x$, $y$, $z$ Cartesian coordinates, i.e., for an elliptic motion in the $x$-$y$ plane, it is the true anomaly}, $i$ is the angle of inclination between the real orbit and the observation plane, $\Omega$ is the angle of the ascending node, and $\omega$ is the argument of pericenter. It is worth noting that the $Z$-axis of the coordinate system is defined by the vector pointing from the Solar System to the Galactic center. {Reproduced from \cite{2020A&A...641A..34B} with permission from Astronomy \& Astrophysics, Copyright ESO}.}\label{fig:orbital_parameters}%
\end{figure}

When a telescope measures the motion of a star, it does not measure the real dynamics but rather an apparent one, i.e., it measures the orbit and velocity data projected on the plane that lies perpendicular to the line of sight of the star. For this reason, to compare the theoretical orbit with the observational data, we must project the real orbit on the observation plane in the sky as shown in Fig.~\ref{fig:orbital_parameters}. This plane-of-sky is described in coordinates $(X, Y)$ defined by the observed angular positions (the declination $\delta$ and the right ascension $\alpha$), where $X=R_\odot(\delta-\delta_{\rm SgrA*})$ and $Y=R_\odot(\alpha-\alpha_{\rm SgrA*})$ being $R_\odot$ the distance to the Galactic center \cite{2008ApJ...689.1044G,2018ApJ...854...12C,2019Sci...365..664D}. According to the above description, the apparent theoretical orbit can then be obtained from:
    \begin{eqnarray}
        X &=& x \ \cos(X,x) + y \ \cos(X,y), \\
        Y &=& x \ \cos(Y,x) + y \ \cos(Y,y),
    \end{eqnarray}

\noindent    where the coordinates $(x, y)$ are determined in the plane of the orbit from $x=r\cos\phi$ and $y=r\sin\phi$, while the direction cosines can be found  by Eulerian rotation of axes, this is:
    
    \begin{equation*}
    \left(
    \begin{matrix}
    X \\
    Y \\
    Z \\
    \end{matrix}
    \right) = \left(
    \begin{matrix}
     \cos\Omega & -\sin\Omega & 0 \\
    \sin\Omega & \cos\Omega & 0 \\
    0 & 0 & 1 \\
    \end{matrix}
    \right) 
    \left(
    \begin{matrix}
     1 & 0 & 0 \\
     0 & \cos i & -\sin i \\
    0 & \sin i & \cos i \\
    \end{matrix}
    \right)
    \left(
    \begin{matrix}
    \cos\omega & -\sin\omega & 0 \\
    \sin\omega & \cos\omega & 0 \\
    0 & 0 & 1 \\
    \end{matrix}
    \right)
    \left(
    \begin{matrix}
    x \\
    y \\
    z \\
    \end{matrix}
    \right),
    \end{equation*} 

    \begin{footnotesize}
    \begin{equation*}
    \left(
    \begin{matrix}
     X \\
     Y \\
    Z \\
    \end{matrix}
    \right) = 
    \left(
    \begin{array}{ccc}
    \cos\omega\cos\Omega-\cos i\sin\omega\sin\Omega & -\sin\omega\cos\Omega -\cos i\cos\omega\sin\Omega & \sin i\sin\Omega \\
    \cos i\sin\omega\cos\Omega + \cos\omega\sin\Omega & \cos i\cos\omega\cos\Omega - \sin\omega\sin\Omega & -\sin i\cos\Omega \\
    \sin i\sin\omega & \sin i\cos\omega & \cos i \\
    \end{array}
    \right)
    \left(
    \begin{matrix}
    x \\
    y \\
    z \\
    \end{matrix}
    \right),
    \end{equation*} 
    \end{footnotesize}
\noindent being
    \begin{eqnarray}
    \cos(X,x) &=& \cos\Omega \cos\omega - \sin\Omega \sin\omega \cos i,\\
    \cos(X,y) &=& -\cos\Omega \sin\omega - \sin\Omega \cos\omega \cos i,\\
    \cos(Y,x) &=& \sin\Omega \cos\omega + \cos\Omega \sin\omega \cos i,\\
    \cos(Y,y) &=& -\sin\Omega \sin\omega + \cos\Omega \cos\omega \cos i,
    \end{eqnarray}

 \noindent  finally we find that orbit on plane-of-sky is given by:
   
    \begin{eqnarray}
        X &= & r \left[\cos(\phi + \omega) \cos\Omega - \sin(\phi + \omega) \sin\Omega \cos i\right], \\
        Y &= & r \left[\cos(\phi + \omega) \sin\Omega  + \sin(\phi + \omega) \cos\Omega \cos i\right].
    \end{eqnarray}

\

    Similarly to orbit, the star's radial velocity must also be projected onto the plane of the sky. The radial velocity on the observation plane is defined as the velocity in the observer's direction along the line of sight. Therefore, if we adopt the $X$-$Y$ plane as the observation plane, then the line of sight is in the direction of the $Z$-axis given by

    \begin{equation}
        Z \ = \ x \ \cos(Z,x) + y \ \cos(Z,y),
    \end{equation}

\noindent     where
    \begin{eqnarray}
    \cos(Z,x) &=& \sin\omega \sin i,\\
    \cos(Z,y) &=& \cos\omega \sin i,
    \end{eqnarray}

\noindent in terms of $\phi, \omega$ and $i$; $Z$ is given by

    \begin{equation}
        Z = r \sin(\phi+\omega) \sin i.
    \end{equation}\

    The radial velocity is defined as the velocity along the observer's line of sight, and since $Z$ is the direction along the observer's line of sight, the derivative of $Z$ relative to time gives us the apparent radial velocity of the star. This is

    \begin{equation}
    \frac{dZ}{dt} = u_Z = \left[r \dot{\phi} \ \cos(\phi + \omega) + \dot{r} \  \sin(\phi+\omega)\right] \sin(i).
    \end{equation}\

    It is worth noting that in the coordinate system $(X, Y, Z)$, the direction of $Z$-axis is defined by the vector pointing from the Solar System to the Galactic center and that the axes originate at Sgr A*, which is considered the focus of the orbit \cite{2017ApJ...837...30G,2018A&A...615L..15G,2019Sci...365..664D}.
    
%%%%%%%%%%%%%%%%%%%%%%%%%%%%%%%%%%%%%%%%%%
\begin{adjustwidth}{-\extralength}{0cm}

\reftitle{References}

\PublishersNote{}
\end{adjustwidth}
\end{document}